\documentclass[aps,reprint,twocolumns,longbibliography]{revtex4-1}
\usepackage{graphicx}
\usepackage{color}
\usepackage{upgreek}
\usepackage{graphics}
\usepackage{enumitem}
\usepackage{amsmath}
\usepackage{float}
\usepackage{siunitx}
\usepackage[english]{babel}

\begin{document}
\title{Quantitative Digital Microscopy with Deep Learning}
\author{Benjamin Midtvedt}

\author{Saga Helgadottir}

\author{Aykut Argun}

\author{Jes\'us Pineda}

\author{Daniel Midtvedt}

\author{Giovanni Volpe}

\affiliation{Department of Physics, University of Gothenburg, Origov\"agen 6B, SE-41296 Gothenburg, Sweden}

\date{\today}

\begin{abstract}
Video microscopy has a long history of providing insights and breakthroughs for a broad range of disciplines, from physics to biology.
Image analysis to extract quantitative information from video microscopy data has traditionally relied on algorithmic approaches, which are often difficult to implement, time consuming, and computationally expensive.
Recently, alternative data-driven approaches using deep learning have greatly improved quantitative digital microscopy, potentially offering automatized, accurate, and fast image analysis.
However, the combination of deep learning and video microscopy remains underutilized primarily due to the steep learning curve involved in developing custom deep-learning solutions. 
To overcome this issue, we introduce a software, DeepTrack 2.0, to design, train and validate deep-learning solutions for digital microscopy.
We use it to exemplify how deep learning can be employed for a broad range of applications, from particle localization, tracking and characterization to cell counting and classification.
Thanks to its user-friendly graphical interface, DeepTrack 2.0 can be easily customized for user-specific applications, and, thanks to its open-source object-oriented programming, it can be easily expanded to add features and functionalities, potentially introducing deep-learning-enhanced video microscopy to a far wider audience.
\end{abstract}

\maketitle

\section{Introduction}\label{ch:intro}

\begin{figure*}
    \centering
    \includegraphics[width=17.5cm]{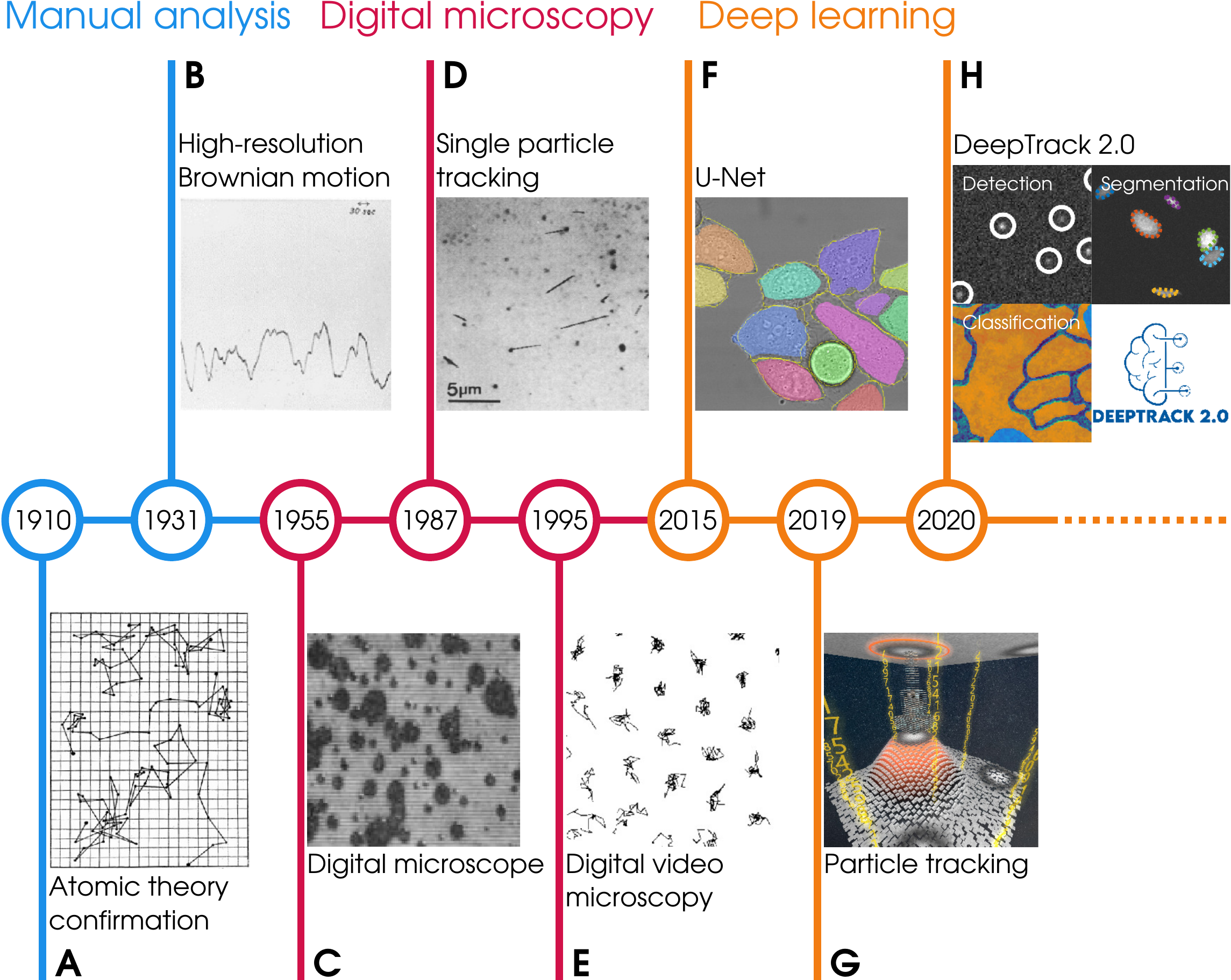}
    \caption{
    {\bf Brief history of quantitative microscopy and particle tracking.}
    {\bf A}-{\bf B} 1910--1950: The manual analysis era. 
    {\bf A} Examples of manually tracked trajectories of colloids in a suspension from Perrin's experiment that convinced the world of the existence of atoms \cite{Perrin1910MouvementMolecules}. The time resolution is 30 seconds.
    {\bf B} Kappler manually tracked the rotational Brownian motion of a suspended micromirror to determine the Avogadro number \cite{Kappler1931VersucheDrehwaage}.
    {\bf C}-{\bf E} 1951-2015: The digital microscopy era.
    {\bf C} Causley and Young developed a computerized microscope to count particles and cells using a flying-spot microscope and an analog analysis circuit  \cite{Causley1955CountingMicroscope}.
    {\bf D} Geerts \emph{et. al.} developed an automatized method to track single gold nanoparticles on the membranes of  living cells \cite{Geerts1987NanovidMicroscopy}.
    {\bf E} Crocker and Grier kickstarted modern particle tracking, achieving high accuracy using a largely setup-agnostic approach \cite{Crocker1996MethodsStudies}.
    {\bf F}-{\bf I} 2015-2020: The deep-learning-enhanced microscopy era.
    {\bf F} Ronnerberger \emph{et. al.} developed the U-Net, a variation of a convolutional neural network that is particularly suited for image segmentation and has been very successful for biomedical applications \cite{Ronneberger2015U-net:Segmentation}.
    {\bf G} Helgadottir \emph{et. al.} developed a software to track particles using  convolutional neural networks (DeepTrack 1.0) and demonstrated that it can achieve higher tracking accuracy than traditional algorithmic approaches \cite{Helgadottir2019DigitalLearning}.
    {\bf J} This article presents DeepTrack 2.0, which provides an integrated environment to design, train and validate deep learning solutions for quantitative digital microscopy.}
    \label{fig1}
\end{figure*}

During the last century, the quantitative analysis of microscopy images has provided important insights for various disciplines, ranging from physics to biology. 
An early example is the pioneering experiment performed by Jean Perrin in 1910 that demonstrated beyond any doubt the physical existence of atoms \cite{Perrin1910MouvementMolecules}: he  manually tracked the positions of microscopic colloidal particles in a solution by projecting their image on a sheet of paper (Fig.~\ref{fig1}A) and, despite a time resolution of just 30 seconds, he managed to quantify their Brownian motion and connect it to the atomic nature of matter.
In the following decades, several scientists followed in Perrin's footsteps, improving the time resolution of the experiment down to seconds 
\cite{Nordlund2017EineQuecksilberkugelchen, Kappler1931VersucheDrehwaage} (Fig.~\ref{fig1}B).
Despite these improvements, manual tracking of particles intrinsically limits the time resolution of conceivable experiments.

In the 1950s, analog electronics provided some tools to increase acquisition and analysis speed. According to Preston \cite{Preston1976DigitalStates}, the history of digital microscopy begins in Britain in 1951 with an unlikely actor: the British National Coal Committee convened to investigate ``the possibility of making a machine to replace the human observer'' to measure coal dust in mining operations \cite{Walton1952AutomaticParticles}. In 1955, Causley and Young developed a flying-spot microscope to count and size particles and cells \cite{Causley1955CountingMicroscope}:
The flying-spot microscope used a cathode-ray tube to scan a sample pixel by pixel, while the cells were counted and sized by a simple analog integrated circuit (Fig.~\ref{fig1}C). 
This device allowed over an order of magnitude faster counting than human operators, while maintaining the same accuracy.

During the 1950s and in earnest in the 1960s, researchers started employing digital computers to add speed and functionalities to microscopic image analysis, with a growing focus on biomedical applications. In 1965, Prewitt  and  Mendelsohn managed to distinguish cells in a blood smear by analyzing with a computer their images obtained by a flying-spot microscope and recorded as 8-bit values on a magnetic tape \cite{PrewittTHEIMAGES}.
In the following years, digital microscopy went from research labs to clinical settings with the development of the computerized tomography scanner (CT-scanner) in 1972  \cite{Hounsfield1973ComputerizedSystem} and of the automated flow cytometer in 1974 \cite{Mansberg1974TheSystem.}.

In soft matter physics, despite the early success of Perrin's experiment \cite{Perrin1910MouvementMolecules}, most studies focused on the ensemble behavior of colloidal particles employing methods such as selective photobleaching and image correlation \cite{Magde1972ThermodynamicSpectroscopy, Axelrod1976LateralFibers, Manzo2015AApproaches}. 
These methods can resolve fast dynamics, but they can only measure the average behavior of a homogeneous colloidal solution \cite{Manzo2015AApproaches}. 
To overcome these limitations, Geerts {\it et al.} automated particle tracking in 1987, developing what is now known as \emph{single particle tracking}, and used it to track individual gold nanoparticles on the surface of living cells from images acquired with a differential interference contrast (DIC) microscope \cite{Geerts1987NanovidMicroscopy} (Fig.~\ref{fig1}D).
In the following decades, researchers have also used fluorescent molecules \cite{Schmidt1996ImagingDiffusion, Schutz2000PropertiesMicroscopy., Alcor2009Single-particleDynamics} and quantum dots \cite{Dahan2003DiffusionTracking, Pinaud2010ProbingTime} as tracers within biological systems.

It became quickly evident that highly accurate tracking algorithms were needed to analyze the collected data. 
In 1996, Crocker and Grier proposed an algorithm to determine particle positions based on the measurement of the centroids of their images \cite{Crocker1996MethodsStudies} (Fig.~\ref{fig1}E).
The main advantage of this algorithm is that it is largely setup-agnostic, i.e., it does not depend on the specific properties of the imaging system and of the particle. 
Other setup-agnostic approaches have been proposed in more recent years analyzing, e.g., the Fourier transform of the particle image \cite{Yu2011FastPrecision}, or its radial symmetry \cite{Parthasarathy2012RapidCenters}. 
Other algorithms, instead, made a model of the image based on the properties of the imaging system and of the particle \cite{Thompson2002PreciseProbes, Ober2004LocalizationMicroscopy, Zhang2007GaussianModels, Abraham2009QuantitativeTechniques, Stallinga2010AccuracyMicroscopy,  Stallinga2012PositionModel, Manzo2015AApproaches}. 
These alternative methods were less general and often more computationally expensive, but they often achieved higher accuracy and could also provide quantitative information about the particle, such as its size \cite{Lee2007CharacterizingMicroscopy} or its out-of-plane position \cite{Holtzer2007NanometricCells, Deschout2014PreciselyMicroscopy, Godinez2015TrackingAssociation}.
Despite the large number of methods being introduced, digital video microscopy remained a hard problem, requiring the development of ad hoc algorithms tuned to the needs of each experiment.  
In fact, a 2014 comparison of 14 tracking methods found that, when compared on several different simulated scenarios, no single algorithm performed best in all scenarios \cite{Chenouard2014ObjectiveMethods}.

Only in the last few years has machine learning started to be employed for the analysis of images obtained from digital microscopy.
This comes in the wake of the deep learning revolution \cite{Lecun2015DeepLearning}, thanks to which computer-vision task such as image recognition \cite{Ciresan2012Multi-columnClassification}, semantic segmentation \cite{ Shelhamer2017FullySegmentation}, and image generation \cite{Li2016ConvolutionalGeneration}, are now automatized with relative ease.
Recent results have demonstrated the potential of applying deep learning to microscopy, vastly improving techniques for particle tracking \cite{Helgadottir2019DigitalLearning, Hannel2018Machine-learningParticles, Newby2018Convolutional3D}, cell segmentation and classification \cite{Ronneberger2015U-net:Segmentation, ChenDeepClassification, CoudrayClassification3, ZhangDeepPap:Classification, Falk2019}, particle characterization \cite{Midtvedt2020HolographicLearning, Altman2020CATCH:Networks, Hannel2018Machine-learningParticles}, object counting \cite{Xie2018MicroscopyNetworks}, depth-of-field extension \cite{Wu2018ExtendedRecovery}, and image resolution \cite{Nehme2018Deep-STORM:Learning, Ouyang2018DeepMicroscopy}. 
In 2015, Ronnerberger \emph{et. al.} developed a special kind of neural network (U-Net) for the segmentation of cell-images \cite{Ronneberger2015U-net:Segmentation} (Fig.~\ref{fig1}F), which is now widely used for the segmentation of biomedical images.
In particle tracking, Hannel \emph{et. al.} employed deep learning to track and measure colloids from their holographic images \cite{Hannel2018Machine-learningParticles}, Newby \emph{et. al.} demonstrated how deep learning can be used for the simultaneous tracking of multiple particles \cite{Newby2018Convolutional3D}, and Helgadottir \emph{et. al.} achieved tracking accuracy surpassing standard methods \cite{Helgadottir2019DigitalLearning} (Fig.~\ref{fig1}G).
These early successes clearly demonstrate the potential of deep learning to analyze microscopy data.
However, they also point to a key limiting factor for the development and deployment of deep-learning solutions to microscopy: the availability of high-quality training data.
In fact, training data often need to be experimentally acquired specifically for each application and, especially for biomedical applications, to be manually annotated by experts, which are expensive, time-consuming and potentially biased processes \cite{Xing2018DeepSurvey}.

In this article, we provide a brief review of the applications of deep learning to digital microscopy and we introduce a comprehensive software (DeepTrack 2.0, Fig.~\ref{fig1}H) to design, train and validate deep-learning solutions for quantitative digital microscopy.
In section~\ref{ch:review}, we review the main applications of deep learning to microscopy and the most frequently employed neural-network architectures.
In section~\ref{ch:deeptrack}, we introduce DeepTrack 2.0, which greatly expands the functionalities of the particle-tracking software DeepTrack 1.0 \cite{Helgadottir2019DigitalLearning}, and features a user-friendly graphical interface and a modular (object-oriented) architecture that can be easily expanded and customized for specific applications.
Finally, in section~\ref{ch:results}, we demonstrate the versatility and power of deep learning and DeepTrack 2.0 by using it to tackle a variety of physical and biological quantitative digital microscopy challenges, from particle localization, tracking and characterization to cell counting and classification.  

\section{Deep Learning for Microscopy}\label{ch:review}

In this section, we will start by providing an overview of machine learning and deep learning, in particular introducing and comparing the deep-learning models that are most commonly used in microscopy: \emph{fully connected neural networks}, \emph{convolutional neural networks}, \emph{convolutional encoder-decoders}, \emph{U-Nets}, and \emph{generative adversarial networks}.
Subsequently, we will review some key applications of deep learning in microscopy focusing on three key areas: image segmentation, image enhancement, and particle tracking.

Image segmentation partitions an image into multiple segments each corresponding to a specific object (e.g., cells of different kinds).
In this context, deep learning has been very successful, especially in the segmentation of biological and biomedical images. 
However, one limiting factor is the need for high-quality training datasets, which often need to be annotated by experts manually, a time-consuming and tedious task.

Image enhancement includes tasks such as noise reduction, deaberration, refocusing and superresolution.
Also in this case, deep learning has been widely employed in the last few years, especially in the context of computational microscopy.
Differently from image segmentation, image enhancement can often utilize training datasets that are directly acquired from experiments.

Particle tracking deals with the localization of objects (often microscopic colloidal particles or tracer molecules) in 2D or 3D. Deep-learning-powered solutions are more accurate than algorithmic approaches, work in extremely difficult environments with poor signal-to-noise ratio, and can extract quantitative information about the particles.
Particle-tracking algorithms can often be trained using simulated data by employing physical simulations of the required images.

\subsection{Deep learning}

In contrast to standard computer algorithms where the user is required to define explicit rules to process the data, machine-learning algorithms can learn patterns and rules to perform specific tasks directly from series of data. 
In supervised learning, machine-learning algorithms learn by adjusting their behavior according to a set of input data and corresponding desired outputs (the ground truth). 
These input--output pairs constitute the training dataset, which can be obtained either from experiments or from simulations.

Deep learning is a kind of machine learning built on artificial neural networks (ANN) \cite{Lecun2015DeepLearning}.
ANNs were originally conceived to emulate the capabilities of the brain, specifically its ability to learn \cite{mehlig2019artificial}.
They are constituted by interconnected artificial neurons (simple computing units often just returning a non-linear function of their inputs).
Often, these artificial neurons are organized in layers (typically with tens or hundreds of artificial neurons).
In the most commonly employed architectures, each layer receives the output of the previous layer, computes some transformation, and feeds the result into the next layer.
In many machine vision applications, the number of layers is in the tens (this number is the ``depth'' of the ANN; hence the term ``deep learning'').

The weights of the connections between artificial neurons and layers are the parameters that are adjusted in the training process.
The training can be broken down into the following steps (referred to as the \emph{error back-propagation algorithm} \cite{rumelhart1986learning}):
First, the ANN receives an input and calculates a predicted output based on its current weights.
Second, the output is compared to the true, desired output, and the error is measured using a \emph{loss function}.
Third, the ANN propagates this error backwards, calculating for each weight whether it should be increased or decreased in order to reduce the error (and a local estimate of the rate of change of the error depending on that weight).
Finally, the weights are updated using an \emph{optimiser}, which determines how much each weight should be changed.
By feeding the network additional training data, it typically improves its performance, gradually converging to some optimum weight configuration.

In microscopy applications, the most commonly employed ANN architectures are \emph{dense neural networks}, \emph{convolutional neural networks}, \emph{convolutional encoder-decoders}, \emph{U-Nets}, and \emph{generative adversarial networks} (Table~\ref{tab1}). 

\begin{table*}
    \begin{tabular}{p{6cm}p{6cm}p{6cm}}
        \begin{center}
            {\bf Architecture} 
        \end{center}
    & 
        \begin{center}
            {\bf Advantages}
        \end{center}
    & 
        \begin{center}
            {\bf Disadvantages}
        \end{center}
    \\
    \hline
        \begin{center}
            {\bf Dense Neural Network (DNN)}
            \includegraphics[width=6cm]{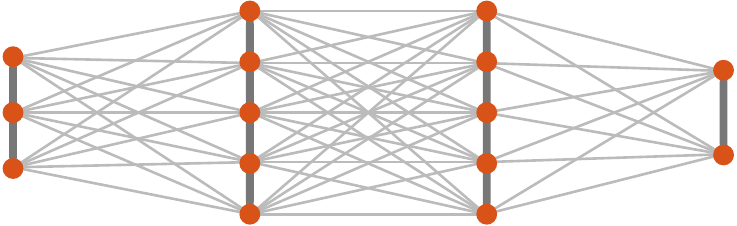}
        \end{center}
    &
        \begin{itemize}
        \item Can use all available information.
        \item Can represent any transformation between the input and output.
        \item Input and output can easily have any dimensions.
        \end{itemize}
    &
        \begin{itemize}
        \item The number of weights increases quickly with the number of layers and the dimension of the input.
        \item The input and output dimensions must be known in advance.
        \end{itemize}
    \\
    \hline
        \begin{center}
            {\bf Convolutional Neural Network (CNN)}
            \includegraphics[width=5cm]{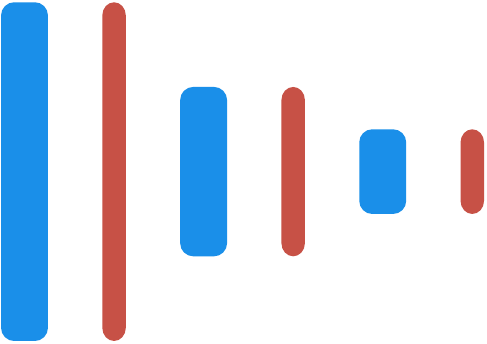}
        \end{center}
    &
        \begin{itemize}
        \item Can be constructed with a limited number of weights.
        \item Highly effective at extracting local information from images.  
        \item Analysis is position-independent.
        \end{itemize}
    &
        \begin{itemize}
        \item Cannot access global information.
        \item Can be computationally expensive.
        \item Difficult to retain an exact output shape.
        \end{itemize}     
    \\
    \hline
        \begin{center}
            {\bf Convolutional Encoder-Decoder (CED)}
            \includegraphics[width=6cm]{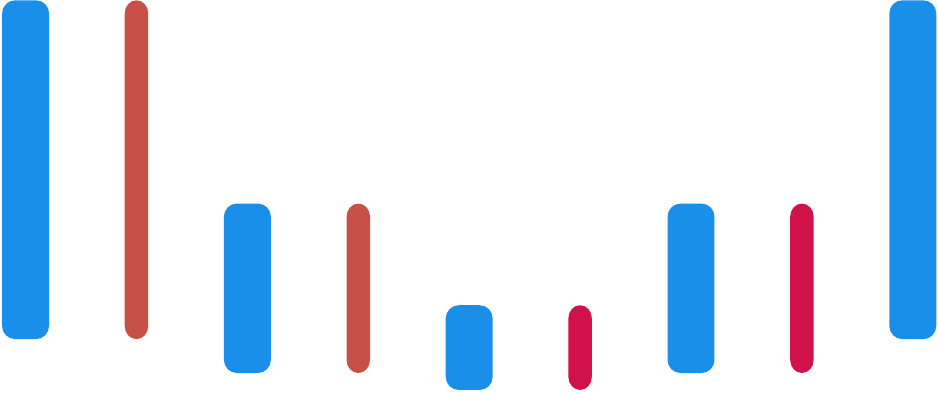}
        \end{center}
    &
        \begin{itemize}
        \item Only the number of features needs to be known in advance.
        \item Returns an image in output, which can be more interpretable by humans.
        \item Can be trained as an auto-encoder without annotated data.
        \end{itemize} 
    &
        \begin{itemize}
        \item Positional information is lost during downsampling.
        \item Can be hard to annotate data.
        \end{itemize} 
    \\
    \hline
        \begin{center}
            {\bf U-Net}
            \includegraphics[width=6cm]{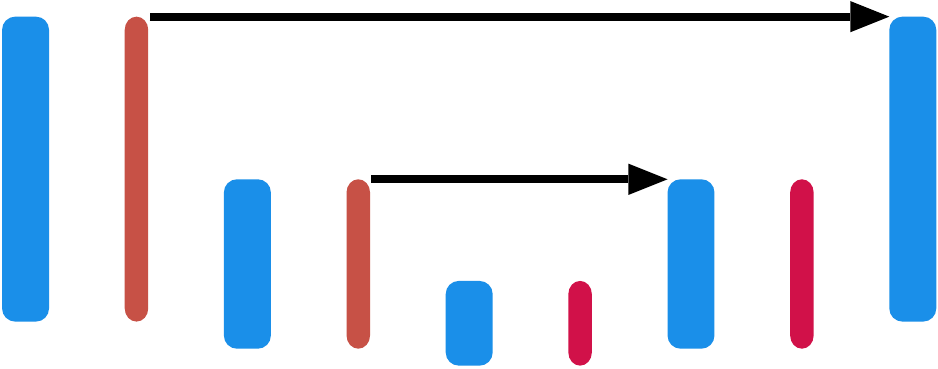}
        \end{center}
    &
        \begin{itemize}
        \item Retains positional information.
        \item Only the number of features needs to be known in advance.
        \item Returns an image in output, which can be more interpretable by humans.
        \end{itemize} 
    &
        \begin{itemize}
        \item Can quickly grow large.
        \item Forward concatenation layers disallow use as an auto-encoder.
        \end{itemize} 
    \\
    \hline
        \begin{center}
            {\bf Generative Adversarial Network (GAN)}
            \includegraphics[width=6cm]{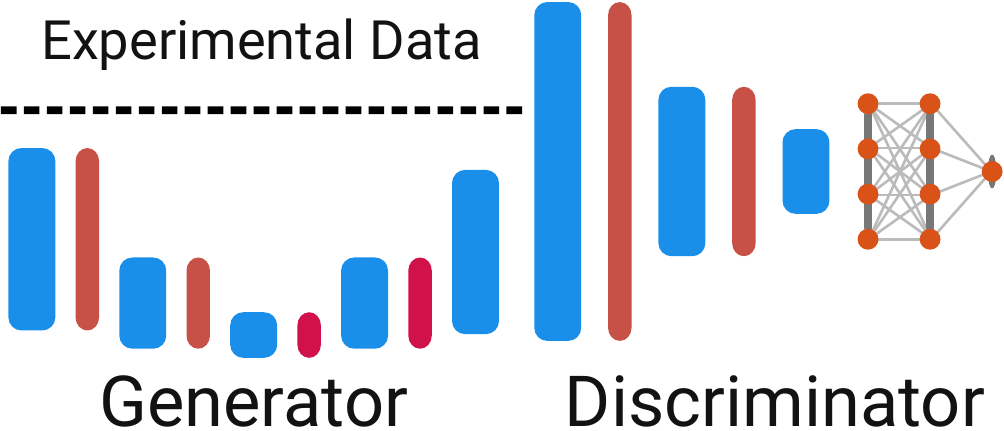}
        \end{center}
    &
        \begin{itemize}
        \item Can create very realistic images.
        \item Encourages high-frequency predictions.
        \item Can be trained without annotated data.
        \end{itemize} 
    &
        \begin{itemize}
        \item Very hard to train.
        \item The outputs are designed to \emph{look} correct, not to \emph{be} correct.
        \item Output quality very sensitive to the details of the architecture.
        \end{itemize} 
    \\
    \hline    
    
    \end{tabular}
    
    \caption{{\bf A comparison of common deep learning architectures.} 
    Advantages and disadvantages of deep learning architectures commonly employed for microscopy, i.e., the \textit{dense neural network} (DNN), the \textit{convolutional neural network} (CNN),
    the \textit{convolutional encoder-decoder} (CED), the \textit{U-Net}, and the \textit{generative adversarial network} (GAN). For each model, we also show a miniature example of the architecture, where gray lines with orange circles represent dense layers, blue rectangles represent convolutional layers, red rectangles represent pooling layers, and magenta rectangles represent deconvolutional layers. The arrows depict the forward concatenation steps.}
    \label{tab1}
\end{table*}

The workhorse of ANNs are \emph{dense neural networks} (DNNs), which consist of a series of layers fully connected in sequence.
While sufficiently large DNNs can approximate any function \cite{Cybenko1989ApproximationFunction}, the number of weights required quickly grows to unmanageable levels, especially for large inputs such as images. 
Furthermore, they present a rigid structure, where the dimensions of both the input and output are fixed.
Therefore, when analyzing images, they are rarely used on their own, while they are often employed as the final steps of some other network to generate the final output from already pre-processed data.

In contrast, \emph{convolutional neural networks} (CNNs) are particularly useful to analyze images.
They are primarily built upon convolutional layers.
In each convolutional layer, a series of 2D filters are convolved with the input image, producing as output a series of \emph{feature maps}. 
The size of the filters with respect to the input image determines the features that can be detected in each layer. 
To gradually detect larger features, the feature maps are downsampled after each convolutional layer. 
The downsampled feature maps are then fed as input to the next network layer. 
There is often a dense top after the last convolutional layer, i.e., a relatively small DNN that integrates the information contained in the output feature maps of the last layer to determine the sought-after result.

\emph{Convolutional encoder-decoders} are convolutional neural networks constituted by two paths.
First, there is the \emph{encoder path}, which reduces the dimensionality of the input through a series of convolutional layers and downsampling layers, therefore encoding the information about the original image.
Then, there is the \emph{decoder path}, which uses the encoded information to reconstruct either the original image or some transformed version of it (e.g., in segmentation tasks). 
Therefore, when trained to reconstruct the input image at the output, the information at the end of the encoder path can serve as a compressed version of the input image.
When trained to reconstruct a transformed version of the input image, the encoded information can serve as a powerful representation of the input image useful for the specific task at hand.

\emph{U-Nets} are an especially useful evolution of convolutional encoder-decoders. 
In addition to the encoder and decoder convolutional paths, U-Nets feature also forward concatenation steps between corresponding levels of these two paths.
This permits them to preserve positional information lost when the image resolution is reduced.
They have been particularly successful in analyzing and segmenting biological and biomedical images.

Differently from the previous case, \emph{generative adversarial networks} (GANs) combine two networks, a \emph{generator} and a \emph{discriminator}, regardless of their specific architectures \cite{goodfellow2014generative}. 
The generator manufactures data, usually images, from some input. 
The discriminator, in turn, classifies its input as either real data or synthetic data created by the generator. 
The term \emph{adversarial} refers to the fact that these two networks compete against each other: The generator tries to fool the discriminator with manufactured data, while the discriminator tries to expose the generator.
The generator can be trained to either transform images by feeding it with a real image as input or to make up images by feeding it with a random input. 
The generator is typically either a convolutional encoder-decoder or a U-Net, while the discriminator is often a convolutional neural network. 
While GANs are a breakthrough for data generation and offer many benefits, they are difficult to train and highly sensitive to hyperparameter tuning: Slight changes in their overall architecture can lead to vanishing gradients, lack of convergence, and uncorrelated generator loss and image quality \cite{yadav2017stabilizing,foster2019generative}.

\subsection{Image segmentation}

\begin{figure*}
    \centering
    \includegraphics[width=17cm, clip, trim={0 18cm 0 0}]{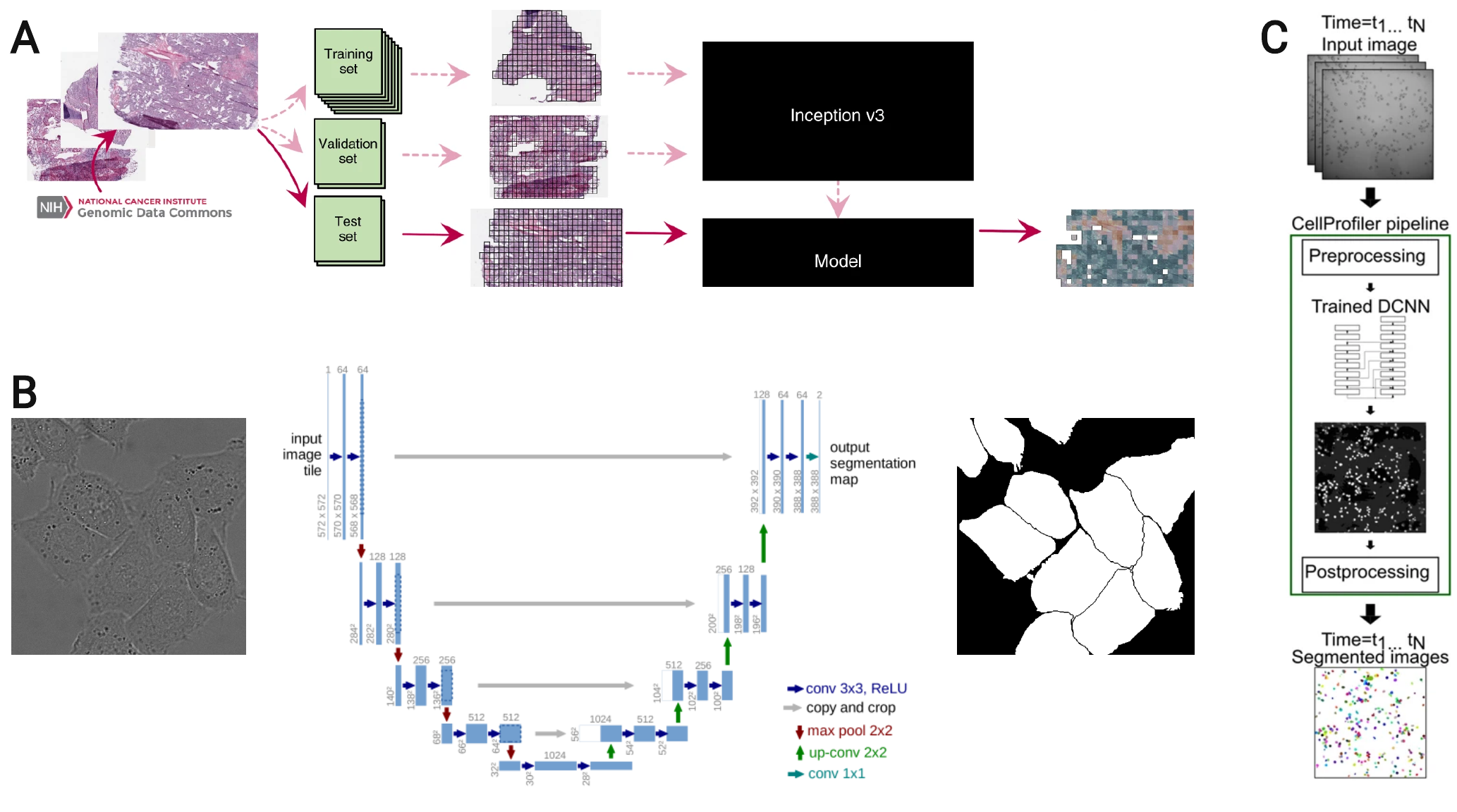}
    \caption{
    {\bf Image segmentation with deep learning.}
    {\bf A} Lung cell classification and mutation prediction \cite{CoudrayClassification3}, using Inception v3 \cite{SzegedyRethinkingVision}: Non-overlapping tiles in the input are analyzed, returning a low-resolution segmentation mask, containing either just a binary classification as tumor or healthy, or a complete prediction of the mutation type.
    {\bf B} The U-Net architecture as originally proposed by Ronneberger \cite{Ronneberger2015U-net:Segmentation} differs from the CED by the addition of forward concatenation steps between the encoder and decoder parts, which allow the network to forward positional information lost during encoding. The network is used to segment nearly overlapping cells.
    {\bf C} A cell segmentation software, based on a model closely resembling the U-Net \cite{Sadanandan2017AutomatedSegmentation}, can be automatically retrained by feeding it additional fluorescence images. 
    }
    \label{fig2}
\end{figure*}

Deep learning has been extremely successful at segmentation tasks, especially for biomedical applications \cite{Al-Kofahi2018AImages, Sadanandan2017AutomatedSegmentation, Song2017AccurateImages, Akram2016CellAnalysis, Arbelle2017MicroscopyNetworks, Hatipoglu2017CellRelationships, Arbelle2018MicroscopyNetworks, Lugagne2020DeLTA:Learning, Raza2017MIMO-Net:Images, Falk2019}, but also in material science \cite{Ma2018DeepImages, Azimi2018AdvancedMethods}.
Image segmentation is typically used to locate objects and boundaries in images. 
More precisely, image segmentation assigns a label to every pixel in an image such that pixels with the same label share certain characteristics (e.g., represent objects of the same type).

Generally, deep-learning models performing image segmentation are trained using experimental data that need to be manually annotated by experts.
In some cases, to alleviate the need for annotated images, pretrained neural networks are employed (i.e., neural networks that have been trained for classification tasks on a large dataset of different images, often not directly related to the task at hand) and fine-tuned using a relatively small set of manually annotated data \cite{Falk2019, Sadanandan2017AutomatedSegmentation}. 

If the exact topography of the sample is not needed, one can downsample the image using several convolutional layers, obtaining a coarse classification of its various regions.
For example, this approach was used by Coudray \emph{et. al.} \cite{CoudrayClassification3} to distinguish cancerous lung cells from normal tissue by fine-tuning a pre-trained neural network for image analysis and object detection (Inception v3 \cite{SzegedyRethinkingVision}) (Fig.~\ref{fig2}A). 

Much more frequently, image segmentation uses convolutional encoder-decoders and U-Nets \cite{lateef2019survey}.
In fact, U-Nets were originally developed for cell segmentation, where one of the key requirements is to clearly mark the cell edges, such that neighboring cells can be distinguished \cite{Ronneberger2015U-net:Segmentation} (Fig.~\ref{fig2}B).

High-quality image segmentation annotations are time-consuming to obtain, which is why many researches opt to design networks that can be retrained for a specific task using a much smaller dataset. For example, Sadanandan \emph{et. al.} developed a neural network that can be automatically be retrained using fluorescently labeled cells \cite{Sadanandan2017AutomatedSegmentation} (Fig.~\ref{fig2}C).
With such an approach, the neural network can relatively easily be adapted to different experimental setups, even though the process requires some experimental effort in acquiring the additional training data.

Segmentation has also been used for three-dimensional images.
For example, Li \emph{et. al.} used a three-dimensional convolutional neural network to reconstruct the interconnections between biological neurons \cite{Li2017DeepReconstruction}.
Similar approaches have been employed also for the volumetric reconstruction of organs  \cite{Cicek20163DAnnotation, Kleesiek2016DeepStripping}.

\subsection{Image enhancement}

\begin{figure*}
    \centering
    \includegraphics[width=16cm, clip, trim={0 18cm 1cm 0}]{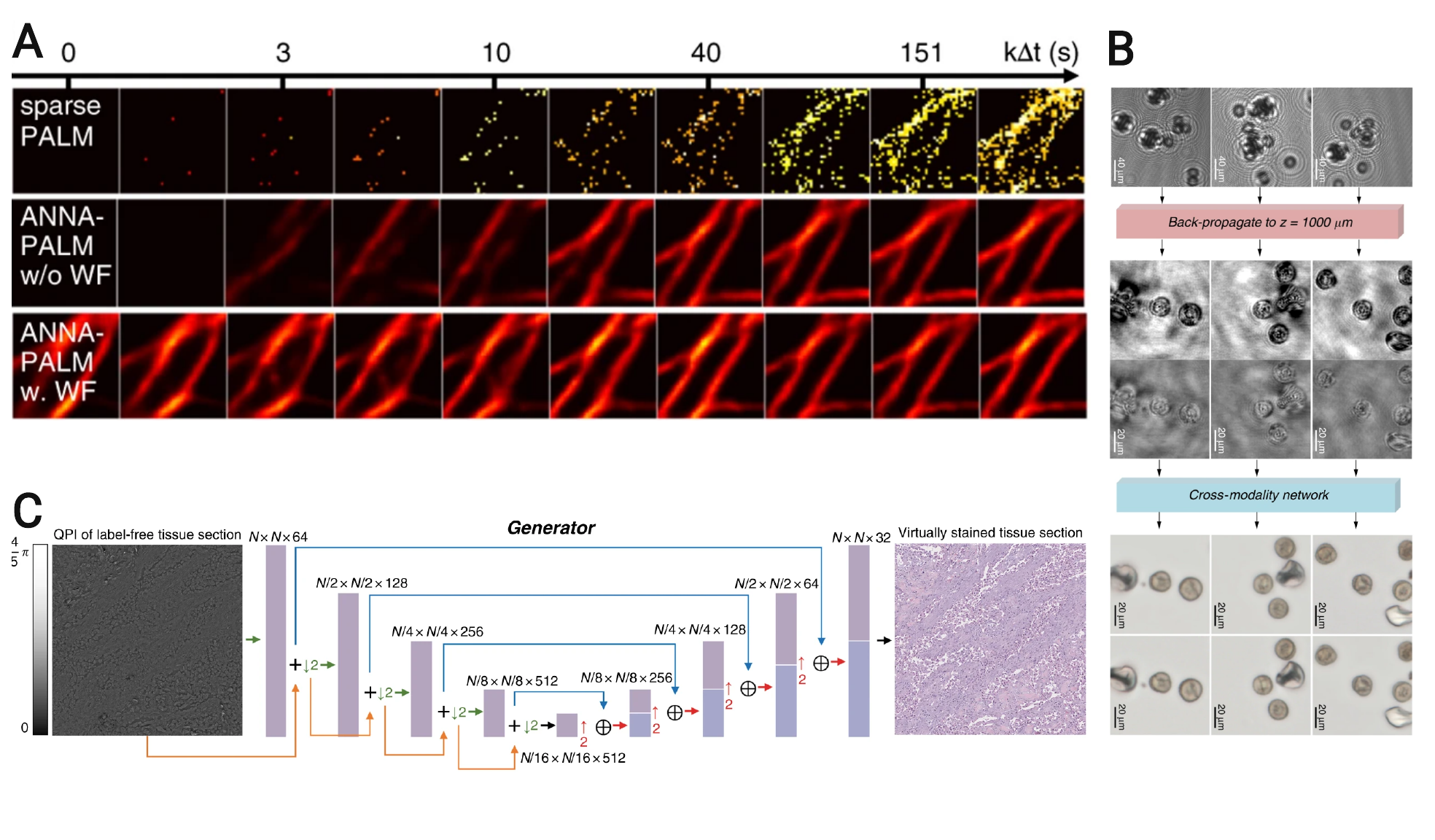}
    \caption{
    {\bf Image enhancement with deep learning.}
    {\bf A} Fluorescence superresolution localization microscopy using deep learning \cite{Ouyang2018DeepMicroscopy}: It uses sparse PALM \cite{Betzig2006ImagingResolution} (optionally together with a widefield (WF) image), to construct a super-resolved image. We see how the quality of the produced image increases with the acquisition time.
    {\bf B} A GAN is used to transform holography images to brightfield \cite{Wu2019}. From the top to bottom, we see holographic images of pollen, backpropagated to the focal plane, and finally transformed into brightfield images. The bottom set of images are the real brightfield images for comparison.
    {\bf C} A GAN is used to convert quantitative phase images (in-line holography) into virtual tissue stainings, mimicking histologically stained brightfield images \cite{Rivenson2019PhaseStain:Learning}.
    }
    \label{fig3}
\end{figure*}

Deep learning has been widely employed for image enhancement. 
This is particularly interesting because it permits also to perform tasks on images that would be extremely difficult or impossible to do with microscopy because of intrinsic physical limitations.

Deep learning has been employed to achieve super-resolution by using diffraction-limited images to reconstruct images beyond the diffraction limit.
For example, Ouyang \emph{et. al.} trained a GAN to imitate the output of the standard super-resolution method PALM \cite{Betzig2006ImagingResolution}  significantly improving the resolution of fluorescence images \cite{Ouyang2018DeepMicroscopy} (Fig.~\ref{fig3}A).

An interesting application of deep learning is to realize cross-modality analysis, where a neural network learns how to translate the output of a certain optical device to that of another.
For example, Wu \emph{et. al.} used a U-Net to translate between holography and brightfield microscopy, enabling volumetric imaging without the speckle and artifacts associated with holography \cite{Wu2019} (Fig.~\ref{fig3}B).
This method uses experimental pairs of images collected simultaneously by two different optical devices.

Going one step further, deep learning can also be used to generate images that cannot be directly obtained from the sample using optical devices, but would require some other kind of analysis of the sample.
For example, Rivenson \emph{et al.} used phase information obtained from holography to create a virtually stained sample corresponding to a histologically stained brightfield image \cite{Rivenson2019PhaseStain:Learning} (Fig.~\ref{fig3}C). 

Image-enhancement techniques typically train networks using experimental images that do not need any manual annotation. 
Either the target is calculated using known methods, or it is collected simultaneously using an alternate path for the light. 
While this reduces the amount of manual labor required, both approaches have their drawbacks. 
A network trained to imitate a traditional method is unlikely to improve upon it on primary metrics. Instead, it can improve by allowing less ideal inputs, or decrease execution time. On the other hand, using a dual microscope will lead to networks specialized for the optical devices used to acquire the training images. Moreover, such a dual-purpose microscope is usually non-standard, so the user need to alter and customize their setup. 

\subsection{Particle tracking}

\begin{figure*}
    \centering
    \includegraphics[width=17cm]{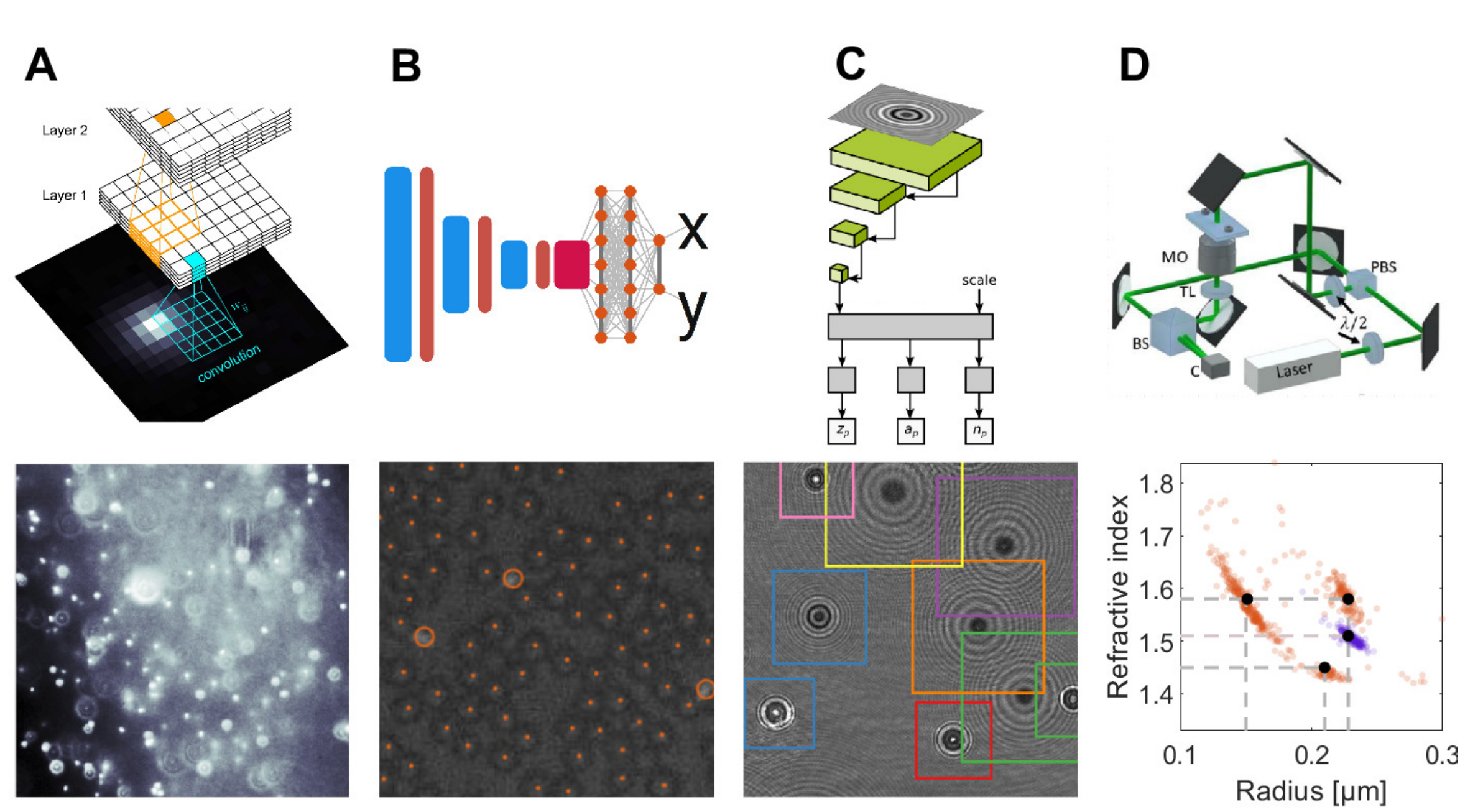}
    \caption{
    {\bf Particle tracking with deep learning.}
    {\bf A} Particle detection in dense images of varying diffraction patterns \cite{Newby2018Convolutional3D}, where a relatively small network of three convolutional layers estimates a pixel-by-pixel probability map of background versus particle.
    {\bf B} High-accuracy single particle localization using a CNN with a dense top \cite{Helgadottir2019DigitalLearning}. The network is scanned across the image to detect and localize all particles (dots) and bacteria (circles) in the image.
    {\bf C} Particle tracking and characterization in terms of radius and refractive index using in-line holography images \cite{Altman2020CATCH:Networks}, where bounding boxes for each particle in the field of view are extracted and fed to a CNN. They showcase accurate measurements on data for particles between \SI{0.5}{\micro\meter} and \SI{1.5}{\micro\meter}. 
    {\bf D} Particle characterization in terms of radius and refractive index using off-axis holography images \cite{Midtvedt2020HolographicLearning}. A CNN with latent space temporal averaging is used to measure multiple observations of a single particle to improve accuracy. This allows characterization of particles down to around \SI{0.2}{\micro\meter}.
    }
    \label{fig4}
\end{figure*}

Single particle tracking has become a crucial tool for probing the microscopic world. Standard approaches are typically limited by the complexity of the system: Higher particle densities, higher levels of noise, and more complex point-spread functions often lead to worse results. Developments using deep learning have shown that it is possible to largely overcome these limitations.
A big advantage of deep-learning solutions for particle tracking is that often simulated data can be used to train the networks. 

Newby \emph{et. al.} demonstrated that deep learning can be used for the detection of particles in high-density, low-signal-to-noise-ratio images \cite{Newby2018Convolutional3D}. 
Their method uses a small CNN to construct a pixel-by-pixel classification probability map of background versus particle (Fig.~\ref{fig4}A).
Standard algorithms can then be applied to this probability map to track the particles.

Helgadottir \emph{et. al.} achieved a tracking accuracy surpassing that of traditional methods using a convolutional neural network with a dense top to detect particle centroids in challenging conditions \cite{Helgadottir2019DigitalLearning} (Fig.~\ref{fig4}B).

Along with particle localization, deep learning has also been used to measure other characteristics of particles. 
For example, Altman \emph{et. al.} used a convolutional neural network to measure the radius and refractive index of images of colloids acquired by an in-line holographic microscope \cite{Altman2020CATCH:Networks} (Fig.~\ref{fig4}C). 
Midtvedt \emph{et al.} used an off-axis microscope and a time-average convolutional neural network to measure the radius and refractive index of even smaller particles \cite{Midtvedt2020HolographicLearning} (Fig.~\ref{fig4}D).

Moreover, deep learning has been used for micro-tubule tracking \cite{MasoudiInstance-LevelTracking}, 3d tracking of fluorescence images \cite{Zelger2018Three-dimensionalLearning, Franchini2020CutSetups}, intra-cellular particle detection \cite{Wollmann2019Detnet:Images, Ritter2020DeepImages}, nanoparticle sizing in transmission electron microscopy (TEM) \cite{Oktay2019AutomaticImages}, frame-to-frame linking \cite{Spilger2020ADetections}, and single-particle anomalous diffusion characterization \cite{Granik2019Single-ParticleLearning, Bo2019MeasurementNetworks, Kowalek2019ClassificationApproach}.

\section{DeepTrack 2.0}\label{ch:deeptrack}

In this section, we introduce DeepTrack 2.0, which is an integrated software environment to design, train, and validate deep-learning solutions for digital microscopy \cite{deeptrackgithub}.
DeepTrack 2.0 builds on the particle-tracking software package DeepTrack, which we introduced in 2019 \cite{Helgadottir2019DigitalLearning}, and greatly expands it beyond particle tracking towards a whole new range of quantitative microscopy applications, such as classification, segmentation, and cell counting.

To accommodate users with any level of experience in programming and deep learning, we provide access to the software through several channels, from a high-level graphical user interface that can be used without any programming knowledge, to scripts that can be adapted for specific applications, to a low-level set of abstract classes to implement new functionalities.
Furthermore, we provide various tutorials to use the software at each level of complexity, including several video tutorials to guide the user through each step of a deep-learning analysis for microscopy: from defining the training image generation routine, to deciding the neural network model, to training and validating the network, to applying the trained network to real data.

As main entry point, we provide a completely stand-alone graphical user interface, which delivers all the power of DeepTrack 2.0 without requiring programming knowledge.
This is available both for Windows and for MacOS \cite{appgithub}.
In fact, we recommend all users to start with the graphical user interface, which provides a visual approach to deep learning and an intuitive feel for how the various software components interact. 

As more precise control is desired, we recommend the users to peruse the available Jupyter notebooks \cite{deeptrackgithub}, which provide complete examples of how to write scripts for DeepTrack 2.0. 

For most applications, DeepTrack 2.0 already includes all necessary components.
However, if more advanced functionalities that are not already included are required, it is easy to extend DeepTrack 2.0 building on its framework of abstract objects and its native communication with the popular deep learning package Keras \cite{chollet2015keras}.
In fact, we expect the most advanced users to expand the functionalities of DeepTrack 2.0 according to their needs. 

All users are also welcome to report any bugs and to propose additions to DeepTrack 2.0 through its GitHub page \cite{deeptrackgithub}.

\subsection{Graphical user interface}

\begin{figure*}
    \centering
    \includegraphics[width=17cm]{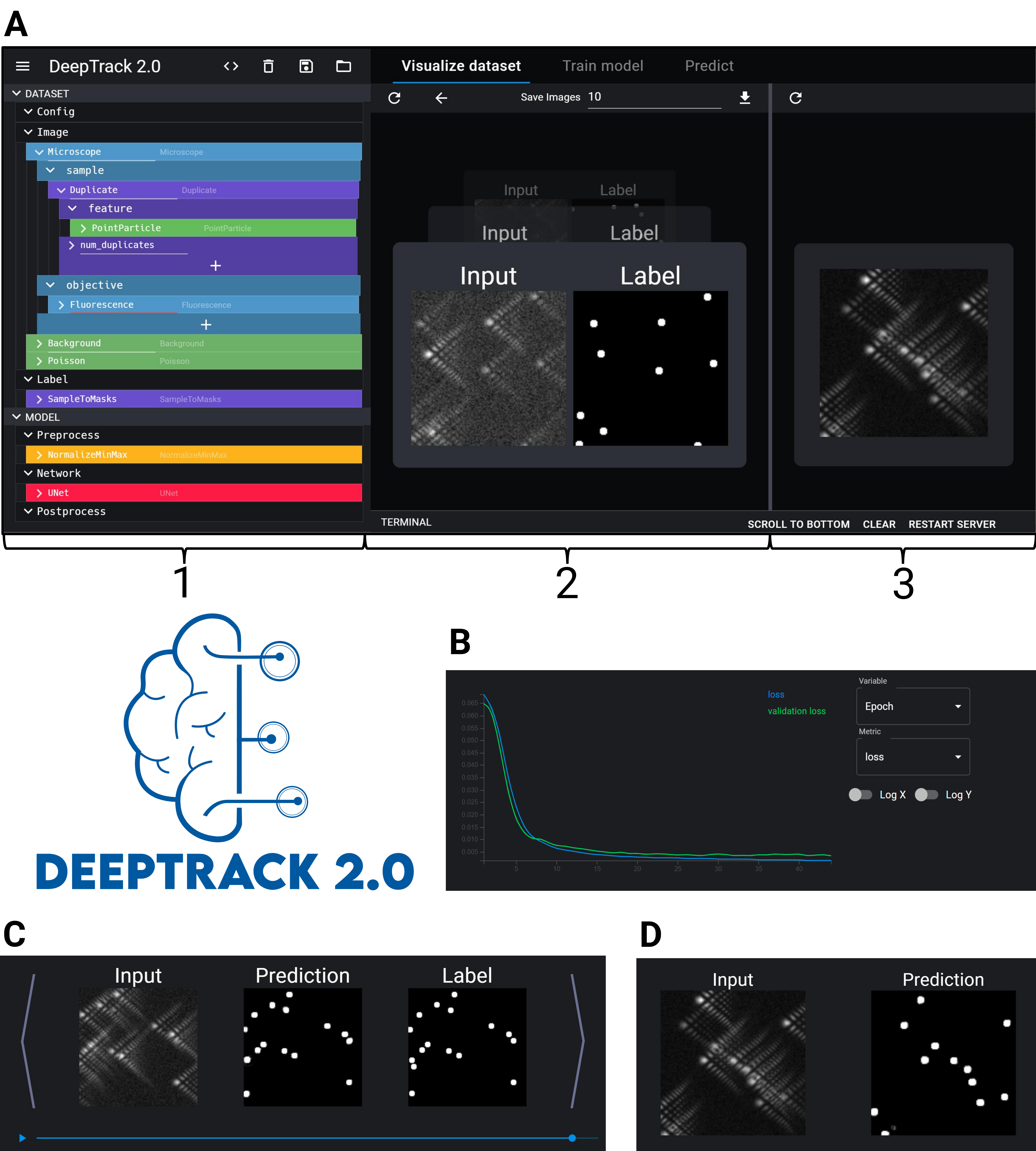}
    \caption{
    {\bf DeepTrack 2.0 graphical user interface.}
    {\bf A} The main interface: 1: The image generation pipeline is defined using drag and drop components. 2: An image created using the pipeline and the corresponding label are shown. 3: A comparison image is also shown to help ensure that the generated image is similar to experimental images.
    {\bf B} The training loss and validation loss over time can be monitored in real time during training. It is also possible to monitor custom metrics, or any metric as a function of some property of the image (e.g., particle size, signal-to-noise ratio, aberration strength).
    {\bf C} The model prediction on individual images in the validation set can be compared to the corresponding target in real time during training, providing another way to concretely visualize the improvement of the model performance over time.
    {\bf D} Finally, the model can be evaluated on experimental images also during training, which can help quickly hone in on a model that correctly handles specific experimental data.
    }
    \label{fig5}
\end{figure*}

The graphical user interface of DeepTrack 2.0 provides an intuitive way to perform the various steps that are necessary for the realization of a deep-learning analysis for microscopy.
Through the graphical user interface, users can define and visualize image generation pipelines (Fig.~\ref{fig5}A), train models (Fig.~\ref{fig5}B-C), and analyze experimental data (Fig.~\ref{fig5}D).

A typical workflow is the following:
\begin{enumerate}
    \item Define the image generation pipeline, e.g., a pipeline to generate images of a particle corrupted by noise.
    \item Define the ground-truth training target, e.g., the particle image without noise (image target), or the particle position (numeric target).
    \item Define a deep learning model, e.g., U-Net.
    \item Train and evaluate the deep learning model.
    \item Apply the deep learning model to the user's experimental data.
\end{enumerate}

Projects realized with DeepTrack 2.0 can be saved and subsequently loaded, which is useful for archival purposes as well as to share deep learning models and results between users and platforms. 
Furthermore, projects can be automatically exported as Python code, which can then be executed to train a network command-line or imported into an existing project.

At a more advanced level, it is possible to extend the capabilities of the graphical user interface by adding new Python-coded objects.
We envision that this possibility will motivate users to create and share additional software compatible with DeepTrack 2.0.

\subsection{Scripts}

We provide several Jupyter notebooks both as examples of how to write scripts using DeepTrack 2.0 and as a foundation to create customized solutions.
To facilitate this, we provide several video tutorials \cite{deeptrackgithub}, which detail how the solutions are constructed and how they can be modified.

\subsection{Code}

\begin{figure*}
    \centering
    \includegraphics[width=17cm]{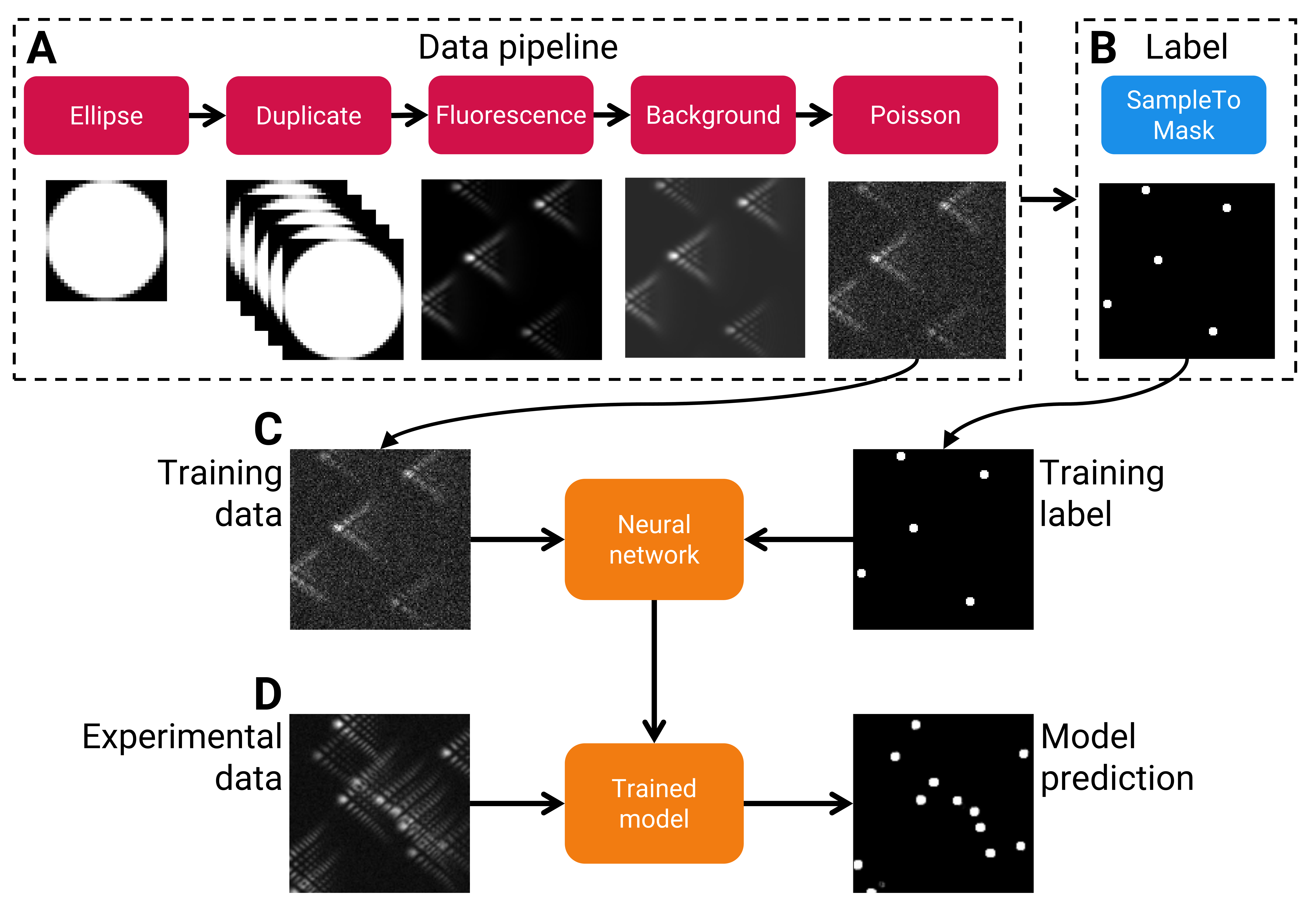}
    \caption{
    {\bf DeepTrack 2.0 framework.}
    \textbf{A} An example image-generation pipeline composed of five features. Five distinct ellipses are generated and passed to a fluorescence microscope simulator, which produces an image to which a constant background and Poisson noise are added.
    \textbf{B} The position of the ellipses are imprinted on the image, allowing us to create a ground-truth label, where each particle is represented by a small circle.
    \textbf{C} A generator that can continuously create images and corresponding labels is used to train a deep-learning model. Typically many thousands of images are created to train the neural network.
    \textbf{D} The trained model is then able to analyze experimental data. (Image shown here is not experimental, and only used for demonstration purposes.)
    }
    \label{fig6}
\end{figure*}

The software architecture of DeepTrack 2.0 (Fig.~\ref{fig6}) is built on four main components: features, properties, images, and deep-learning models:
\begin{description}
    \item[Features] They are the foundations on which DeepTrack 2.0 is built. They receive a list of images as input, and either apply some transformation to all of them (e.g., adding noise), or add a new image to the list (e.g., adding a scatterer), or merge them into a single image (e.g., imaging a list of scatterers through an optical device). By defining a set of features and how they connect, we produce a single feature that defines the entire image creation process.
    \item[Properties] They are the parameters that determine how features operate. For example, a property can control the position of a particle, the intensity of the added noise, or the amount by which an image is rotated. They can have a constant value or be defined by a function (e.g., to place a particle at a random position), which can depend on the value of other properties, either from the same feature or from other features.
    \item[Images] They are the objects on which the features operate. They behave like n-dimensional NumPy arrays (\texttt{ndarray}) and can therefore be directly used with most Python packages. They contain a list of the property values used by the features that created the image, which can be used to generate ground-truth labels for training neural networks as well as to evaluate how the error in deep-learning models depends on the properties defining the image (e.g., signal-to-noise ratio, background gradient, illumination wavelength).
    \item[Models] They are the deep-learning models. A series of standard models is already provided in DeepTrack 2.0, including models for DNNs, CNNs, U-Nets, and GANs. In each case, the parameters of the model (e.g., number of layers and number of artificial neurons) can be defined by the user.
\end{description}

DeepTrack 2.0 solutions depend on the interactions between these objects. In general, there are three distinct typical operations a feature can perform. The first operation is to add an image to a list of images (notably Scatterers):
\[
    [\begin{gathered}[t]
        \mathrm{I}_1,\\
        \uparrow \\
        [P_1]
    \end{gathered}
    \;...,
    \begin{gathered}[t]
        \mathrm{I}_n\\
        \uparrow \\
        [P_n]
    \end{gathered}
    ] \rightarrow
    F^\prime(P^\prime)
    \rightarrow
    [\begin{gathered}[t]
        \mathrm{I}^\prime_1,\\
        \uparrow \\
        [P_1]
    \end{gathered}
    \;...,
    \begin{gathered}[t]
        \mathrm{I}_n,\\
        \uparrow \\
        [P_n]
    \end{gathered}
    \begin{gathered}[t]
        \mathrm{I^\prime}\\
        \uparrow \\
        [P^\prime]
    \end{gathered}
    ]
\]
Here, a list of $n$ images are fed to the feature $F^\prime$. Each of these images has a list of properties $P_i$, which describe the process used to create that image. The feature is controlled by some properties $P^\prime$, and returns a new list of images. The first $n$ images are unchanged, but a new image $I^\prime$ is appended to the end, on which the properties $P^\prime$ are imprinted.

The second operation is to transform all images in the list in some way (the standard behavior of features, including noise, augmentations and most mathematical operations):
\[
    [\begin{gathered}[t]
        \mathrm{I}_1,\\
        \uparrow \\
        [P_1]
    \end{gathered}
    \;...,
    \begin{gathered}[t]
        \mathrm{I}_n\\
        \uparrow \\
        [P_n]
    \end{gathered}
    ] \rightarrow
    F^\prime(P^\prime)
    \rightarrow
    [\begin{gathered}[t]
        \mathrm{I}^\prime_1,\\
        \uparrow \\
        [P_1, P^\prime]
    \end{gathered}
    \;...,
    \begin{gathered}[t]
        \mathrm{I}^\prime_n\\
        \uparrow \\
        [P_n, P^\prime]
    \end{gathered}
    ]
\]
Here, the feature returns a list of the same length, but each image is altered (e.g., some noise is added or it is rotated). The properties characterizing this alteration $P^\prime$ are imprinted on all images.

The third operation is to merge several images into a single image:
\[
    [\begin{gathered}[t]
        \mathrm{I}_1,\\
        \uparrow \\
        [P_1]
    \end{gathered}
    \;...,
    \begin{gathered}[t]
        \mathrm{I}_n\\
        \uparrow \\
        [P_n]
    \end{gathered}
    ] \rightarrow
    F^\prime(P^\prime)
    \rightarrow
    \begin{gathered}[t]
        [\mathrm{I}^\prime]\\
        \uparrow \\
        [P_1,\;...,\;P_n,\;P^\prime]
    \end{gathered}
\]
Here, all the properties of the input images, as well as the feature's own properties, are imprinted on the resulting image (notably optical devices). 

A typical complete image generation pipeline can look something like:
\begin{align*}
    [] & \rightarrow
    F_{s1}(P_{s1})
    \rightarrow
    [\begin{gathered}[t]
        \mathrm{I_{s1}}\\
        \uparrow \\
        [P_{s1}]
    \end{gathered}
    ]\\&\rightarrow
    F_{s2}(P_{s2})
    \rightarrow
    [\begin{gathered}[t]
        \mathrm{I_{s1}},\\
        \uparrow \\
        [P_{s1}]
    \end{gathered}
    \begin{gathered}[t]
        \mathrm{I_{s2}}\\
        \uparrow \\
        [P_{s2}]
    \end{gathered}
    ] \\
    \\
    & \rightarrow
    F^\prime(P^\prime)
    \rightarrow
    [\begin{gathered}[t]
        \mathrm{I^\prime_{s1}},\\
        \uparrow \\
        [P_{s1},\;P^\prime]
    \end{gathered}
    \begin{gathered}[t]
        \mathrm{I^\prime_{s2}}\\
        \uparrow \\
        [P_{s2},\;P^\prime]
    \end{gathered}
    ]\\& \rightarrow
    F_o(P_o)
    \rightarrow
    \begin{gathered}[t]
        [\mathrm{I}_o]\\
        \uparrow \\
        [P_{s1},\;P_{s2},\;P^\prime,\;P_o]
    \end{gathered}
\end{align*}
Here, the start is an empty list. Two initial features ($F_{s1}$, $F_{s2}$) append images to that list, creating a list of two images ($I_{s1}$, $I_{s2}$) (e.g., these could be two scattering particles int eh field of view). Each such image is modified by a feature $F^\prime$ (e.g., by adding some noise), before being merged into a single image by $F_o$ (e.g., representing the output of a microscope). Note that $P^\prime$ is not added to the list of properties twice; the list is in fact a set, and cannot contain duplicate properties.

We show an even more concrete example in Fig. \ref{fig6}A. Here, we have an initial feature \texttt{Ellipse} which creates a single image of an ellipse. We follow this by the feature \texttt{Duplicate}, which creates a fixed number of duplicates (here five). (Note that \texttt{Duplicate}  duplicates the feature \texttt{Ellipse}, not the generated image, which is why it can create several different ellipses, i.e., with different radius, intensity, or in-image position.) This list of images is sent to the feature \texttt{Fluorescence}, which images them through a simulated fluorescence microscope. After this, a background offset is added, and Poisson noise is introduced.

Since the positions of all ellipses are stored as properties, they are imprinted on the final image. This allows us to create a segmented mask, shown in Fig.~\ref{fig6}B, which we can use as ground-truth label to train the deep-learning model. These two image creation pipelines (one for the data, one for the label), are passed to a generator which continuously creates new images by updating the properties that control the features using user-defined update rules. These images are fed to a neural network to train it (Fig.~\ref{fig6}C), which results in a trained model that can analyze experimental data (Fig.~\ref{fig6}D).

Writing code directly using the DeepTrack 2.0 framework allows the user to extend the capabilities of the package. 
In most cases, it is sufficient to use the \texttt{Lambda} feature, which allows any external function to be incorporated into the framework. 
However, certain scenarios may require the user to write custom features. 
For example, the user can extend the feature \texttt{Optics} (features that simulate optical devices) to create a new imaging modality, the feature \texttt{Scatterer} (features that represent some object in the sample) to create a custom scatter, or the feature \texttt{Augmentation} (features that augment an image to cheaply broaden the input space) to expand the range of available augmentations.
It is also straightforward to add new neural-network models: any Keras model can be directly merged with DeepTrack 2.0 without any configuration, while models from other packages can easily be wrapped.

To help users get started writing code using DeepTrack 2.0, we provide several comprehensive video tutorials \cite{deeptrackgithub}, ranging in scope from implementing custom features to writing complete solutions.

\section{Case Studies}\label{ch:results}

In this section, we use DeepTrack 2.0 to exemplify how deep learning can be employed for a broad range of microscopy applications.
We start with a standard benchmark for image classification: the MNIST digit recognition challenge \cite{lecun2010mnist} (Section~\ref{mnist}).
Afterwards, we employ Deeptrack 2.0 to analyze microscopy images. First, we develop a model to track particles whose images are acquired by brightfield microscopy, training a single-particle tracker whose accuracy surpasses standard algorithmic approaches especially in noisy imaging conditions \cite{Helgadottir2019DigitalLearning} (Section~\ref{tracking}). 
Then, we expand this example to also extract quantitative information about the particle, namely its size and refractive index (Section~\ref{characterization}). 
Deep learning is especially powerful in tracking multiple particles in noisy environments. As a demonstration of this, we develop a model that can detect quantum dots on a living cell imaged by fluorescence microscopy (Section~\ref{multiparticle}). 
Again, we expand this example to demonstrate three-dimensional tracking of multiple particles whose images are acquired using holography (Section~\ref{3d}). 
We also develop a neural network to count the number of cells in fluorescence images (Section~\ref{counting}).
Finally, we train a GAN to create synthetic images of cells from a semantic mask (Section~\ref{mitogan}).
All these examples are available both as project files for DeepTrack 2.0 graphical user interface \cite{deeptrackgithub} and as Jupyter notebooks \cite{appgithub}, and they are complemented by video tutorials.

\subsection{MNIST digit recognition \label{mnist}}

\begin{figure*}
    \centering
    \includegraphics[width=11.5cm, clip, trim={0 8cm 0 0}]{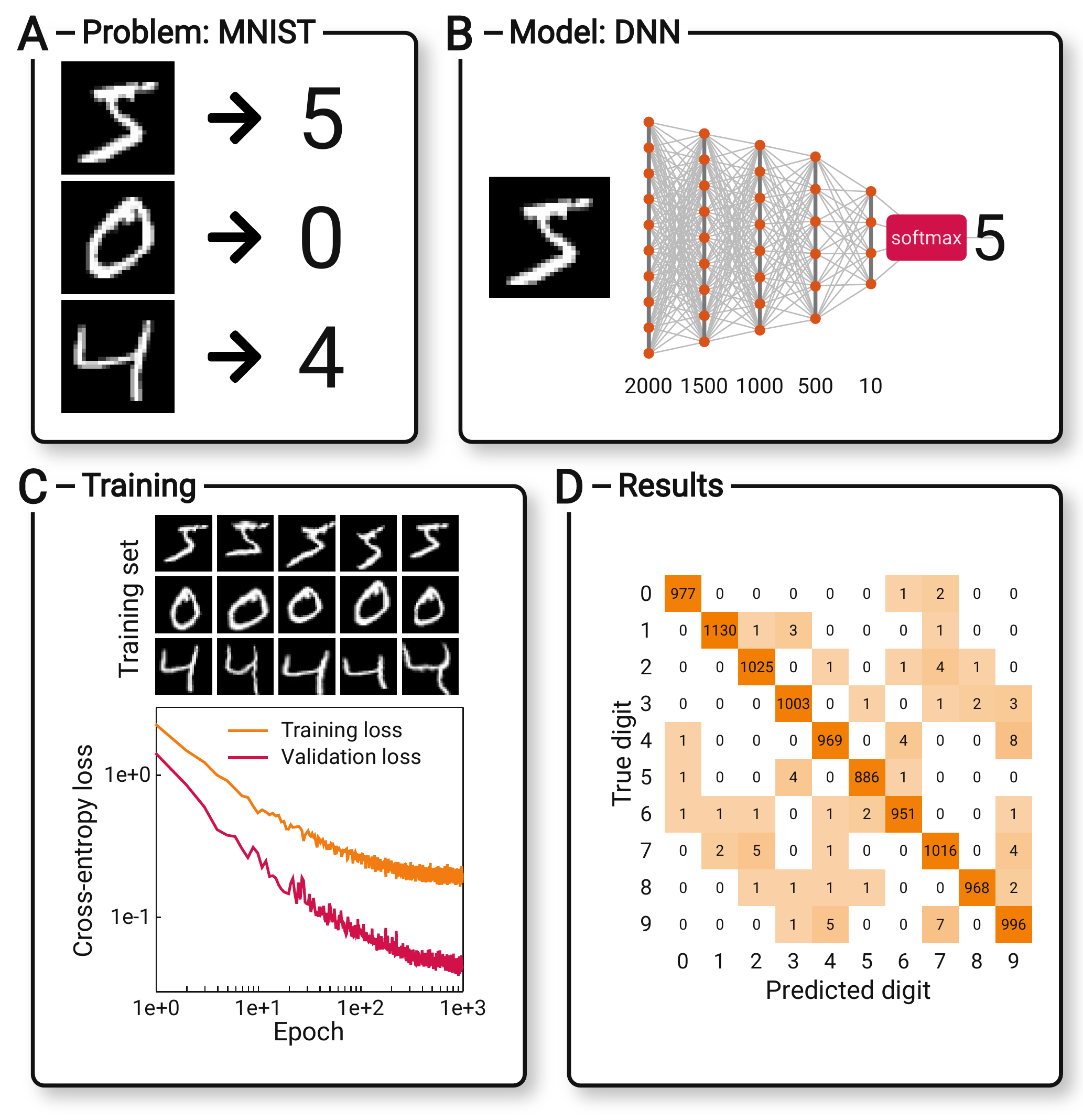}
    \caption{
    {\bf A dense neural network to classify hand-written digits.} 
    {\bf A} Three example images from the MNIST dataset with their corresponding labels.
    {\bf B} The network architecture consists of five fully connected layers of decreasing size, with the final layer having 10 nodes, whose outputs correspond to classification probabilities.
    {\bf C} Examples of augmented training images: The network is trained on a set of $6\cdot 10^{4}$ $28 \times 28$ pixel images augmented by translations, rotations, shear, and elastic distortions, using a categorical cross-entropy loss. The validation loss (magenta line) is significantly lower than the training loss (orange line), likely due to augmentations making the training set harder than the validation set.
    {\bf D} Confusion matrix showing how the $1\cdot 10^{4}$ validation images are classified by the network: The diagonal represents the correctly classified digits, constituting the vast majority of digits. The off-diagonal cells represent incorrectly classified digits.
    }
    \label{fig7}
\end{figure*}

Recognizing hand-written digits of the MNIST dataset is a classical benchmark for machine learning \cite{lecun2010mnist}. 
The task consists of recognizing handwritten digits from 0 to 9 in $28 \times 28$ pixel images.
In the dataset, there are $6 \cdot 10^4$ training images and $1 \cdot 10^4$ validation images, some examples of which are provided in Fig.~\ref{fig7}A.

Since is a relatively simple task, we employ a DNN (Fig.~\ref{fig7}B). The architecture of the networks is that of Ciersan \emph{et. al.}, which has achieved the best results using DNNs amongst the attempts listed on the MNIST webpage \cite{lecun2010mnist}. As a loss function, we use categorical cross-entropy, which is a standard loss function for classification tasks. 

We train the network using the $6\cdot 10^{4}$ training images augmented using affine transformations and elastic distortions, exemplified in Fig.~\ref{fig7}C. We train it for 1000 epochs, where one epoch represents one pass through all training images. This results in a validation loss of 0.05, as compared to a training loss of 0.20 (Fig.~\ref{fig7}C). The higher training loss is likely due to the augmentations, which make the training data harder than the validation data. It is, as such, unlikely that the network has overfitted the training set.

Finally, we validate the trained network on the $1\cdot 10^{4}$ validation images. The network achieves an accuracy of $99.34\%$, which is in line with the best performance achieved by DNN on the MNIST digit recognition task \cite{lecun2010mnist}.
The confusion matrix (Fig.~\ref{fig7}D) shows that the incorrectly classified digits consist mainly of 9s classified as 7s and of 4s classified as 9s. 

\subsection{Particle localization \label{tracking}}

\begin{figure*}
    \centering
    \includegraphics[width=17cm, clip, trim={0 14cm 0 0}]{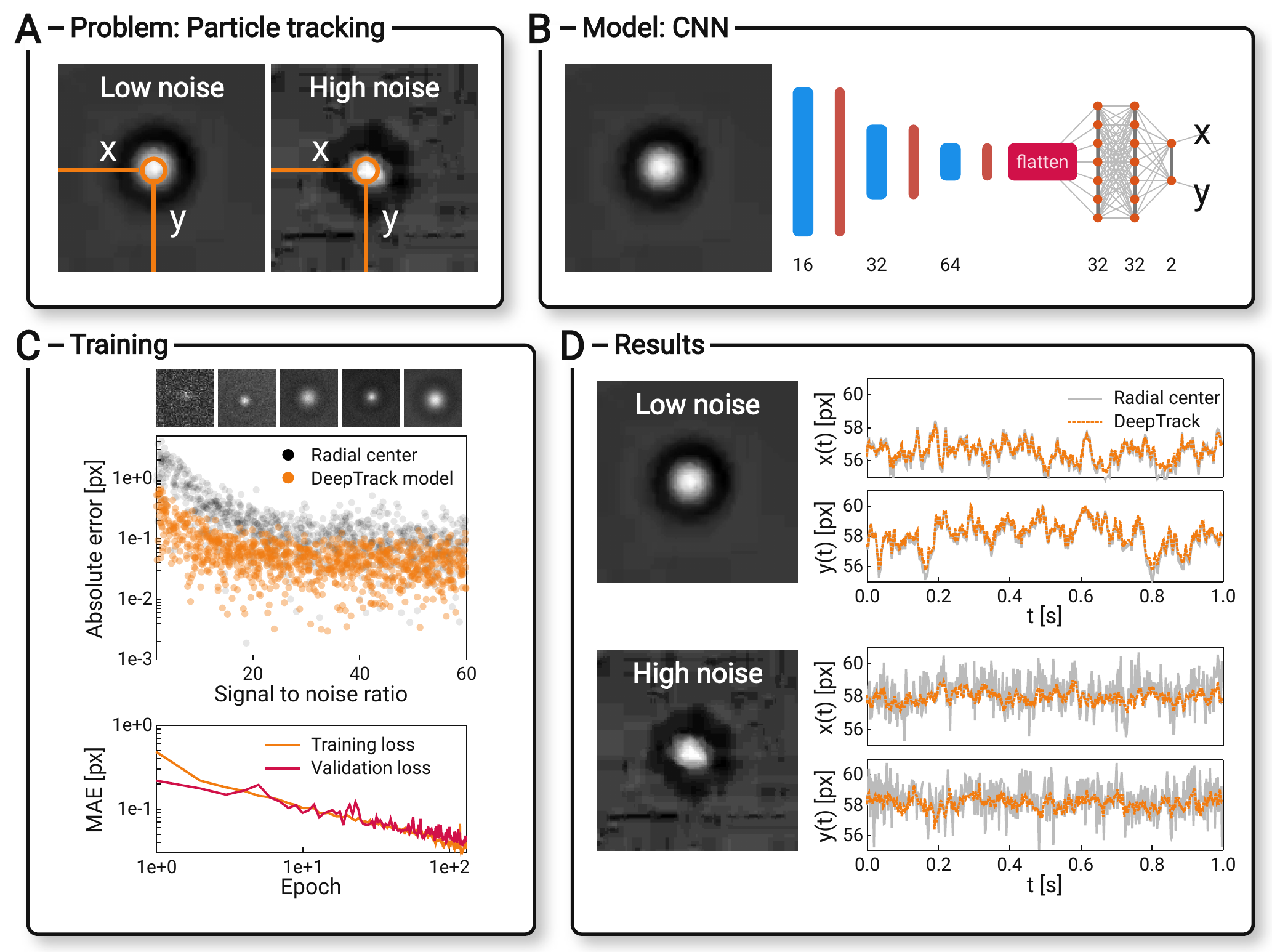}
    \caption{
    {\bf A convolutional neural network to track a single particle.} 
    {\bf A} Frames of the same particle held in the same optical trap, but with different illumination which results in a low-noise video (left) and a high-noise video (right).
    {\bf B} The network architecture consists of 3 convolutional layers, each followed by a pooling layer. The resulting tensor is flattened and passed through three fully connected layers, which return the predicted $x$ and $y$ position of the particle.
    {\bf C} Five examples of the outputs of the image generation pipeline at increasing signal-to-noise ratio (SNR). The pixel tracking error for 1000 images using the DeepTrack model (orange markers) and the radial-center algorithm (gray markers). The DeepTrack model systematically outperforms radial center, especially for low SNR. The model was trained for 110 epochs, on a set of $1\cdot10^4$ synthetic images. The validation loss (magenta line) and the training loss (orange line) remain similar for the whole training session.
    {\bf D} The predicted position of the particle in the low-noise video (top panel) and the high-noise video (bottom panel) as found by the radial center algorithm (gray line) and by the DeepTrack model (dotted orange line). In the low-noise case, they overlap within a fraction of a pixel, while for the high-noise case, the radial center algorithm produces erratic predictions. 
    }
    \label{fig8}
\end{figure*}

Determining the position of objects within an image is a fundamental task for digital microscopy. In this example, we aim at localizing with very high accuracy the position of an optically trapped particle. Two videos are captured of the same particle in the same optical trap, one with good image quality and one with poor image quality, from which we want to extract the particle's $x$ and $y$ positions (Fig.~\ref{fig8}A). 

To analyze these images, we first use a CNN to transform the
$51 \times 51$ pixel input image to a $6 \times 6 \times 64$ tensor. Subsequently, we pass this result to a DNN, which outputs an estimate of the particle's in-plane position (Fig.~\ref{fig8}B). This model is based on the one described by Helgadottir \cite{Helgadottir2019DigitalLearning}. We use mean absolute error (MAE) as the loss function. (Alternatively, we could also use mean squared error (MSE), which delivers equally accurate results.)

The network is trained purely on synthetic data generated using DeepTrack 2.0. The generation of synthetic microscopy data for training a network generally entails the following steps. First, the optical properties of the instrument used to capture the data is replicated (e.g., NA, illumination spectra, magnification, and pixel size). This ensures that the simulated point-spread function of the simulated optical system closely matches the experimental setup. Second, the properties of the sample are specified, including the radius and refractive index of the particle. As a final step, noise is added to the simulated images to be representative of experimental data. During training, each parameter of the simulation (e.g., the optical properties, the sample properties, and the noise strength) is stochastically varied around the expected experimental values to make the network more robust. In Fig.~\ref{fig8}C, we show a few outputs from the image generation pipeline with SNR increasing from left to right. We also demonstrate that the network outperforms the radial center algorithm \cite{Parthasarathy2012RapidCenters}. This is achieved by training the network for 110 epochs, on a set of $1\cdot10^4$ synthetic images. The validation set consisted of 1000 images. We can see that the loss is still decreasing at this point, but the gain is minimal (Fig~\ref{fig8}C). No signs of overfitting can be seen.

Finally, we use the network to track the two videos of the optically trapped particle. In Fig.~\ref{fig8}D, we see that in the low-noise video, the radial center method and the DeepTrack model agree, while, for the high noise-video, the radial center method makes large, sporadic jumps. Since the videos are of the same particle in the same optical trap, we expect the dynamics of the particle to be similar. Since only the DeepTrack method gives consistent dynamics in the two cases, it indicates that DeepTrack is better able to track this more difficult case.
A more detailed discussion of this example can be found in Ref.~\cite{Helgadottir2019DigitalLearning}.

\subsection{Particle characterization \label{characterization}}

\begin{figure*}
    \centering
    \includegraphics[width=11.5cm, clip, trim={0 9cm 0 0}]{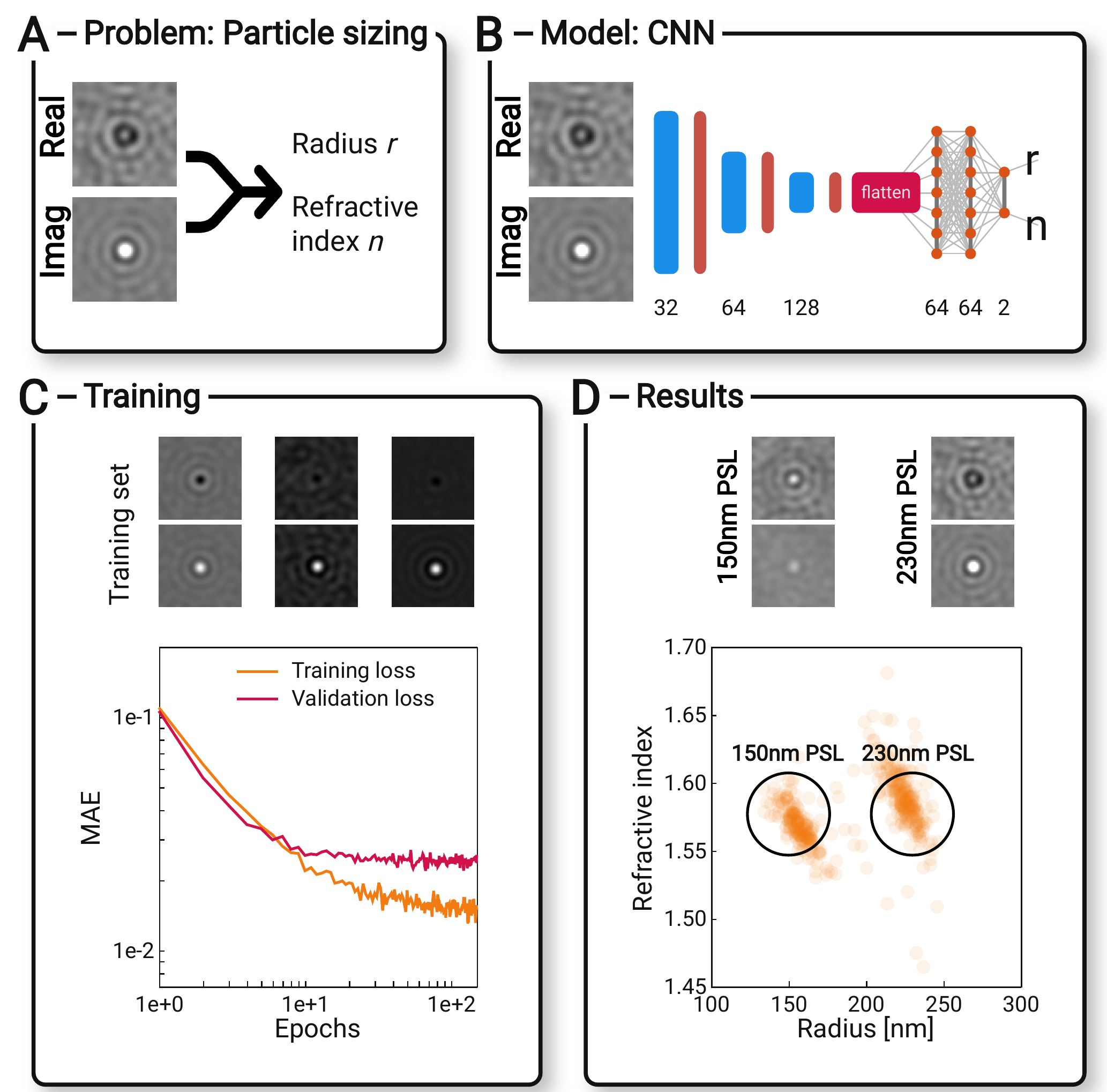}
    \caption{
    {\bf A convolutional neural network to measure the radius and refractive index of a single particle.} 
    {\bf A} The real and imaginary parts of the scattered field are used to measure the radius and refractive index of a single particle. The field is captured using an off-axis holographic microscope and numerically propagated such that the particle is in focus. The total datasets consists of roughly $8\cdot 10^3$ such observations, belonging to 352 individual polystyrene particles with \SI{150}{\nano\meter} or \SI{230}{\nano\meter} radius.
    {\bf B} The network architecture consists of 3 convolutional layers, each followed by a pooling layer. The resulting tensor is flattened and passed through three fully connected layers which return the predicted radius and refractive index of the particle.
    {\bf C} Three pairs of real and imaginary parts of the scattered field from a single particle. The network is trained for 110 epochs on 1000 $64 \times 64$ pixel images, using MAE loss. The validation loss (magenta line) stops decreasing significantly after only 20 epochs, while the training loss (orange line) keeps decreasing.
    {\bf D} Measured radius versus measured refractive index for an ensemble of particles. There are two clearly distinguished populations, which closely match the modal characteristics of the particles (shown by the two circles).
    }
    \label{fig9}
\end{figure*}

Microscopy images contain quantitative information about the morphology and optical properties of the sample. However, extracting this information using conventional image analysis techniques is extremely demanding. Deep learning has proven to offer a remedy to this \cite{Midtvedt2020HolographicLearning, Altman2020CATCH:Networks}. In this example, we employ DeepTrack 2.0 to develop a model to quantify the radius and refractive index of particles based on their complex-valued scattering patterns. As experimental verification, we record the scattering patterns of a heterogeneous mix of two particle populations (\SI{150}{\nano\meter} polystyrene and \SI{230}{\nano\meter} polystyrene bead flowing in a microfluidic channel) using an off-axis holographic microscope.

In line with the previous example, we use a CNN to downsample the $64 \times 64 \times 2$ pixel input (the two channels corresponding to the real and the imaginary parts of the field) to a $8 \times 8 \times 128$ tensor. Subsequently, we pass this tensor to a DNN, which outputs an estimate of the particle's radius and refractive index (Fig.~\ref{fig9}B). The number of channels in each layer is doubled compared to the previous example, which may help capture subtle changes in the scattered field. We used MAE loss.

To account for imperfections in the experimental system, we approximate the experimental PSF for the simulated images by adding coma aberrations with random strength. In Fig.~\ref{fig9}C we show three examples of outputs from the image generation pipeline. The network is trained for 110 epochs on a set of $1\cdot10^4$ synthetic images. The validation set consists of 1000 images. The validation loss diverges from the training loss after only 20 epochs suggesting that the training could be terminated earlier, or that a larger training set could be beneficial.

Finally, we evaluate the model on the experimental dataset.
In each frame, all particles are roughly localized using a standard tracking algorithm, and focused using the criteria described in Ref.~\cite{Midtvedt2020}. These observations are subsequently linked frame to frame to form traces. We use the fact that we observe each particle multiple times to improve the accuracy of the sizing. Specifically, we predict the size and refractive index of a particle using an image from each observation of that particle. We then average the result to obtain the final prediction for that particle. (This deviates slightly from the method proposed in \cite{Midtvedt2020HolographicLearning} where the averaging is performed in the latent space, which result in a more complex and accurate method.) We can see the results in Fig.~\ref{fig9}D, showing the radius versus the refractive index of each measured particle. We clearly distinguish two populations, which closely match the modal characteristics of the particles (shown by the two circles).

\subsection{Multiparticle tracking \label{multiparticle}}

\begin{figure*}
    \centering
    \includegraphics[width=11.5cm, clip, trim={0 9cm 0 0}]{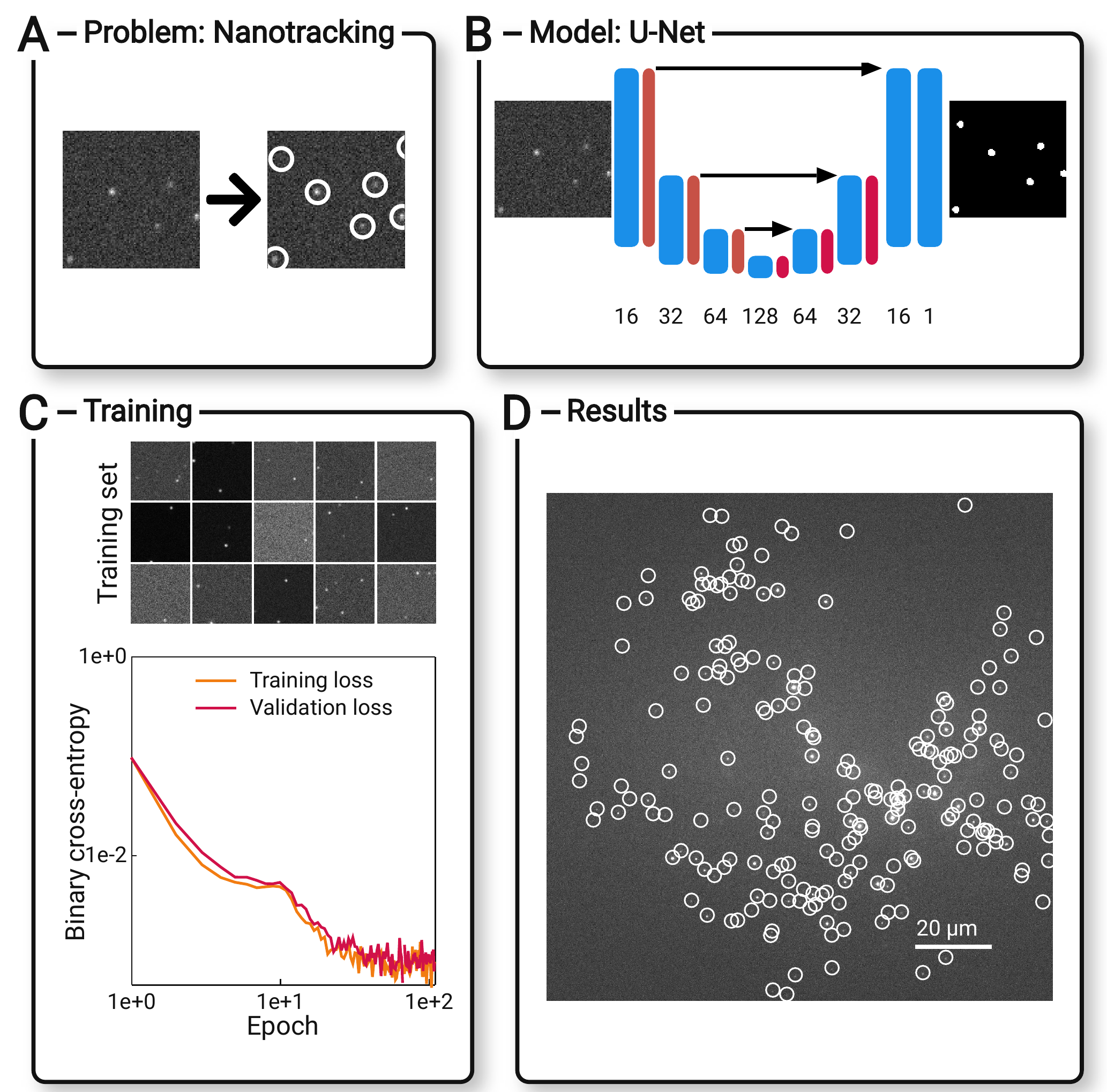}
    \caption{
    {\bf A U-Net to detect quantum dots in fluorescence images.} 
    {\bf A} A small slice of an image depicting quantum dots situated on a living cell, imaged through a fluorescence microscope (data kindly provided by Carlo Manzo). The quantum dots in the image are circled in white.
    {\bf B} The network architecture is a small U-Net. A final convolutional layer outputs a single image, where each particle in the input is represented by a circle of 1s.
    {\bf C} Examples of synthetic images used in the training process. The network is trained on 2000 $128 \times 128$ pixel image for 110 epochs using binary cross-entropy loss. The validation loss (magenta line) and the training loss (orange line) are similar in magnitude for the entire training session. After 10 epochs both losses start decreasing more rapidly, which is explained by a change in the weighting in the loss function, which is explained in the text.
    {\bf D} A single frame tracked using the trained model. It detects all obvious particles, as well as a few that are hard to conclusively verify as real observations.
    }
    \label{fig10}
\end{figure*}

The previous examples have been focused on analyzing a single particle at a time. Frequently, however, microscopy involves detecting multiple particles at once. In this example, we extract the positions of quantum dots situated on the surface of a living cell. A small slice of an image is shown in Fig.~\ref{fig10}A, with each particle circled in white.

We train a U-Net model to transform the input into a binarized representation, where each pixel within 3 pixels of a particle in the input is set to 1, and every other pixel is set to 0, as shown in Fig.~\ref{fig10}B. The network returns a probability for each pixel, which is thresholded into the binary representation. (Note that in this example we can use a network that is  smaller than the original U-Net because the information is highly localized; however, if, for example, the data were aberrated, a deeper network would be better.) The network is compiled with a binary cross-entropy loss.

The network is trained purely on synthetic data, simulating the appearance of a quantum dot as the PSF of the optical device. In Fig.~\ref{fig10}C, we show several examples of the outputs from the image generation  pipeline. The network is trained on 2000 $128 \times 128$ pixel images in two sessions. The first session consists of 10 epochs where the loss is weighted such that setting a pixel value of 1 to 0 is penalized 10 times more than setting a value of 0 to 1. This helps the network avoid the very simple local optimum of setting every pixel to 0. For the following 100 epochs, the two types of errors are penalized equally. This explains the sudden change of training rate after 10 epochs seen in Fig.~\ref{fig10}C. The validation set consists of 256 images, and shows no signs of overfitting.

In Fig.~\ref{fig10}D, we show a single image tracked using the trained network. It detects all obvious particles, as well as a few that are hard to verify as real observations. However, for most such cases, they are detected again the next frame, indicating that it is a real observation instead of just random noise. (However, this method to verify the tracking is not conclusive, since quantum dots are known to frequently flicker, meaning that they are not guaranteed to be visible in the next frame. Conversely, two observations in a row do not necessarily mean that it is a real observation. It can be a product of optical effects that are consistent between frames.)

\subsection{3D multiparticle tracking \label{3d}}

\begin{figure*}
    \centering
    \includegraphics[width=17cm, clip, trim={0 14cm 0 0}]{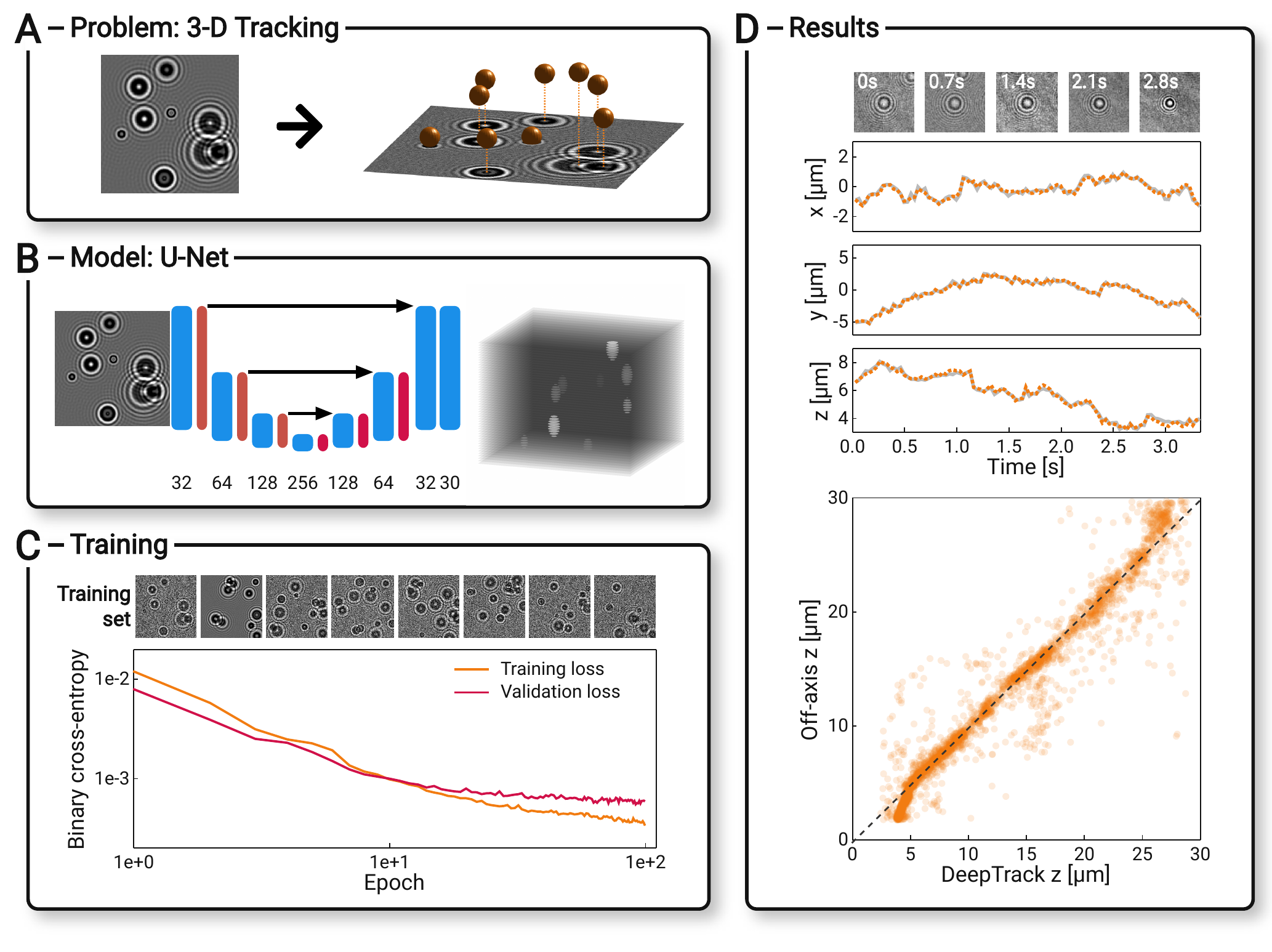}
    \caption{
    {\bf A U-Net to track spherical particles in three dimensions.} 
    {\bf A} A sample network input, consisting of scattering patterns of several spherical particles. The sample contains a mixture  of \SI{150}{\nano\meter} and \SI{230}{\nano\meter} polystyrene particles.
    {\bf B} The network architecture is a small U-Net. A final convolutional layer outputs a volume, where each particle in the input is represented by a sphere of 1s. The out-of-plane direction spans \SIrange{2}{30}{\micro\meter}
    {\bf C} Examples of synthetic images used in the training process. The network is trained on 2000 $256 \times 256$ pixel images for 100 epochs using a binary cross-entropy loss. The validation loss (magenta line) diverges from the training loss (orange line) after roughly 10 epochs.
    {\bf D} A single particle tracked using the DeepTrack model (dotted orange line) and off-axis holography (gray line), showing the $x$, $y$, and $z$ positions over time. The two methods almost perfectly overlap. Moreover, we show the predicted out-of-plane position of all detections as found using the DeepTrack model versus the off-axis holography. Most observations fall close to the central line, with a few outliers and some deviations near the edges of the range.
    }
    \label{fig11}
\end{figure*}

Similarly to single particle analysis, multi-particle analysis can be extended to extract quantitative information about the particles. In this example, we locate spherical particles in 3D space from the intensity of the scattered field captured by an in-line holographic microscope (Fig.~\ref{fig11}A). In order to be able to validate the out-of-plane positioning with ground-truth experimental data, we capture the experimental data using an off-axis holographic microscope. This allows us to accurately track the particles in 3D space using standard methods \cite{Midtvedt2020}. Off-axis holographic microscopes, unlike the inline counterpart, retrieve the entire complex field instead of just the field intensity. We approximate the conversion from off-axis to in-line holographic microscopy by squaring the amplitude of the field.
Similarly to the previous example, we represent each particle in the input by a region of pixel values of 1s in the output. The difference is that this network returns a volume, with each particle instead represented by a sphere with radius 3 pixels (Fig.~\ref{fig11}B). The last dimension of the output represents the out-of-plane position of the particle, ranging from \SIrange{2}{30}{\micro\meter}. The network is slightly larger than the previous example, since it needs to extract more information about the particles. Just as in the previous example, binary cross-entropy is used as loss function.

The network is trained purely on synthetic data. We replicate the optical properties of the instrument used to capture the data, simulating the appearance of a particle using Mie theory. Each parameter of the simulation is stochastically varied around the experimentally expected values, making the network more robust. Additionally, we approximate the experimental PSF by adding coma aberrations with random strength. In Fig.~\ref{fig11}C, we show a few images from this pipeline. The network is trained for 100 epochs on a set of $1\cdot10^3$ synthetic images. The validation set consisted of 256 images, and diverges from the training loss after 10 epochs, suggesting that it could be terminated earlier, or that a larger training set could be beneficial (Fig.~\ref{fig11}C).

In Fig.~\ref{fig11}D, we show a single particle tracked in three dimensions, with the position found using the trained network in orange and the off-axis method in gray. The two methods overlap almost exactly. Moreover, we also show the out-of-plane positioning found by the off-axis method and the trained model for all detections. We see that most observations fall very close to the central line, with few outliers. Moreover, we see a divergence from the central line at the edges of the range. This is due a limitation of how the position is extracted from the binarized image: A sphere close to the edge of the volume will not be entirely contained within the image, so that its centroid will not be the center of the sphere, resulting in a bias.

\subsection{Cell counting \label{counting}}

\begin{figure*}
    \centering
    \includegraphics[width=17cm, clip, trim={0 14cm 0 0}]{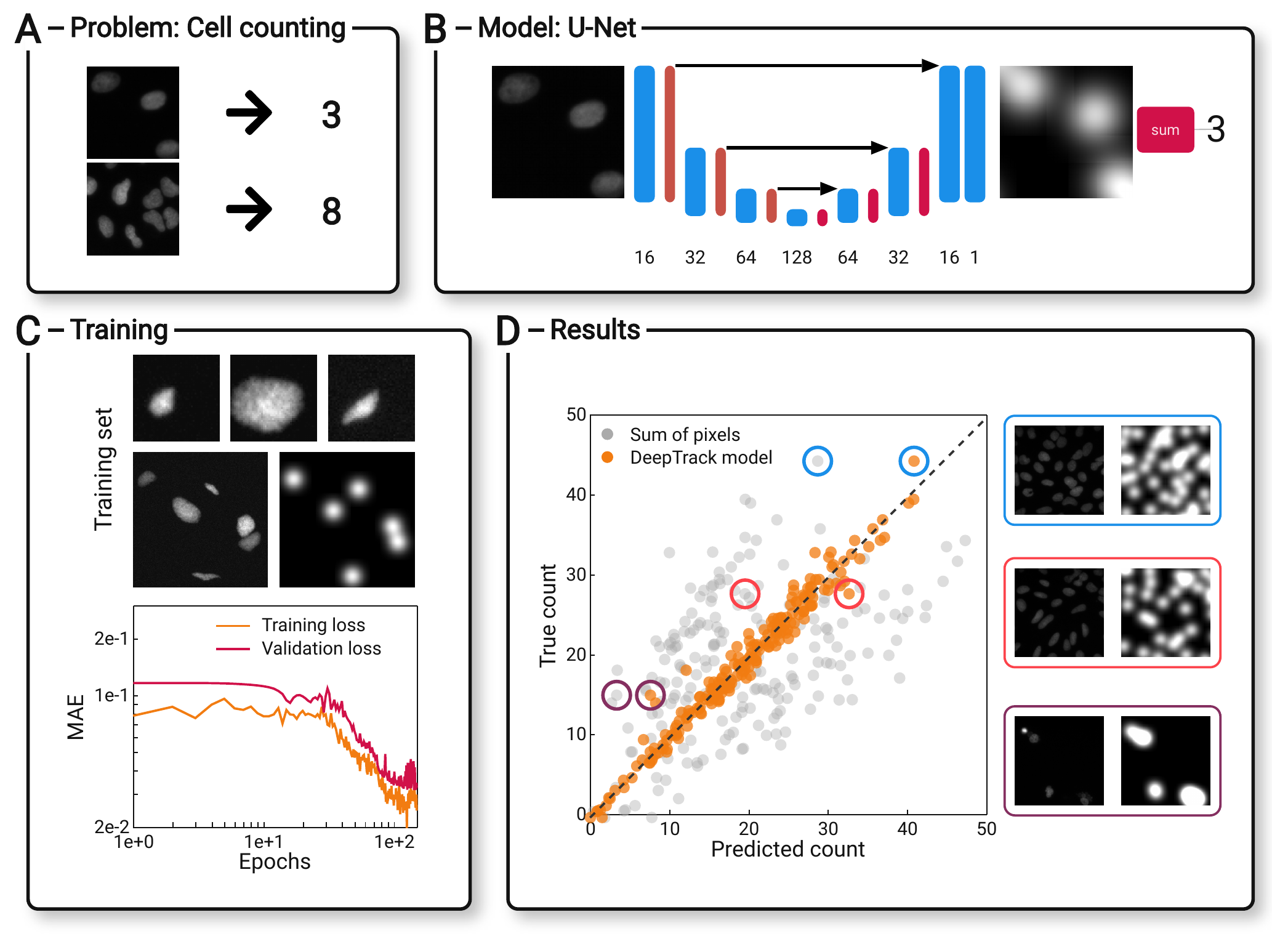}
    \caption{
    {\bf A U-Net to count cells in a fluorescence image.} 
    {\bf A} Two slices from the BBBC039 dataset, with the corresponding number of cells in the image.
    {\bf B} The network architecture is a small U-Net. A final convolutional layer outputs an image with a single feature, where each cell in the input is represented by a Gaussian distribution with a standard deviation of 10 pixels and whose intensity integrates to one. Thus, the integration of the intensity of the output corresponds to the number of cells in the image.
    {\bf C} Examples of the cell images created by the image simulation pipeline, followed by a sample input-output pair containing six cells. The network is trained on 1000 $256 \times 256$ pixel images, and validated on 150 $512 pixel \times 688 pixel$ images, using MAE loss. The validation loss (magenta line) is consistently higher than the training loss (orange line), but follows a similar curve.
    {\bf D} The number of cells as found by the DeepTrack model compared to a naive approach based on the summation of the values of the pixels of the image. Each data point represents a $256 \times 256$ pixel slice of one of the 50 images in the test set. Three points are circled and have their corresponding input-output pair shown on the right.
    }
    \label{fig12}
\end{figure*}

DeepTrack 2.0 is not limited to particle analysis. Counting the number of cells in an image has traditionally been a tedious task performed manually by trained experts. In this example, we count the number of U2OS cells (cells cultivated from the bone tissue of a patient suffering from osteosarcoma \cite{PontenU2OS1967}) in fluorescence images shown in Fig.~\ref{fig12}A. We use the BBBC039 dataset for evaluation \cite{Ljosa2012}.

We once again use a U-net model. This time, we represent each cell by a Gaussian distribution with a standard deviation of 10 pixels, whose intensity values integrate to one (Fig.~\ref{fig12}B). In this way, the integral of the output intensity corresponds to the number of cells in the image. By representing the cell by a Gaussian profile, we also reduce the emphasis on the absolute positioning of the cell, while retaining the ability for a human to validate the output visually. We compile the network using MAE loss.

The training data consists of synthetic data generated by imaging cell-like objects through a simulated fluorescence microscope. A few example cells, as well as a single training input-output pair is shown in Fig.~\ref{fig12}C. The network is trained for 190 epochs on a set of 1000 synthetic images. Since the training set of the BBBC039 dataset is not used for training, we merge the training set and the validation set and use the merged set for validation. The validation loss is consistently higher than the training loss, but follows a very similar curve (Fig.~\ref{fig12}C). This suggests that the synthetic data is a decent approximation of the experimental images. The offset can be largely explained by a few images in the validation set that are particularly hard for the network.

For large images, errors can average out, which can result in deceptively accurate counting. To eliminate this concern, we show the predicted number of cells versus the true number of cells for smaller slices of images ($256 \times 256$ pixels) in the BBBC039 dataset in Fig.~\ref{fig12}D. The network predicts the correct number of cells within just a few percent. As a comparison, we show that the images cannot be analyzed by simply integrating the intensity of the input images (Fig.~\ref{fig12}D). In order to show a best case scenario for the sum-of-pixels method, we transformed each sum by an affine transformation that minimizes the square error on the test set itself. It is apparent that this is not sufficient to achieve an acceptable counting accuracy.

\subsection{GAN image generation \label{mitogan}}

\begin{figure*}
    \centering
    \includegraphics[width=17cm, clip, trim={0 14cm 0 0}]{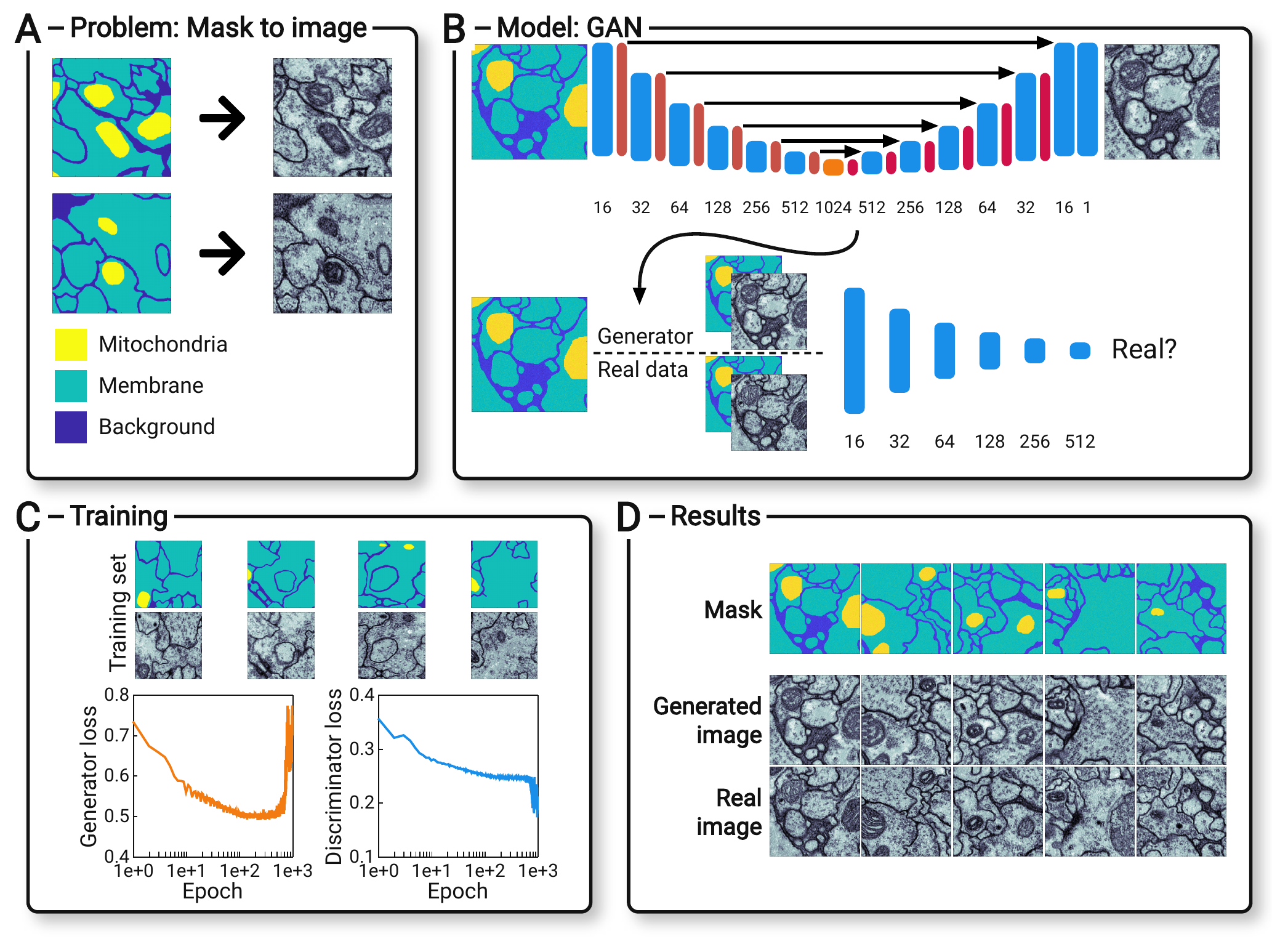}
    \caption{
    {\bf A conditional GAN is used to create cell images from a semantic mask.} 
    {\bf A} Example masks (left) from which images of \emph{drosophila melanogaster} third instar larva ventral nerve cord (right) are generated using the segmented anisotropic ssTEM dataset \cite{gerhard2013segmented}.
    {\bf B} The network architecture is a conditional generative adversarial network. The generator transforms an input semantic mask into a realistic cell image, using a U-Net architecture with the most condensed layer being replaced by two residual network blocks \cite{He2016DeepRecognition}. The discriminator is designed similar to the PatchGan discriminator \cite{isola2017image}, and receives both the mask and an image as an input. The generator is trained using a MAE loss between the experimental image and the generated image, as well as a MSE loss on the discriminator output. Conversely, the discriminator is trained with a MSE loss.
    {\bf C} Examples of masks and corresponding experimental images. The loss of the generator (left) and of the discriminator (right) are shown over 1000 training epochs, each of which consists of 16 mini-batches of 7 samples. We see that the generator loss increases towards the end of the training, a signature that continuing training beyond this point destabilizes the generator.
    {\bf D} Masks images from a validation set, and corresponding generated image and real image. The generated images are qualitatively similar to the real images.
    }
    \label{fig13}
\end{figure*}

DeepTrack 2.0 can also efficiently handle more cases where the training set is derived directly from experimentally captured data, instead of being simulated. In this example, we combine the two approaches, by using experimental data to train a GAN to create new data from a semantic representation of the image (Fig.~\ref{fig13}A). More specifically, the GAN creates images of the \emph{drosophila melanogaster} third instar larva ventral nerve cord from a semantic representation of background, membrane, and mitochondria. This GAN, once trained, can subsequently be used as a part of an image simulation pipeline, just as any other DeepTrack feature. 

The architecture of the neural network we employ is shown in Fig.~\ref{fig13}B. The model is composed of a generator that learns the mapping relation between the input mask and its corresponding cell-image, and of a discriminator that, given the semantic segmentation, determines if the generated image plausibly could have been drawn from a real sample.  

The generator follows a U-Net design with symmetric encoder and decoder paths connected through skip connections. The encoder consists of convolutional blocks followed by strided convolutions for downsampling.  Each convolutional block contains two sequences of $3 \times 3$ convolutional layers.  At each step of the encoding path, we increase the number of feature channels by a factor of 2.

The encoder connects to the decoder through two residual network (ResNet) blocks \cite{He2016DeepRecognition}, each with 1024 feature channels. For upsampling, we use bilinear interpolations, followed by a convolutional layer (stride = 1). This operation is followed by concatenation with the corresponding feature map from the encoding path. Furthermore, we add two convolutional blocks with 16 feature channels at the final layer of the decoder. We use a $1 \times 1$ convolutional layer to map each 16-component feature vector to the output image. Herein, the hyperbolic tangent activation (tanh) is employed to transform the output to the range $[-1, 1]$.  Every layer in the generator, except the last layer, is followed by an instance normalization (alpha = 2) and a LeakyRelu activation layer.

The discriminator follows a PatchGan architecture \cite{isola2017image}, which divides the input images into overlapping patches and classifies each path as real or fake, rather than using a single descriptor.  This splitting arises naturally as a consequence of the discriminator's convolutional architecture \cite{foster2019generative}.  The discriminator's convolutional blocks consist of $4 \times 4$ convolutional layers with a stride of 2, which decrease the input resolution to half the width and height.  In all layers, we use instance normalization (with no learnable parameters) and LeakyRelu activation.  Finally, the network outputs an $8 \times  8$ single-channel tensor containing the predicted probability for each path.

The training data consists of experimental data from the segmented anisotropic ssTEM dataset \cite{gerhard2013segmented}. Each sample is normalized between -1 and 1, and augmented by mirroring, rotating, shearing, and scaling. Moreover, a Gaussian random noise with standard distribution randomly sampled (per image) between 0 and 0.1 is added to the mask. Adding noise to the mask qualitatively improves the image quality. Specifically, without adding noise, the network is prone to tiling very similar internal structures, especially far away from the border of a mask. This occurs because there is no internal structure in the input, making two nearby regions of the input virtually identical from the point of view of the network. By introducing some internal structure to the mask in the form of noise, we help the network distinguish otherwise very similar regions in the input. An additional benefit is that it is the possible to generate many images from a single mask, just by varying the noise.  A few example training input--output pairs are shown in Fig.~\ref{fig12}C. For this example, we define the loss functions of the generator as,
$l_{\textup{G}}  = \gamma  \cdot \textup{MAE}\left \{ z_{\textup{label}}, z_{\textup{output}} \right \} +  \left ( 1 - \textup{D}(z_{\textup{output}}) \right)^2$ ,
and discriminator as,
$l_\textup{D}  = \textup{D}\left ( z_{\textup{output}} \right )^2 + \left ( 1 - \textup{D}(z_{\textup{label}}) \right)^2$ ,
where $\textup{D}(\cdot)$ denotes the discriminator network prediction, $z_{\textup{label}}$ refers to the ground truth cell-image, and  $z_{\textup{output}}$ is the generated image.  Note that the generator loss function, $l_{\textup{G}}$, aims to minimize the MAE between the generator output image and its target, based on the regularization parameter $\gamma$ set to 0.8. For training, we use the Adam optimizer with a learning rate of $0.0002$ and $\beta_1 = 0.5$ for 1000 epochs, each of which consisting of 16 mini-batches.

The resulting model is able to create new images from masks it has never seen before. We show five such cases in Fig.~\ref{fig13}D. The generated images are not identical to the real cell images in terms of texture and
appearance, which is expected since the masks only contain spatial information about the cells' structures. However, the generated images are qualitatively similar to images from the experimental dataset.

\section{Outlook}

The adoption of new deep-learning methods for the analysis of microscopy data is extremely promising, but it has been hampered by difficulties in generating high-quality training datasets. 
While manually annotated experimental data ensures that the training set is representative of the validation set, it is not guaranteed that the trained network can correctly analyze data obtained with another setup or annotated by another operator. Moreover, it limits the network to human-level accuracy, which is not sufficient for tasks requiring higher level of accuracy, such as single-particle tracking. Synthetically generated data bypasses these issues because the ground truth can be known exactly, and the networks can be trained with parameters that exactly match each user's setup.

Thanks to the increasing available inference speed, it will become easier to perform real-time analysis of microscopy data. This can be used to make real-time decisions, from simple experiment control (e.g., such as controlling the sample flow speed) to more complex decisions (e.g., real-time sorting and optical force feedback systems). For example, one could imagine a completely automated experimental feedback system that applies optical forces to optimize imaging parameters and to acquire the best possible measurements of the quantities of interest.

In this article, we have introduced DeepTrack 2.0, which provides a software environment to develop neural-network models for quantitative digital microscopy, from the generation of training datasets to the deployment of deep-learning solutions tailored to the needs of each user.
We have shown that DeepTrack 2.0 is capable of training neural networks that perform a broad range of tasks using purely synthetic training data. For tasks where it is infeasible to simulate the training set, DeepTrack 2.0 can augment images on the fly to expand the available training set. 
Moreover, DeepTrack 2.0 is complemented by a graphical user interface, allowing users with minimal programming experience to explore and create deep learning models.

We envision DeepTrack 2.0 as a open-source project, where contributors with different areas of expertise can help improve and expand the framework to cover the users' needs. 
Interesting possible directions for the future expansion of  DeepTrack 2.0 can, for example, provide tools for the analysis of time sequences using recurrent neural networks, understand physical processes using reservoir computing, and even support physical implementations of neural networks for greater execution speed and higher energy efficiency. 

Deep-learning has the potential to revolutionize how we do microscopy. However, there are still many challenges to overcome, not least of which figuring out how to obtain enough training data for the model to generalize. We believe that physical simulations will play a crucial part in overcoming this roadblock. As such, we strongly encourage researchers and community collaborators to contribute with objects and models in their area of expertise: from specialized in-sample structures and improved optics simulation methods, to new and exciting neural network architectures.

\begin{acknowledgements}
The authors would like to thank Carlo Manzo for providing the experimental images used in the fourth case study, Jose Alvarez for designing the logo of DeepTrack 2.0, as well as the European Research Council (grant number 677511), the Knut and Alice Wallenberg Foundation, and Vetenskapsr{\aa}det (grant numbers 2016-03523 and 2019-05071) for funding this research.
\end{acknowledgements}


\begin{thebibliography}{96}%
\makeatletter
\providecommand \@ifxundefined [1]{%
 \@ifx{#1\undefined}
}%
\providecommand \@ifnum [1]{%
 \ifnum #1\expandafter \@firstoftwo
 \else \expandafter \@secondoftwo
 \fi
}%
\providecommand \@ifx [1]{%
 \ifx #1\expandafter \@firstoftwo
 \else \expandafter \@secondoftwo
 \fi
}%
\providecommand \natexlab [1]{#1}%
\providecommand \enquote  [1]{``#1''}%
\providecommand \bibnamefont  [1]{#1}%
\providecommand \bibfnamefont [1]{#1}%
\providecommand \citenamefont [1]{#1}%
\providecommand \href@noop [0]{\@secondoftwo}%
\providecommand \href [0]{\begingroup \@sanitize@url \@href}%
\providecommand \@href[1]{\@@startlink{#1}\@@href}%
\providecommand \@@href[1]{\endgroup#1\@@endlink}%
\providecommand \@sanitize@url [0]{\catcode `\\12\catcode `\$12\catcode
  `\&12\catcode `\#12\catcode `\^12\catcode `\_12\catcode `\%12\relax}%
\providecommand \@@startlink[1]{}%
\providecommand \@@endlink[0]{}%
\providecommand \url  [0]{\begingroup\@sanitize@url \@url }%
\providecommand \@url [1]{\endgroup\@href {#1}{\urlprefix }}%
\providecommand \urlprefix  [0]{URL }%
\providecommand \Eprint [0]{\href }%
\providecommand \doibase [0]{http://dx.doi.org/}%
\providecommand \selectlanguage [0]{\@gobble}%
\providecommand \bibinfo  [0]{\@secondoftwo}%
\providecommand \bibfield  [0]{\@secondoftwo}%
\providecommand \translation [1]{[#1]}%
\providecommand \BibitemOpen [0]{}%
\providecommand \bibitemStop [0]{}%
\providecommand \bibitemNoStop [0]{.\EOS\space}%
\providecommand \EOS [0]{\spacefactor3000\relax}%
\providecommand \BibitemShut  [1]{\csname bibitem#1\endcsname}%
\let\auto@bib@innerbib\@empty
\bibitem [{\citenamefont {Perrin}(1910)}]{Perrin1910MouvementMolecules}%
  \BibitemOpen
  \bibfield  {author} {\bibinfo {author} {\bibfnamefont {J.}~\bibnamefont
  {Perrin}},\ }\bibfield  {title} {\enquote {\bibinfo {title} {{Mouvement
  brownien et mol{\'{e}}cules}},}\ }\href {\doibase
  10.1051/jphystap:0191000900500{\"{i}}} {\bibfield  {journal} {\bibinfo
  {journal} {J. Phys. Theor. Appl.}\ }\textbf {\bibinfo {volume} {9}},\
  \bibinfo {pages} {5--39} (\bibinfo {year} {1910})}\BibitemShut {NoStop}%
\bibitem [{\citenamefont {Kappler}(1931)}]{Kappler1931VersucheDrehwaage}%
  \BibitemOpen
  \bibfield  {author} {\bibinfo {author} {\bibfnamefont {E.}~\bibnamefont
  {Kappler}},\ }\bibfield  {title} {\enquote {\bibinfo {title} {{Versuche zur
  Messung der Avogadro-Loschmidtschen Zahl aus der Brownschen Bewegung einer
  Drehwaage}},}\ }\href {\doibase 10.1002/andp.19314030208} {\bibfield
  {journal} {\bibinfo  {journal} {Ann. Phys.}\ }\textbf {\bibinfo {volume}
  {403}},\ \bibinfo {pages} {233--256} (\bibinfo {year} {1931})}\BibitemShut
  {NoStop}%
\bibitem [{\citenamefont {Causley}\ and\ \citenamefont
  {Young}(1955)}]{Causley1955CountingMicroscope}%
  \BibitemOpen
  \bibfield  {author} {\bibinfo {author} {\bibfnamefont {D.}~\bibnamefont
  {Causley}}\ and\ \bibinfo {author} {\bibfnamefont {J.~Z.}\ \bibnamefont
  {Young}},\ }\bibfield  {title} {\enquote {\bibinfo {title} {{Counting and
  sizing of particles with the flying-spot microscope}},}\ }\href {\doibase
  10.1038/176453a0} {\bibfield  {journal} {\bibinfo  {journal} {Nature}\
  }\textbf {\bibinfo {volume} {176}},\ \bibinfo {pages} {453--454} (\bibinfo
  {year} {1955})}\BibitemShut {NoStop}%
\bibitem [{\citenamefont {Geerts}\ \emph {et~al.}(1987)\citenamefont {Geerts},
  \citenamefont {De~Brabander}, \citenamefont {Nuydens}, \citenamefont
  {Geuens}, \citenamefont {Moeremans}, \citenamefont {De~Mey},\ and\
  \citenamefont {Hollenbeck}}]{Geerts1987NanovidMicroscopy}%
  \BibitemOpen
  \bibfield  {author} {\bibinfo {author} {\bibfnamefont {H.}~\bibnamefont
  {Geerts}}, \bibinfo {author} {\bibfnamefont {M.}~\bibnamefont
  {De~Brabander}}, \bibinfo {author} {\bibfnamefont {R.}~\bibnamefont
  {Nuydens}}, \bibinfo {author} {\bibfnamefont {S.}~\bibnamefont {Geuens}},
  \bibinfo {author} {\bibfnamefont {M.}~\bibnamefont {Moeremans}}, \bibinfo
  {author} {\bibfnamefont {J.}~\bibnamefont {De~Mey}}, \ and\ \bibinfo {author}
  {\bibfnamefont {P.}~\bibnamefont {Hollenbeck}},\ }\bibfield  {title}
  {\enquote {\bibinfo {title} {{Nanovid tracking: a new automatic method for
  the study of mobility in living cells based on colloidal gold and video
  microscopy}},}\ }\href {\doibase 10.1016/S0006-3495(87)83271-X} {\bibfield
  {journal} {\bibinfo  {journal} {Biophysical Journal}\ }\textbf {\bibinfo
  {volume} {52}},\ \bibinfo {pages} {775--782} (\bibinfo {year}
  {1987})}\BibitemShut {NoStop}%
\bibitem [{\citenamefont {Crocker}\ and\ \citenamefont
  {Grier}(1996)}]{Crocker1996MethodsStudies}%
  \BibitemOpen
  \bibfield  {author} {\bibinfo {author} {\bibfnamefont {John~C.}\ \bibnamefont
  {Crocker}}\ and\ \bibinfo {author} {\bibfnamefont {David~G.}\ \bibnamefont
  {Grier}},\ }\bibfield  {title} {\enquote {\bibinfo {title} {{Methods of
  digital video microscopy for colloidal studies}},}\ }\href {\doibase
  10.1006/jcis.1996.0217} {\bibfield  {journal} {\bibinfo  {journal} {Journal
  of Colloid and Interface Science}\ }\textbf {\bibinfo {volume} {179}},\
  \bibinfo {pages} {298--310} (\bibinfo {year} {1996})}\BibitemShut {NoStop}%
\bibitem [{\citenamefont {Ronneberger}\ \emph {et~al.}(2015)\citenamefont
  {Ronneberger}, \citenamefont {Fischer},\ and\ \citenamefont
  {Brox}}]{Ronneberger2015U-net:Segmentation}%
  \BibitemOpen
  \bibfield  {author} {\bibinfo {author} {\bibfnamefont {Olaf}\ \bibnamefont
  {Ronneberger}}, \bibinfo {author} {\bibfnamefont {Philipp}\ \bibnamefont
  {Fischer}}, \ and\ \bibinfo {author} {\bibfnamefont {Thomas}\ \bibnamefont
  {Brox}},\ }\bibfield  {title} {\enquote {\bibinfo {title} {{U-net:
  Convolutional networks for biomedical image segmentation}},}\ }in\ \href
  {\doibase 10.1007/978-3-319-24574-4{\_}28} {\emph {\bibinfo {booktitle}
  {Lecture Notes in Computer Science (including subseries Lecture Notes in
  Artificial Intelligence and Lecture Notes in Bioinformatics)}}},\ Vol.\
  \bibinfo {volume} {9351}\ (\bibinfo  {publisher} {Springer Verlag},\ \bibinfo
  {year} {2015})\ pp.\ \bibinfo {pages} {234--241}\BibitemShut {NoStop}%
\bibitem [{\citenamefont {Helgadottir}\ \emph {et~al.}(2019)\citenamefont
  {Helgadottir}, \citenamefont {Argun},\ and\ \citenamefont
  {Volpe}}]{Helgadottir2019DigitalLearning}%
  \BibitemOpen
  \bibfield  {author} {\bibinfo {author} {\bibfnamefont {Saga}\ \bibnamefont
  {Helgadottir}}, \bibinfo {author} {\bibfnamefont {Aykut}\ \bibnamefont
  {Argun}}, \ and\ \bibinfo {author} {\bibfnamefont {Giovanni}\ \bibnamefont
  {Volpe}},\ }\bibfield  {title} {\enquote {\bibinfo {title} {{Digital video
  microscopy enhanced by deep learning}},}\ }\href {\doibase
  10.1364/OPTICA.6.000506} {\bibfield  {journal} {\bibinfo  {journal} {Optica}\
  }\textbf {\bibinfo {volume} {6}},\ \bibinfo {pages} {506} (\bibinfo {year}
  {2019})}\BibitemShut {NoStop}%
\bibitem [{\citenamefont
  {Nordlund}(2017)}]{Nordlund2017EineQuecksilberkugelchen}%
  \BibitemOpen
  \bibfield  {author} {\bibinfo {author} {\bibfnamefont {I.}~\bibnamefont
  {Nordlund}},\ }\bibfield  {title} {\enquote {\bibinfo {title} {{Eine neue
  Bestimmung der Avogadroschen Konstante aus der Brownschen Bewegung kleiner,
  in Wasser suspendierten Quecksilberk{\"{u}}gelchen}},}\ }\href {\doibase
  10.1515/zpch-1914-8703} {\bibfield  {journal} {\bibinfo  {journal} {Z. Phys.
  Chemie}\ }\textbf {\bibinfo {volume} {87U}},\ \bibinfo {pages} {40--62}
  (\bibinfo {year} {2017})}\BibitemShut {NoStop}%
\bibitem [{\citenamefont {Preston}(1976)}]{Preston1976DigitalStates}%
  \BibitemOpen
  \bibfield  {author} {\bibinfo {author} {\bibfnamefont {K.}~\bibnamefont
  {Preston}},\ }\bibfield  {title} {\enquote {\bibinfo {title} {{Digital Image
  Processing in the United States}},}\ }in\ \href {\doibase
  10.1007/978-1-4684-0769-3{\_}1} {\emph {\bibinfo {booktitle} {Digital
  Processing of Biomedical Images}}}\ (\bibinfo  {publisher} {Springer US},\
  \bibinfo {year} {1976})\ pp.\ \bibinfo {pages} {1--10}\BibitemShut {NoStop}%
\bibitem [{\citenamefont {Walton}(1952)}]{Walton1952AutomaticParticles}%
  \BibitemOpen
  \bibfield  {author} {\bibinfo {author} {\bibfnamefont {W.~H.}\ \bibnamefont
  {Walton}},\ }\bibfield  {title} {\enquote {\bibinfo {title} {{Automatic
  Counting of Microscopic Particles}},}\ }\href {\doibase 10.1038/169518a0}
  {\bibfield  {journal} {\bibinfo  {journal} {Nature}\ }\textbf {\bibinfo
  {volume} {169}},\ \bibinfo {pages} {518--520} (\bibinfo {year}
  {1952})}\BibitemShut {NoStop}%
\bibitem [{\citenamefont {Prewitt}\ and\ \citenamefont
  {Mendelsohn}(1965)}]{PrewittTHEIMAGES}%
  \BibitemOpen
  \bibfield  {author} {\bibinfo {author} {\bibfnamefont {J.~M.~S.}\
  \bibnamefont {Prewitt}}\ and\ \bibinfo {author} {\bibfnamefont {M.~L.}\
  \bibnamefont {Mendelsohn}},\ }\bibfield  {title} {\enquote {\bibinfo {title}
  {The analysis of cell images},}\ }\href@noop {} {\bibfield  {journal}
  {\bibinfo  {journal} {Ann. New York Acad. Sci.}\ }\textbf {\bibinfo {volume}
  {128}},\ \bibinfo {pages} {1035--1053} (\bibinfo {year} {1965})}\BibitemShut
  {NoStop}%
\bibitem [{\citenamefont
  {Hounsfield}(1973)}]{Hounsfield1973ComputerizedSystem}%
  \BibitemOpen
  \bibfield  {author} {\bibinfo {author} {\bibfnamefont {G.~N.}\ \bibnamefont
  {Hounsfield}},\ }\href@noop {} {\emph {\bibinfo {title} {British J.
  Radiol.}}},\ \bibinfo {type} {Tech. Rep.}\ (\bibinfo  {institution} {{Central
  Research Laboratories of EMI Limited, Hayes, Middlesex}},\ \bibinfo {year}
  {1973})\BibitemShut {NoStop}%
\bibitem [{\citenamefont {Mansberg}\ \emph {et~al.}(1974)\citenamefont
  {Mansberg}, \citenamefont {Saunders},\ and\ \citenamefont
  {Groner}}]{Mansberg1974TheSystem.}%
  \BibitemOpen
  \bibfield  {author} {\bibinfo {author} {\bibfnamefont {H.~P.}\ \bibnamefont
  {Mansberg}}, \bibinfo {author} {\bibfnamefont {A.~M.}\ \bibnamefont
  {Saunders}}, \ and\ \bibinfo {author} {\bibfnamefont {W.}~\bibnamefont
  {Groner}},\ }\bibfield  {title} {\enquote {\bibinfo {title} {{The Hemalog D
  white cell differential system}},}\ }\href {\doibase 10.1177/22.7.711}
  {\bibfield  {journal} {\bibinfo  {journal} {J. Histochem. Cytochem.}\
  }\textbf {\bibinfo {volume} {22}},\ \bibinfo {pages} {711--24} (\bibinfo
  {year} {1974})}\BibitemShut {NoStop}%
\bibitem [{\citenamefont {Magde}\ \emph {et~al.}(1972)\citenamefont {Magde},
  \citenamefont {Elson},\ and\ \citenamefont
  {Webb}}]{Magde1972ThermodynamicSpectroscopy}%
  \BibitemOpen
  \bibfield  {author} {\bibinfo {author} {\bibfnamefont {D.}~\bibnamefont
  {Magde}}, \bibinfo {author} {\bibfnamefont {E.}~\bibnamefont {Elson}}, \ and\
  \bibinfo {author} {\bibfnamefont {W.~W.}\ \bibnamefont {Webb}},\ }\bibfield
  {title} {\enquote {\bibinfo {title} {{Thermodynamic fluctuations in a
  reacting system measurement by fluorescence correlation spectroscopy}},}\
  }\href {\doibase 10.1103/PhysRevLett.29.705} {\bibfield  {journal} {\bibinfo
  {journal} {Phys. Rev. Lett.}\ }\textbf {\bibinfo {volume} {29}},\ \bibinfo
  {pages} {705--708} (\bibinfo {year} {1972})}\BibitemShut {NoStop}%
\bibitem [{\citenamefont {Axelrod}\ \emph {et~al.}(1976)\citenamefont
  {Axelrod}, \citenamefont {Ravdin}, \citenamefont {Koppel}, \citenamefont
  {Schlessinger}, \citenamefont {Webb}, \citenamefont {Elson},\ and\
  \citenamefont {Podleski}}]{Axelrod1976LateralFibers}%
  \BibitemOpen
  \bibfield  {author} {\bibinfo {author} {\bibfnamefont {D.}~\bibnamefont
  {Axelrod}}, \bibinfo {author} {\bibfnamefont {P.}~\bibnamefont {Ravdin}},
  \bibinfo {author} {\bibfnamefont {D.~E.}\ \bibnamefont {Koppel}}, \bibinfo
  {author} {\bibfnamefont {J.}~\bibnamefont {Schlessinger}}, \bibinfo {author}
  {\bibfnamefont {W.~W.}\ \bibnamefont {Webb}}, \bibinfo {author}
  {\bibfnamefont {E.~L.}\ \bibnamefont {Elson}}, \ and\ \bibinfo {author}
  {\bibfnamefont {T.~R.}\ \bibnamefont {Podleski}},\ }\bibfield  {title}
  {\enquote {\bibinfo {title} {{Lateral motion of fluorescently labeled
  acetylcholine receptors in membranes of developing muscle fibers}},}\ }\href
  {\doibase 10.1073/pnas.73.12.4594} {\bibfield  {journal} {\bibinfo  {journal}
  {Proc. Natl. Acad. Sci. U.S.A.}\ }\textbf {\bibinfo {volume} {73}},\ \bibinfo
  {pages} {4594--4598} (\bibinfo {year} {1976})}\BibitemShut {NoStop}%
\bibitem [{\citenamefont {Manzo}\ \emph {et~al.}(2015)\citenamefont {Manzo},
  \citenamefont {Garcia-Parajo},\ and\ \citenamefont
  {Torreno-Pina}}]{Manzo2015AApproaches}%
  \BibitemOpen
  \bibfield  {author} {\bibinfo {author} {\bibfnamefont {C.}~\bibnamefont
  {Manzo}}, \bibinfo {author} {\bibfnamefont {M.~F.}\ \bibnamefont
  {Garcia-Parajo}}, \ and\ \bibinfo {author} {\bibfnamefont {J.~A.}\
  \bibnamefont {Torreno-Pina}},\ }\bibfield  {title} {\enquote {\bibinfo
  {title} {{A review of progress in single particle tracking: from methods to
  biophysical insights}},}\ }\href {\doibase 10.1088/0034-4885/78/12/124601}
  {\bibfield  {journal} {\bibinfo  {journal} {Rep. Prog. Phys.}\ }\textbf
  {\bibinfo {volume} {78}},\ \bibinfo {pages} {124601} (\bibinfo {year}
  {2015})}\BibitemShut {NoStop}%
\bibitem [{\citenamefont {Schmidt}\ \emph {et~al.}(1996)\citenamefont
  {Schmidt}, \citenamefont {Sch{\"{u}}tz}, \citenamefont {Baumgartner},
  \citenamefont {Gruber},\ and\ \citenamefont
  {Schindler}}]{Schmidt1996ImagingDiffusion}%
  \BibitemOpen
  \bibfield  {author} {\bibinfo {author} {\bibfnamefont {Th.}\ \bibnamefont
  {Schmidt}}, \bibinfo {author} {\bibfnamefont {G.~J.}\ \bibnamefont
  {Sch{\"{u}}tz}}, \bibinfo {author} {\bibfnamefont {W.}~\bibnamefont
  {Baumgartner}}, \bibinfo {author} {\bibfnamefont {H.~J.}\ \bibnamefont
  {Gruber}}, \ and\ \bibinfo {author} {\bibfnamefont {H.}~\bibnamefont
  {Schindler}},\ }\bibfield  {title} {\enquote {\bibinfo {title} {{Imaging of
  single molecule diffusion}},}\ }\href {\doibase 10.1073/pnas.93.7.2926}
  {\bibfield  {journal} {\bibinfo  {journal} {Proc. Natl. Acad. Sci. U.S.A.}\
  }\textbf {\bibinfo {volume} {93}},\ \bibinfo {pages} {2926--2929} (\bibinfo
  {year} {1996})}\BibitemShut {NoStop}%
\bibitem [{\citenamefont {Sch{\"{u}}tz}\ \emph {et~al.}(2000)\citenamefont
  {Sch{\"{u}}tz}, \citenamefont {Kada}, \citenamefont {Pastushenko},\ and\
  \citenamefont {Schindler}}]{Schutz2000PropertiesMicroscopy.}%
  \BibitemOpen
  \bibfield  {author} {\bibinfo {author} {\bibfnamefont {G.~J.}\ \bibnamefont
  {Sch{\"{u}}tz}}, \bibinfo {author} {\bibfnamefont {G.}~\bibnamefont {Kada}},
  \bibinfo {author} {\bibfnamefont {V.}~\bibnamefont {Pastushenko}}, \ and\
  \bibinfo {author} {\bibfnamefont {H.}~\bibnamefont {Schindler}},\ }\bibfield
  {title} {\enquote {\bibinfo {title} {{Properties of lipid microdomains in a
  muscle cell membrane visualized by single molecule microscopy.}}}\
  }\href@noop {} {\bibfield  {journal} {\bibinfo  {journal} {EMBO J.}\ }\textbf
  {\bibinfo {volume} {19}},\ \bibinfo {pages} {892--901} (\bibinfo {year}
  {2000})}\BibitemShut {NoStop}%
\bibitem [{\citenamefont {Alcor}\ \emph {et~al.}(2009)\citenamefont {Alcor},
  \citenamefont {Gouzer},\ and\ \citenamefont
  {Triller}}]{Alcor2009Single-particleDynamics}%
  \BibitemOpen
  \bibfield  {author} {\bibinfo {author} {\bibfnamefont {D.}~\bibnamefont
  {Alcor}}, \bibinfo {author} {\bibfnamefont {G.}~\bibnamefont {Gouzer}}, \
  and\ \bibinfo {author} {\bibfnamefont {A.}~\bibnamefont {Triller}},\
  }\bibfield  {title} {\enquote {\bibinfo {title} {{Single-particle tracking
  methods for the study of membrane receptors dynamics}},}\ }\href {\doibase
  10.1111/j.1460-9568.2009.06927.x} {\bibfield  {journal} {\bibinfo  {journal}
  {Eur. J. Neurosci.}\ }\textbf {\bibinfo {volume} {30}},\ \bibinfo {pages}
  {987--997} (\bibinfo {year} {2009})}\BibitemShut {NoStop}%
\bibitem [{\citenamefont {Dahan}\ \emph {et~al.}(2003)\citenamefont {Dahan},
  \citenamefont {L{\'{e}}vi}, \citenamefont {Luccardini}, \citenamefont
  {Rostaing}, \citenamefont {Riveau},\ and\ \citenamefont
  {Triller}}]{Dahan2003DiffusionTracking}%
  \BibitemOpen
  \bibfield  {author} {\bibinfo {author} {\bibfnamefont {M.}~\bibnamefont
  {Dahan}}, \bibinfo {author} {\bibfnamefont {S.}~\bibnamefont {L{\'{e}}vi}},
  \bibinfo {author} {\bibfnamefont {C.}~\bibnamefont {Luccardini}}, \bibinfo
  {author} {\bibfnamefont {P.}~\bibnamefont {Rostaing}}, \bibinfo {author}
  {\bibfnamefont {B.}~\bibnamefont {Riveau}}, \ and\ \bibinfo {author}
  {\bibfnamefont {A.}~\bibnamefont {Triller}},\ }\bibfield  {title} {\enquote
  {\bibinfo {title} {{Diffusion Dynamics of Glycine Receptors Revealed by
  Single-Quantum Dot Tracking}},}\ }\href {\doibase 10.1126/science.1088525}
  {\bibfield  {journal} {\bibinfo  {journal} {Science}\ }\textbf {\bibinfo
  {volume} {302}},\ \bibinfo {pages} {442--445} (\bibinfo {year}
  {2003})}\BibitemShut {NoStop}%
\bibitem [{\citenamefont {Pinaud}\ \emph {et~al.}(2010)\citenamefont {Pinaud},
  \citenamefont {Clarke}, \citenamefont {Sittner},\ and\ \citenamefont
  {Dahan}}]{Pinaud2010ProbingTime}%
  \BibitemOpen
  \bibfield  {author} {\bibinfo {author} {\bibfnamefont {F.}~\bibnamefont
  {Pinaud}}, \bibinfo {author} {\bibfnamefont {S.}~\bibnamefont {Clarke}},
  \bibinfo {author} {\bibfnamefont {A.}~\bibnamefont {Sittner}}, \ and\
  \bibinfo {author} {\bibfnamefont {M.}~\bibnamefont {Dahan}},\ }\href
  {\doibase 10.1038/nmeth.1444} {\enquote {\bibinfo {title} {{Probing cellular
  events, one quantum dot at a time}},}\ } (\bibinfo {year} {2010})\BibitemShut
  {NoStop}%
\bibitem [{\citenamefont {Yu}\ \emph {et~al.}(2011)\citenamefont {Yu},
  \citenamefont {Chen}, \citenamefont {Qu},\ and\ \citenamefont
  {Niu}}]{Yu2011FastPrecision}%
  \BibitemOpen
  \bibfield  {author} {\bibinfo {author} {\bibfnamefont {B.}~\bibnamefont
  {Yu}}, \bibinfo {author} {\bibfnamefont {D.}~\bibnamefont {Chen}}, \bibinfo
  {author} {\bibfnamefont {J.}~\bibnamefont {Qu}}, \ and\ \bibinfo {author}
  {\bibfnamefont {H.}~\bibnamefont {Niu}},\ }\bibfield  {title} {\enquote
  {\bibinfo {title} {{Fast Fourier domain localization algorithm of a single
  molecule with nanometer precision}},}\ }\href {\doibase 10.1364/ol.36.004317}
  {\bibfield  {journal} {\bibinfo  {journal} {Opt. Lett.}\ }\textbf {\bibinfo
  {volume} {36}},\ \bibinfo {pages} {4317--4319} (\bibinfo {year}
  {2011})}\BibitemShut {NoStop}%
\bibitem [{\citenamefont
  {Parthasarathy}(2012)}]{Parthasarathy2012RapidCenters}%
  \BibitemOpen
  \bibfield  {author} {\bibinfo {author} {\bibfnamefont {R.}~\bibnamefont
  {Parthasarathy}},\ }\bibfield  {title} {\enquote {\bibinfo {title} {{Rapid,
  accurate particle tracking by calculation of radial symmetry centers}},}\
  }\href {\doibase 10.1038/nmeth.2071} {\bibfield  {journal} {\bibinfo
  {journal} {Nat. Methods}\ }\textbf {\bibinfo {volume} {9}},\ \bibinfo {pages}
  {724--726} (\bibinfo {year} {2012})}\BibitemShut {NoStop}%
\bibitem [{\citenamefont {Thompson}\ \emph {et~al.}(2002)\citenamefont
  {Thompson}, \citenamefont {Larson},\ and\ \citenamefont
  {Webb}}]{Thompson2002PreciseProbes}%
  \BibitemOpen
  \bibfield  {author} {\bibinfo {author} {\bibfnamefont {R.~E.}\ \bibnamefont
  {Thompson}}, \bibinfo {author} {\bibfnamefont {D.~R.}\ \bibnamefont
  {Larson}}, \ and\ \bibinfo {author} {\bibfnamefont {W.~W.}\ \bibnamefont
  {Webb}},\ }\bibfield  {title} {\enquote {\bibinfo {title} {{Precise nanometer
  localization analysis for individual fluorescent probes}},}\ }\href {\doibase
  10.1016/S0006-3495(02)75618-X} {\bibfield  {journal} {\bibinfo  {journal}
  {Biophys. J.}\ }\textbf {\bibinfo {volume} {82}},\ \bibinfo {pages}
  {2775--2783} (\bibinfo {year} {2002})}\BibitemShut {NoStop}%
\bibitem [{\citenamefont {Ober}\ \emph {et~al.}(2004)\citenamefont {Ober},
  \citenamefont {Ram},\ and\ \citenamefont
  {Ward}}]{Ober2004LocalizationMicroscopy}%
  \BibitemOpen
  \bibfield  {author} {\bibinfo {author} {\bibfnamefont {R.~J.}\ \bibnamefont
  {Ober}}, \bibinfo {author} {\bibfnamefont {S.}~\bibnamefont {Ram}}, \ and\
  \bibinfo {author} {\bibfnamefont {E.~S.}\ \bibnamefont {Ward}},\ }\bibfield
  {title} {\enquote {\bibinfo {title} {{Localization Accuracy in
  Single-Molecule Microscopy}},}\ }\href {\doibase
  10.1016/S0006-3495(04)74193-4} {\bibfield  {journal} {\bibinfo  {journal}
  {Biophys. J.}\ }\textbf {\bibinfo {volume} {86}},\ \bibinfo {pages}
  {1185--1200} (\bibinfo {year} {2004})}\BibitemShut {NoStop}%
\bibitem [{\citenamefont {Zhang}\ \emph {et~al.}(2007)\citenamefont {Zhang},
  \citenamefont {Zerubia},\ and\ \citenamefont
  {Olivo-Marin}}]{Zhang2007GaussianModels}%
  \BibitemOpen
  \bibfield  {author} {\bibinfo {author} {\bibfnamefont {B.}~\bibnamefont
  {Zhang}}, \bibinfo {author} {\bibfnamefont {J.}~\bibnamefont {Zerubia}}, \
  and\ \bibinfo {author} {\bibfnamefont {J.~C.}\ \bibnamefont {Olivo-Marin}},\
  }\bibfield  {title} {\enquote {\bibinfo {title} {{Gaussian approximations of
  fluorescence microscope point-spread function models}},}\ }\bibfield
  {booktitle} {\emph {\bibinfo {booktitle} {Appl. Opt.}},\ }\href {\doibase
  10.1364/AO.46.001819} {\bibfield  {journal} {\bibinfo  {journal} {Applied
  Optics}\ }\textbf {\bibinfo {volume} {46}},\ \bibinfo {pages} {1819--1829}
  (\bibinfo {year} {2007})}\BibitemShut {NoStop}%
\bibitem [{\citenamefont {Abraham}\ \emph {et~al.}(2009)\citenamefont
  {Abraham}, \citenamefont {Ram}, \citenamefont {Chao}, \citenamefont {Ward},\
  and\ \citenamefont {Ober}}]{Abraham2009QuantitativeTechniques}%
  \BibitemOpen
  \bibfield  {author} {\bibinfo {author} {\bibfnamefont {A.~V.}\ \bibnamefont
  {Abraham}}, \bibinfo {author} {\bibfnamefont {S.}~\bibnamefont {Ram}},
  \bibinfo {author} {\bibfnamefont {J.}~\bibnamefont {Chao}}, \bibinfo {author}
  {\bibfnamefont {E.~S.}\ \bibnamefont {Ward}}, \ and\ \bibinfo {author}
  {\bibfnamefont {R.~J.}\ \bibnamefont {Ober}},\ }\bibfield  {title} {\enquote
  {\bibinfo {title} {{Quantitative study of single molecule location estimation
  techniques}},}\ }\href {\doibase 10.1364/oe.17.023352} {\bibfield  {journal}
  {\bibinfo  {journal} {Opt. Express}\ }\textbf {\bibinfo {volume} {17}},\
  \bibinfo {pages} {23352} (\bibinfo {year} {2009})}\BibitemShut {NoStop}%
\bibitem [{\citenamefont {Stallinga}\ and\ \citenamefont
  {Rieger}(2010)}]{Stallinga2010AccuracyMicroscopy}%
  \BibitemOpen
  \bibfield  {author} {\bibinfo {author} {\bibfnamefont {S.}~\bibnamefont
  {Stallinga}}\ and\ \bibinfo {author} {\bibfnamefont {B.}~\bibnamefont
  {Rieger}},\ }\bibfield  {title} {\enquote {\bibinfo {title} {{Accuracy of the
  Gaussian Point Spread Function model in 2D localization microscopy}},}\
  }\href {\doibase 10.1364/oe.18.024461} {\bibfield  {journal} {\bibinfo
  {journal} {Opt. Express}\ }\textbf {\bibinfo {volume} {18}},\ \bibinfo
  {pages} {24461} (\bibinfo {year} {2010})}\BibitemShut {NoStop}%
\bibitem [{\citenamefont {Stallinga}\ and\ \citenamefont
  {Rieger}(2012)}]{Stallinga2012PositionModel}%
  \BibitemOpen
  \bibfield  {author} {\bibinfo {author} {\bibfnamefont {S.}~\bibnamefont
  {Stallinga}}\ and\ \bibinfo {author} {\bibfnamefont {B.}~\bibnamefont
  {Rieger}},\ }\bibfield  {title} {\enquote {\bibinfo {title} {{Position and
  orientation estimation of fixed dipole emitters using an effective Hermite
  point spread function model}},}\ }\href {\doibase 10.1364/oe.20.005896}
  {\bibfield  {journal} {\bibinfo  {journal} {Opt. Express}\ }\textbf {\bibinfo
  {volume} {20}},\ \bibinfo {pages} {5896} (\bibinfo {year}
  {2012})}\BibitemShut {NoStop}%
\bibitem [{\citenamefont {Lee}\ \emph {et~al.}(2007)\citenamefont {Lee},
  \citenamefont {Roichman}, \citenamefont {Yi}, \citenamefont {Kim},
  \citenamefont {Yang}, \citenamefont {van Blaaderen}, \citenamefont {van
  Oostrum},\ and\ \citenamefont {Grier}}]{Lee2007CharacterizingMicroscopy}%
  \BibitemOpen
  \bibfield  {author} {\bibinfo {author} {\bibfnamefont {S.-H.}\ \bibnamefont
  {Lee}}, \bibinfo {author} {\bibfnamefont {Y.}~\bibnamefont {Roichman}},
  \bibinfo {author} {\bibfnamefont {G.-R.}\ \bibnamefont {Yi}}, \bibinfo
  {author} {\bibfnamefont {S.-H.}\ \bibnamefont {Kim}}, \bibinfo {author}
  {\bibfnamefont {S.-M.}\ \bibnamefont {Yang}}, \bibinfo {author}
  {\bibfnamefont {A.}~\bibnamefont {van Blaaderen}}, \bibinfo {author}
  {\bibfnamefont {P.}~\bibnamefont {van Oostrum}}, \ and\ \bibinfo {author}
  {\bibfnamefont {D.~G.}\ \bibnamefont {Grier}},\ }\bibfield  {title} {\enquote
  {\bibinfo {title} {{Characterizing and tracking single colloidal particles
  with video holographic microscopy}},}\ }\href {\doibase 10.1364/oe.15.018275}
  {\bibfield  {journal} {\bibinfo  {journal} {Opt. Express}\ }\textbf {\bibinfo
  {volume} {15}},\ \bibinfo {pages} {18275} (\bibinfo {year}
  {2007})}\BibitemShut {NoStop}%
\bibitem [{\citenamefont {Holtzer}\ \emph {et~al.}(2007)\citenamefont
  {Holtzer}, \citenamefont {Meckel},\ and\ \citenamefont
  {Schmidt}}]{Holtzer2007NanometricCells}%
  \BibitemOpen
  \bibfield  {author} {\bibinfo {author} {\bibfnamefont {L.}~\bibnamefont
  {Holtzer}}, \bibinfo {author} {\bibfnamefont {T.}~\bibnamefont {Meckel}}, \
  and\ \bibinfo {author} {\bibfnamefont {T.}~\bibnamefont {Schmidt}},\
  }\bibfield  {title} {\enquote {\bibinfo {title} {{Nanometric
  three-dimensional tracking of individual quantum dots in cells}},}\ }\href
  {\doibase 10.1063/1.2437066} {\bibfield  {journal} {\bibinfo  {journal}
  {Appl. Phys. Lett.}\ }\textbf {\bibinfo {volume} {90}},\ \bibinfo {pages}
  {053902} (\bibinfo {year} {2007})}\BibitemShut {NoStop}%
\bibitem [{\citenamefont {Deschout}\ \emph {et~al.}(2014)\citenamefont
  {Deschout}, \citenamefont {Cella~Zanacchi}, \citenamefont {Mlodzianoski},
  \citenamefont {Diaspro}, \citenamefont {Bewersdorf}, \citenamefont {Hess},\
  and\ \citenamefont {Braeckmans}}]{Deschout2014PreciselyMicroscopy}%
  \BibitemOpen
  \bibfield  {author} {\bibinfo {author} {\bibfnamefont {H.}~\bibnamefont
  {Deschout}}, \bibinfo {author} {\bibfnamefont {F.}~\bibnamefont
  {Cella~Zanacchi}}, \bibinfo {author} {\bibfnamefont {M.}~\bibnamefont
  {Mlodzianoski}}, \bibinfo {author} {\bibfnamefont {A.}~\bibnamefont
  {Diaspro}}, \bibinfo {author} {\bibfnamefont {J.}~\bibnamefont {Bewersdorf}},
  \bibinfo {author} {\bibfnamefont {S.~T.}\ \bibnamefont {Hess}}, \ and\
  \bibinfo {author} {\bibfnamefont {K.}~\bibnamefont {Braeckmans}},\ }\bibfield
   {title} {\enquote {\bibinfo {title} {{Precisely and accurately localizing
  single emitters in fluorescence microscopy}},}\ }\bibfield  {booktitle}
  {\emph {\bibinfo {booktitle} {Nat. Methods}},\ }\href {\doibase
  10.1038/nmeth.2843} {\bibfield  {journal} {\bibinfo  {journal} {Nature
  Methods}\ }\textbf {\bibinfo {volume} {11}},\ \bibinfo {pages} {253--266}
  (\bibinfo {year} {2014})}\BibitemShut {NoStop}%
\bibitem [{\citenamefont {Godinez}\ and\ \citenamefont
  {Rohr}(2015)}]{Godinez2015TrackingAssociation}%
  \BibitemOpen
  \bibfield  {author} {\bibinfo {author} {\bibfnamefont {W.~J.}\ \bibnamefont
  {Godinez}}\ and\ \bibinfo {author} {\bibfnamefont {K.}~\bibnamefont {Rohr}},\
  }\bibfield  {title} {\enquote {\bibinfo {title} {{Tracking multiple particles
  in fluorescence time-lapse microscopy images via probabilistic data
  association}},}\ }\href {\doibase 10.1109/TMI.2014.2359541} {\bibfield
  {journal} {\bibinfo  {journal} {IEEE Trans. Medical Imaging}\ }\textbf
  {\bibinfo {volume} {34}},\ \bibinfo {pages} {415--432} (\bibinfo {year}
  {2015})}\BibitemShut {NoStop}%
\bibitem [{\citenamefont {Chenouard}\ \emph {et~al.}(2014)\citenamefont
  {Chenouard}, \citenamefont {Smal}, \citenamefont {De~Chaumont}, \citenamefont
  {Ma{\v{s}}ka}, \citenamefont {Sbalzarini}, \citenamefont {Gong},
  \citenamefont {Cardinale}, \citenamefont {Carthel}, \citenamefont
  {Coraluppi}, \citenamefont {Winter}, \citenamefont {Cohen}, \citenamefont
  {Godinez}, \citenamefont {Rohr}, \citenamefont {Kalaidzidis}, \citenamefont
  {Liang}, \citenamefont {Duncan}, \citenamefont {Shen}, \citenamefont {Xu},
  \citenamefont {Magnusson}, \citenamefont {Jald{\'{e}}n}, \citenamefont
  {Blau}, \citenamefont {Paul-Gilloteaux}, \citenamefont {Roudot},
  \citenamefont {Kervrann}, \citenamefont {Waharte}, \citenamefont {Tinevez},
  \citenamefont {Shorte}, \citenamefont {Willemse}, \citenamefont {Celler},
  \citenamefont {Van~Wezel}, \citenamefont {Dan}, \citenamefont {Tsai},
  \citenamefont {De~Sol{\'{o}}rzano}, \citenamefont {Olivo-Marin},\ and\
  \citenamefont {Meijering}}]{Chenouard2014ObjectiveMethods}%
  \BibitemOpen
  \bibfield  {author} {\bibinfo {author} {\bibfnamefont {N.}~\bibnamefont
  {Chenouard}}, \bibinfo {author} {\bibfnamefont {I.}~\bibnamefont {Smal}},
  \bibinfo {author} {\bibfnamefont {F.}~\bibnamefont {De~Chaumont}}, \bibinfo
  {author} {\bibfnamefont {M.}~\bibnamefont {Ma{\v{s}}ka}}, \bibinfo {author}
  {\bibfnamefont {I.~F.}\ \bibnamefont {Sbalzarini}}, \bibinfo {author}
  {\bibfnamefont {Y.}~\bibnamefont {Gong}}, \bibinfo {author} {\bibfnamefont
  {J.}~\bibnamefont {Cardinale}}, \bibinfo {author} {\bibfnamefont
  {C.}~\bibnamefont {Carthel}}, \bibinfo {author} {\bibfnamefont
  {S.}~\bibnamefont {Coraluppi}}, \bibinfo {author} {\bibfnamefont
  {M.}~\bibnamefont {Winter}}, \bibinfo {author} {\bibfnamefont {A.~R.}\
  \bibnamefont {Cohen}}, \bibinfo {author} {\bibfnamefont {W.~J.}\ \bibnamefont
  {Godinez}}, \bibinfo {author} {\bibfnamefont {K.}~\bibnamefont {Rohr}},
  \bibinfo {author} {\bibfnamefont {Y.}~\bibnamefont {Kalaidzidis}}, \bibinfo
  {author} {\bibfnamefont {L.}~\bibnamefont {Liang}}, \bibinfo {author}
  {\bibfnamefont {J.}~\bibnamefont {Duncan}}, \bibinfo {author} {\bibfnamefont
  {H.}~\bibnamefont {Shen}}, \bibinfo {author} {\bibfnamefont {Y.}~\bibnamefont
  {Xu}}, \bibinfo {author} {\bibfnamefont {K.~E.~G.}\ \bibnamefont
  {Magnusson}}, \bibinfo {author} {\bibfnamefont {J.}~\bibnamefont
  {Jald{\'{e}}n}}, \bibinfo {author} {\bibfnamefont {H.~M.}\ \bibnamefont
  {Blau}}, \bibinfo {author} {\bibfnamefont {P.}~\bibnamefont
  {Paul-Gilloteaux}}, \bibinfo {author} {\bibfnamefont {P.}~\bibnamefont
  {Roudot}}, \bibinfo {author} {\bibfnamefont {C.}~\bibnamefont {Kervrann}},
  \bibinfo {author} {\bibfnamefont {F.}~\bibnamefont {Waharte}}, \bibinfo
  {author} {\bibfnamefont {J.~Y.}\ \bibnamefont {Tinevez}}, \bibinfo {author}
  {\bibfnamefont {S.~L.}\ \bibnamefont {Shorte}}, \bibinfo {author}
  {\bibfnamefont {J.}~\bibnamefont {Willemse}}, \bibinfo {author}
  {\bibfnamefont {K.}~\bibnamefont {Celler}}, \bibinfo {author} {\bibfnamefont
  {G.~P.}\ \bibnamefont {Van~Wezel}}, \bibinfo {author} {\bibfnamefont {H.~W.}\
  \bibnamefont {Dan}}, \bibinfo {author} {\bibfnamefont {Y.~S.}\ \bibnamefont
  {Tsai}}, \bibinfo {author} {\bibfnamefont {C.~O.}\ \bibnamefont
  {De~Sol{\'{o}}rzano}}, \bibinfo {author} {\bibfnamefont {J.~C.}\ \bibnamefont
  {Olivo-Marin}}, \ and\ \bibinfo {author} {\bibfnamefont {E.}~\bibnamefont
  {Meijering}},\ }\bibfield  {title} {\enquote {\bibinfo {title} {{Objective
  comparison of particle tracking methods}},}\ }\href {\doibase
  10.1038/nmeth.2808} {\bibfield  {journal} {\bibinfo  {journal} {Nat.
  Methods}\ }\textbf {\bibinfo {volume} {11}},\ \bibinfo {pages} {281--289}
  (\bibinfo {year} {2014})}\BibitemShut {NoStop}%
\bibitem [{\citenamefont {Lecun}\ \emph {et~al.}(2015)\citenamefont {Lecun},
  \citenamefont {Bengio},\ and\ \citenamefont
  {Hinton}}]{Lecun2015DeepLearning}%
  \BibitemOpen
  \bibfield  {author} {\bibinfo {author} {\bibfnamefont {Y.}~\bibnamefont
  {Lecun}}, \bibinfo {author} {\bibfnamefont {Y.}~\bibnamefont {Bengio}}, \
  and\ \bibinfo {author} {\bibfnamefont {G.}~\bibnamefont {Hinton}},\
  }\bibfield  {title} {\enquote {\bibinfo {title} {{Deep learning}},}\
  }\bibfield  {booktitle} {\emph {\bibinfo {booktitle} {Nature}},\ }\href
  {\doibase 10.1038/nature14539} {\bibfield  {journal} {\bibinfo  {journal}
  {Nature}\ }\textbf {\bibinfo {volume} {521}},\ \bibinfo {pages} {436--444}
  (\bibinfo {year} {2015})}\BibitemShut {NoStop}%
\bibitem [{\citenamefont {Ciresan}\ \emph {et~al.}(2012)\citenamefont
  {Ciresan}, \citenamefont {Meier},\ and\ \citenamefont
  {Schmidhuber}}]{Ciresan2012Multi-columnClassification}%
  \BibitemOpen
  \bibfield  {author} {\bibinfo {author} {\bibfnamefont {D.}~\bibnamefont
  {Ciresan}}, \bibinfo {author} {\bibfnamefont {U.}~\bibnamefont {Meier}}, \
  and\ \bibinfo {author} {\bibfnamefont {J.}~\bibnamefont {Schmidhuber}},\
  }\bibfield  {title} {{\selectlanguage {English}\enquote {\bibinfo {title}
  {Multi-column deep neural networks for image classification},}\ }}in\
  \href@noop {} {{\selectlanguage {English}\emph {\bibinfo {booktitle} {2012
  IEEE Conf. Computer Vision Pattern Recognition}}}}\ (\bibinfo {year} {2012})\
  pp.\ \bibinfo {pages} {3642--3649}\BibitemShut {NoStop}%
\bibitem [{\citenamefont {Shelhamer}\ \emph {et~al.}(2017)\citenamefont
  {Shelhamer}, \citenamefont {Long},\ and\ \citenamefont
  {Darrell}}]{Shelhamer2017FullySegmentation}%
  \BibitemOpen
  \bibfield  {author} {\bibinfo {author} {\bibfnamefont {Evan}\ \bibnamefont
  {Shelhamer}}, \bibinfo {author} {\bibfnamefont {Jonathan}\ \bibnamefont
  {Long}}, \ and\ \bibinfo {author} {\bibfnamefont {Trevor}\ \bibnamefont
  {Darrell}},\ }\bibfield  {title} {\enquote {\bibinfo {title} {{Fully
  Convolutional Networks for Semantic Segmentation}},}\ }\href {\doibase
  10.1109/TPAMI.2016.2572683} {\bibfield  {journal} {\bibinfo  {journal} {IEEE
  Transactions on Pattern Analysis and Machine Intelligence}\ }\textbf
  {\bibinfo {volume} {39}},\ \bibinfo {pages} {640--651} (\bibinfo {year}
  {2017})}\BibitemShut {NoStop}%
\bibitem [{\citenamefont {Li}\ \emph {et~al.}(2016)\citenamefont {Li},
  \citenamefont {Zuo},\ and\ \citenamefont
  {Zhang}}]{Li2016ConvolutionalGeneration}%
  \BibitemOpen
  \bibfield  {author} {\bibinfo {author} {\bibfnamefont {Mu}~\bibnamefont
  {Li}}, \bibinfo {author} {\bibfnamefont {Wangmeng}\ \bibnamefont {Zuo}}, \
  and\ \bibinfo {author} {\bibfnamefont {David}\ \bibnamefont {Zhang}},\
  }\bibfield  {title} {\enquote {\bibinfo {title} {{Convolutional Network for
  Attribute-driven and Identity-preserving Human Face Generation}},}\ }\href
  {http://arxiv.org/abs/1608.06434} {\  (\bibinfo {year} {2016})}\BibitemShut
  {NoStop}%
\bibitem [{\citenamefont {Hannel}\ \emph {et~al.}(2018)\citenamefont {Hannel},
  \citenamefont {Abdulali}, \citenamefont {O’Brien},\ and\ \citenamefont
  {Grier}}]{Hannel2018Machine-learningParticles}%
  \BibitemOpen
  \bibfield  {author} {\bibinfo {author} {\bibfnamefont {Mark~D.}\ \bibnamefont
  {Hannel}}, \bibinfo {author} {\bibfnamefont {Aidan}\ \bibnamefont
  {Abdulali}}, \bibinfo {author} {\bibfnamefont {Michael}\ \bibnamefont
  {O’Brien}}, \ and\ \bibinfo {author} {\bibfnamefont {David~G.}\
  \bibnamefont {Grier}},\ }\bibfield  {title} {\enquote {\bibinfo {title}
  {{Machine-learning techniques for fast and accurate feature localization in
  holograms of colloidal particles}},}\ }\href {\doibase 10.1364/oe.26.015221}
  {\bibfield  {journal} {\bibinfo  {journal} {Optics Express}\ }\textbf
  {\bibinfo {volume} {26}},\ \bibinfo {pages} {15221} (\bibinfo {year}
  {2018})}\BibitemShut {NoStop}%
\bibitem [{\citenamefont {Newby}\ \emph {et~al.}(2018)\citenamefont {Newby},
  \citenamefont {Schaefer}, \citenamefont {Lee}, \citenamefont {Forest},\ and\
  \citenamefont {Lai}}]{Newby2018Convolutional3D}%
  \BibitemOpen
  \bibfield  {author} {\bibinfo {author} {\bibfnamefont {Jay~M.}\ \bibnamefont
  {Newby}}, \bibinfo {author} {\bibfnamefont {Alison~M.}\ \bibnamefont
  {Schaefer}}, \bibinfo {author} {\bibfnamefont {Phoebe~T.}\ \bibnamefont
  {Lee}}, \bibinfo {author} {\bibfnamefont {M.~Gregory}\ \bibnamefont
  {Forest}}, \ and\ \bibinfo {author} {\bibfnamefont {Samuel~K.}\ \bibnamefont
  {Lai}},\ }\bibfield  {title} {\enquote {\bibinfo {title} {{Convolutional
  neural networks automate detection for tracking of submicron-scale particles
  in 2D and 3D}},}\ }\href {\doibase 10.1073/pnas.1804420115} {\bibfield
  {journal} {\bibinfo  {journal} {Proceedings of the National Academy of
  Sciences of the United States of America}\ }\textbf {\bibinfo {volume}
  {115}},\ \bibinfo {pages} {9026--9031} (\bibinfo {year} {2018})}\BibitemShut
  {NoStop}%
\bibitem [{\citenamefont {Chen}\ \emph {et~al.}(2016)\citenamefont {Chen},
  \citenamefont {Mahjoubfar}, \citenamefont {Tai}, \citenamefont {Blaby},
  \citenamefont {Huang}, \citenamefont {Niazi},\ and\ \citenamefont
  {Jalali}}]{ChenDeepClassification}%
  \BibitemOpen
  \bibfield  {author} {\bibinfo {author} {\bibfnamefont {Claire~Lifan}\
  \bibnamefont {Chen}}, \bibinfo {author} {\bibfnamefont {Ata}\ \bibnamefont
  {Mahjoubfar}}, \bibinfo {author} {\bibfnamefont {Li-Chia}\ \bibnamefont
  {Tai}}, \bibinfo {author} {\bibfnamefont {Ian~K.}\ \bibnamefont {Blaby}},
  \bibinfo {author} {\bibfnamefont {Allen}\ \bibnamefont {Huang}}, \bibinfo
  {author} {\bibfnamefont {Kayvan~Reza}\ \bibnamefont {Niazi}}, \ and\ \bibinfo
  {author} {\bibfnamefont {Bahram}\ \bibnamefont {Jalali}},\ }\bibfield
  {title} {\enquote {\bibinfo {title} {Deep learning in label-free cell
  classification},}\ }\href {\doibase 10.1038/srep21471} {\bibfield  {journal}
  {\bibinfo  {journal} {Scientific Reports}\ }\textbf {\bibinfo {volume} {6}},\
  \bibinfo {pages} {21471} (\bibinfo {year} {2016})}\BibitemShut {NoStop}%
\bibitem [{\citenamefont {Coudray}\ \emph {et~al.}(2018)\citenamefont
  {Coudray}, \citenamefont {Ocampo}, \citenamefont {Sakellaropoulos},
  \citenamefont {Narula}, \citenamefont {Snuderl}, \citenamefont {Feny{\"o}},
  \citenamefont {Moreira}, \citenamefont {Razavian},\ and\ \citenamefont
  {Tsirigos}}]{CoudrayClassification3}%
  \BibitemOpen
  \bibfield  {author} {\bibinfo {author} {\bibfnamefont {Nicolas}\ \bibnamefont
  {Coudray}}, \bibinfo {author} {\bibfnamefont {Paolo~Santiago}\ \bibnamefont
  {Ocampo}}, \bibinfo {author} {\bibfnamefont {Theodore}\ \bibnamefont
  {Sakellaropoulos}}, \bibinfo {author} {\bibfnamefont {Navneet}\ \bibnamefont
  {Narula}}, \bibinfo {author} {\bibfnamefont {Matija}\ \bibnamefont
  {Snuderl}}, \bibinfo {author} {\bibfnamefont {David}\ \bibnamefont
  {Feny{\"o}}}, \bibinfo {author} {\bibfnamefont {Andre~L.}\ \bibnamefont
  {Moreira}}, \bibinfo {author} {\bibfnamefont {Narges}\ \bibnamefont
  {Razavian}}, \ and\ \bibinfo {author} {\bibfnamefont {Aristotelis}\
  \bibnamefont {Tsirigos}},\ }\bibfield  {title} {\enquote {\bibinfo {title}
  {Classification and mutation prediction from non--small cell lung cancer
  histopathology images using deep learning},}\ }\href {\doibase
  10.1038/s41591-018-0177-5} {\bibfield  {journal} {\bibinfo  {journal} {Nature
  Medicine}\ }\textbf {\bibinfo {volume} {24}},\ \bibinfo {pages} {1559--1567}
  (\bibinfo {year} {2018})}\BibitemShut {NoStop}%
\bibitem [{\citenamefont {Zhang}\ \emph {et~al.}(2017)\citenamefont {Zhang},
  \citenamefont {Lu}, \citenamefont {Nogues}, \citenamefont {Summers},
  \citenamefont {Liu},\ and\ \citenamefont
  {Yao}}]{ZhangDeepPap:Classification}%
  \BibitemOpen
  \bibfield  {author} {\bibinfo {author} {\bibfnamefont {L}~\bibnamefont
  {Zhang}}, \bibinfo {author} {\bibfnamefont {L}~\bibnamefont {Lu}}, \bibinfo
  {author} {\bibfnamefont {I}~\bibnamefont {Nogues}}, \bibinfo {author}
  {\bibfnamefont {RM}~\bibnamefont {Summers}}, \bibinfo {author} {\bibfnamefont
  {S}~\bibnamefont {Liu}}, \ and\ \bibinfo {author} {\bibfnamefont
  {J}~\bibnamefont {Yao}},\ }\bibfield  {title} {\enquote {\bibinfo {title}
  {{DeepPap: deep convolutional networks for cervical cell classification}},}\
  }\href {https://pubmed.ncbi.nlm.nih.gov/28541229/} {\bibfield  {journal}
  {\bibinfo  {journal} {IEEE journal of biomedical and health informatics}\ }
  (\bibinfo {year} {2017})}\BibitemShut {NoStop}%
\bibitem [{\citenamefont {Falk}\ \emph {et~al.}(2019)\citenamefont {Falk},
  \citenamefont {Mai}, \citenamefont {Bensch}, \citenamefont
  {{\c{C}}i{\c{c}}ek}, \citenamefont {Abdulkadir}, \citenamefont {Marrakchi},
  \citenamefont {B{\"o}hm}, \citenamefont {Deubner}, \citenamefont
  {J{\"a}ckel}, \citenamefont {Seiwald}, \citenamefont {Dovzhenko},
  \citenamefont {Tietz}, \citenamefont {Dal~Bosco}, \citenamefont {Walsh},
  \citenamefont {Saltukoglu}, \citenamefont {Tay}, \citenamefont {Prinz},
  \citenamefont {Palme}, \citenamefont {Simons}, \citenamefont {Diester},
  \citenamefont {Brox},\ and\ \citenamefont {Ronneberger}}]{Falk2019}%
  \BibitemOpen
  \bibfield  {author} {\bibinfo {author} {\bibfnamefont {Thorsten}\
  \bibnamefont {Falk}}, \bibinfo {author} {\bibfnamefont {Dominic}\
  \bibnamefont {Mai}}, \bibinfo {author} {\bibfnamefont {Robert}\ \bibnamefont
  {Bensch}}, \bibinfo {author} {\bibfnamefont {{\"O}zg{\"u}n}\ \bibnamefont
  {{\c{C}}i{\c{c}}ek}}, \bibinfo {author} {\bibfnamefont {Ahmed}\ \bibnamefont
  {Abdulkadir}}, \bibinfo {author} {\bibfnamefont {Yassine}\ \bibnamefont
  {Marrakchi}}, \bibinfo {author} {\bibfnamefont {Anton}\ \bibnamefont
  {B{\"o}hm}}, \bibinfo {author} {\bibfnamefont {Jan}\ \bibnamefont {Deubner}},
  \bibinfo {author} {\bibfnamefont {Zoe}\ \bibnamefont {J{\"a}ckel}}, \bibinfo
  {author} {\bibfnamefont {Katharina}\ \bibnamefont {Seiwald}}, \bibinfo
  {author} {\bibfnamefont {Alexander}\ \bibnamefont {Dovzhenko}}, \bibinfo
  {author} {\bibfnamefont {Olaf}\ \bibnamefont {Tietz}}, \bibinfo {author}
  {\bibfnamefont {Cristina}\ \bibnamefont {Dal~Bosco}}, \bibinfo {author}
  {\bibfnamefont {Sean}\ \bibnamefont {Walsh}}, \bibinfo {author}
  {\bibfnamefont {Deniz}\ \bibnamefont {Saltukoglu}}, \bibinfo {author}
  {\bibfnamefont {Tuan~Leng}\ \bibnamefont {Tay}}, \bibinfo {author}
  {\bibfnamefont {Marco}\ \bibnamefont {Prinz}}, \bibinfo {author}
  {\bibfnamefont {Klaus}\ \bibnamefont {Palme}}, \bibinfo {author}
  {\bibfnamefont {Matias}\ \bibnamefont {Simons}}, \bibinfo {author}
  {\bibfnamefont {Ilka}\ \bibnamefont {Diester}}, \bibinfo {author}
  {\bibfnamefont {Thomas}\ \bibnamefont {Brox}}, \ and\ \bibinfo {author}
  {\bibfnamefont {Olaf}\ \bibnamefont {Ronneberger}},\ }\bibfield  {title}
  {\enquote {\bibinfo {title} {U-net: deep learning for cell counting,
  detection, and morphometry},}\ }\href {\doibase 10.1038/s41592-018-0261-2}
  {\bibfield  {journal} {\bibinfo  {journal} {Nature Methods}\ }\textbf
  {\bibinfo {volume} {16}},\ \bibinfo {pages} {67--70} (\bibinfo {year}
  {2019})}\BibitemShut {NoStop}%
\bibitem [{\citenamefont {Midtvedt}\ \emph
  {et~al.}(2020{\natexlab{a}})\citenamefont {Midtvedt}, \citenamefont
  {Ols{\'{e}}n}, \citenamefont {Eklund}, \citenamefont {H{\"{o}}{\"{o}}k},
  \citenamefont {Adiels}, \citenamefont {Volpe},\ and\ \citenamefont
  {Midtvedt}}]{Midtvedt2020HolographicLearning}%
  \BibitemOpen
  \bibfield  {author} {\bibinfo {author} {\bibfnamefont {Benjamin}\
  \bibnamefont {Midtvedt}}, \bibinfo {author} {\bibfnamefont {Erik}\
  \bibnamefont {Ols{\'{e}}n}}, \bibinfo {author} {\bibfnamefont {Fredrik}\
  \bibnamefont {Eklund}}, \bibinfo {author} {\bibfnamefont {Fredrik}\
  \bibnamefont {H{\"{o}}{\"{o}}k}}, \bibinfo {author} {\bibfnamefont
  {Caroline~Beck}\ \bibnamefont {Adiels}}, \bibinfo {author} {\bibfnamefont
  {Giovanni}\ \bibnamefont {Volpe}}, \ and\ \bibinfo {author} {\bibfnamefont
  {Daniel}\ \bibnamefont {Midtvedt}},\ }\bibfield  {title} {\enquote {\bibinfo
  {title} {{Holographic characterisation of subwavelength particles enhanced by
  deep learning}},}\ }\href {http://arxiv.org/abs/2006.11154} {\  (\bibinfo
  {year} {2020}{\natexlab{a}})}\BibitemShut {NoStop}%
\bibitem [{\citenamefont {Altman}\ and\ \citenamefont
  {Grier}(2020)}]{Altman2020CATCH:Networks}%
  \BibitemOpen
  \bibfield  {author} {\bibinfo {author} {\bibfnamefont {Lauren~E.}\
  \bibnamefont {Altman}}\ and\ \bibinfo {author} {\bibfnamefont {David~G.}\
  \bibnamefont {Grier}},\ }\bibfield  {title} {\enquote {\bibinfo {title}
  {{CATCH: Characterizing and Tracking Colloids Holographically Using Deep
  Neural Networks}},}\ }\href {\doibase 10.1021/acs.jpcb.9b10463} {\bibfield
  {journal} {\bibinfo  {journal} {Journal of Physical Chemistry B}\ }\textbf
  {\bibinfo {volume} {124}},\ \bibinfo {pages} {1602--1610} (\bibinfo {year}
  {2020})}\BibitemShut {NoStop}%
\bibitem [{\citenamefont {Xie}\ \emph {et~al.}(2018)\citenamefont {Xie},
  \citenamefont {Alison~Noble},\ and\ \citenamefont
  {Zisserman}}]{Xie2018MicroscopyNetworks}%
  \BibitemOpen
  \bibfield  {author} {\bibinfo {author} {\bibfnamefont {Weidi}\ \bibnamefont
  {Xie}}, \bibinfo {author} {\bibfnamefont {J}~\bibnamefont {Alison~Noble}}, \
  and\ \bibinfo {author} {\bibfnamefont {Andrew}\ \bibnamefont {Zisserman}},\
  }\bibfield  {title} {\enquote {\bibinfo {title} {{Microscopy cell counting
  and detection with fully convolutional regression networks}},}\ }\href
  {\doibase 10.1080/21681163.2016.1149104} {\bibfield  {journal} {\bibinfo
  {journal} {COMPUTER METHODS IN BIOMECHANICS AND BIOMEDICAL ENGINEERING:
  IMAGING {\&} VISUALIZATION}\ }\textbf {\bibinfo {volume} {6}},\ \bibinfo
  {pages} {283--292} (\bibinfo {year} {2018})}\BibitemShut {NoStop}%
\bibitem [{\citenamefont {Wu}\ \emph {et~al.}(2018)\citenamefont {Wu},
  \citenamefont {Rivenson}, \citenamefont {Zhang}, \citenamefont {Wei},
  \citenamefont {G{\"{u}}naydin}, \citenamefont {Lin},\ and\ \citenamefont
  {Ozcan}}]{Wu2018ExtendedRecovery}%
  \BibitemOpen
  \bibfield  {author} {\bibinfo {author} {\bibfnamefont {Yichen}\ \bibnamefont
  {Wu}}, \bibinfo {author} {\bibfnamefont {Yair}\ \bibnamefont {Rivenson}},
  \bibinfo {author} {\bibfnamefont {Yibo}\ \bibnamefont {Zhang}}, \bibinfo
  {author} {\bibfnamefont {Zhensong}\ \bibnamefont {Wei}}, \bibinfo {author}
  {\bibfnamefont {Harun}\ \bibnamefont {G{\"{u}}naydin}}, \bibinfo {author}
  {\bibfnamefont {Xing}\ \bibnamefont {Lin}}, \ and\ \bibinfo {author}
  {\bibfnamefont {Aydogan}\ \bibnamefont {Ozcan}},\ }\bibfield  {title}
  {\enquote {\bibinfo {title} {{Extended depth-of-field in holographic imaging
  using deep-learning-based autofocusing and phase recovery}},}\ }\href
  {\doibase 10.1364/optica.5.000704} {\bibfield  {journal} {\bibinfo  {journal}
  {Optica}\ }\textbf {\bibinfo {volume} {5}},\ \bibinfo {pages} {704} (\bibinfo
  {year} {2018})}\BibitemShut {NoStop}%
\bibitem [{\citenamefont {Nehme}\ \emph {et~al.}(2018)\citenamefont {Nehme},
  \citenamefont {Weiss}, \citenamefont {Michaeli},\ and\ \citenamefont
  {Shechtman}}]{Nehme2018Deep-STORM:Learning}%
  \BibitemOpen
  \bibfield  {author} {\bibinfo {author} {\bibfnamefont {Elias}\ \bibnamefont
  {Nehme}}, \bibinfo {author} {\bibfnamefont {Lucien~E.}\ \bibnamefont
  {Weiss}}, \bibinfo {author} {\bibfnamefont {Tomer}\ \bibnamefont {Michaeli}},
  \ and\ \bibinfo {author} {\bibfnamefont {Yoav}\ \bibnamefont {Shechtman}},\
  }\bibfield  {title} {\enquote {\bibinfo {title} {{Deep-STORM:
  super-resolution single-molecule microscopy by deep learning}},}\ }\href
  {\doibase 10.1364/optica.5.000458} {\bibfield  {journal} {\bibinfo  {journal}
  {Optica}\ }\textbf {\bibinfo {volume} {5}},\ \bibinfo {pages} {458} (\bibinfo
  {year} {2018})}\BibitemShut {NoStop}%
\bibitem [{\citenamefont {Ouyang}\ \emph {et~al.}(2018)\citenamefont {Ouyang},
  \citenamefont {Aristov}, \citenamefont {Lelek}, \citenamefont {Hao},\ and\
  \citenamefont {Zimmer}}]{Ouyang2018DeepMicroscopy}%
  \BibitemOpen
  \bibfield  {author} {\bibinfo {author} {\bibfnamefont {Wei}\ \bibnamefont
  {Ouyang}}, \bibinfo {author} {\bibfnamefont {Andrey}\ \bibnamefont
  {Aristov}}, \bibinfo {author} {\bibfnamefont {Mickaël}\ \bibnamefont
  {Lelek}}, \bibinfo {author} {\bibfnamefont {Xian}\ \bibnamefont {Hao}}, \
  and\ \bibinfo {author} {\bibfnamefont {Christophe}\ \bibnamefont {Zimmer}},\
  }\bibfield  {title} {\enquote {\bibinfo {title} {{Deep learning massively
  accelerates super-resolution localization microscopy}},}\ }\href {\doibase
  10.1038/nbt.4106} {\bibfield  {journal} {\bibinfo  {journal} {Nature
  Biotechnology}\ }\textbf {\bibinfo {volume} {36}},\ \bibinfo {pages}
  {460--468} (\bibinfo {year} {2018})}\BibitemShut {NoStop}%
\bibitem [{\citenamefont {Xing}\ \emph {et~al.}(2018)\citenamefont {Xing},
  \citenamefont {Xie}, \citenamefont {Su}, \citenamefont {Liu},\ and\
  \citenamefont {Yang}}]{Xing2018DeepSurvey}%
  \BibitemOpen
  \bibfield  {author} {\bibinfo {author} {\bibfnamefont {Fuyong}\ \bibnamefont
  {Xing}}, \bibinfo {author} {\bibfnamefont {Yuanpu}\ \bibnamefont {Xie}},
  \bibinfo {author} {\bibfnamefont {Hai}\ \bibnamefont {Su}}, \bibinfo {author}
  {\bibfnamefont {Fujun}\ \bibnamefont {Liu}}, \ and\ \bibinfo {author}
  {\bibfnamefont {Lin}\ \bibnamefont {Yang}},\ }\bibfield  {title} {\enquote
  {\bibinfo {title} {{Deep Learning in Microscopy Image Analysis: A Survey}},}\
  }\href {\doibase 10.1109/TNNLS.2017.2766168} {\bibfield  {journal} {\bibinfo
  {journal} {IEEE Transactions on Neural Networks and Learning Systems}\
  }\textbf {\bibinfo {volume} {29}},\ \bibinfo {pages} {4550--4568} (\bibinfo
  {year} {2018})}\BibitemShut {NoStop}%
\bibitem [{\citenamefont {Mehlig}(2019)}]{mehlig2019artificial}%
  \BibitemOpen
  \bibfield  {author} {\bibinfo {author} {\bibfnamefont {B.}~\bibnamefont
  {Mehlig}},\ }\bibfield  {title} {\enquote {\bibinfo {title} {Artificial
  neural networks},}\ }\href@noop {} {\bibfield  {journal} {\bibinfo  {journal}
  {arXiv preprint arXiv:1901.05639}\ } (\bibinfo {year} {2019})}\BibitemShut
  {NoStop}%
\bibitem [{\citenamefont {Rumelhart}\ \emph {et~al.}(1986)\citenamefont
  {Rumelhart}, \citenamefont {Hinton},\ and\ \citenamefont
  {Williams}}]{rumelhart1986learning}%
  \BibitemOpen
  \bibfield  {author} {\bibinfo {author} {\bibfnamefont {D.~E.}\ \bibnamefont
  {Rumelhart}}, \bibinfo {author} {\bibfnamefont {G.~E.}\ \bibnamefont
  {Hinton}}, \ and\ \bibinfo {author} {\bibfnamefont {R.~J.}\ \bibnamefont
  {Williams}},\ }\bibfield  {title} {\enquote {\bibinfo {title} {Learning
  representations by back-propagating errors},}\ }\href@noop {} {\bibfield
  {journal} {\bibinfo  {journal} {Nature}\ }\textbf {\bibinfo {volume} {323}},\
  \bibinfo {pages} {533--536} (\bibinfo {year} {1986})}\BibitemShut {NoStop}%
\bibitem [{\citenamefont {Cybenko}(1989)}]{Cybenko1989ApproximationFunction}%
  \BibitemOpen
  \bibfield  {author} {\bibinfo {author} {\bibfnamefont {G.}~\bibnamefont
  {Cybenko}},\ }\bibfield  {title} {\enquote {\bibinfo {title} {{Approximation
  by superpositions of a sigmoidal function}},}\ }\href {\doibase
  10.1007/BF02551274} {\bibfield  {journal} {\bibinfo  {journal} {Mathematics
  of Control, Signals, and Systems}\ }\textbf {\bibinfo {volume} {2}},\
  \bibinfo {pages} {303--314} (\bibinfo {year} {1989})}\BibitemShut {NoStop}%
\bibitem [{\citenamefont {Goodfellow}\ \emph {et~al.}(2014)\citenamefont
  {Goodfellow}, \citenamefont {Pouget-Abadie}, \citenamefont {Mirza},
  \citenamefont {Xu}, \citenamefont {Warde-Farley}, \citenamefont {Ozair},
  \citenamefont {Courville},\ and\ \citenamefont
  {Bengio}}]{goodfellow2014generative}%
  \BibitemOpen
  \bibfield  {author} {\bibinfo {author} {\bibfnamefont {Ian}\ \bibnamefont
  {Goodfellow}}, \bibinfo {author} {\bibfnamefont {Jean}\ \bibnamefont
  {Pouget-Abadie}}, \bibinfo {author} {\bibfnamefont {Mehdi}\ \bibnamefont
  {Mirza}}, \bibinfo {author} {\bibfnamefont {Bing}\ \bibnamefont {Xu}},
  \bibinfo {author} {\bibfnamefont {David}\ \bibnamefont {Warde-Farley}},
  \bibinfo {author} {\bibfnamefont {Sherjil}\ \bibnamefont {Ozair}}, \bibinfo
  {author} {\bibfnamefont {Aaron}\ \bibnamefont {Courville}}, \ and\ \bibinfo
  {author} {\bibfnamefont {Yoshua}\ \bibnamefont {Bengio}},\ }\bibfield
  {title} {\enquote {\bibinfo {title} {Generative adversarial nets},}\ }in\
  \href@noop {} {\emph {\bibinfo {booktitle} {Advances in neural information
  processing systems}}}\ (\bibinfo {year} {2014})\ pp.\ \bibinfo {pages}
  {2672--2680}\BibitemShut {NoStop}%
\bibitem [{\citenamefont {Yadav}\ \emph {et~al.}(2017)\citenamefont {Yadav},
  \citenamefont {Shah}, \citenamefont {Xu}, \citenamefont {Jacobs},\ and\
  \citenamefont {Goldstein}}]{yadav2017stabilizing}%
  \BibitemOpen
  \bibfield  {author} {\bibinfo {author} {\bibfnamefont {Abhay}\ \bibnamefont
  {Yadav}}, \bibinfo {author} {\bibfnamefont {Sohil}\ \bibnamefont {Shah}},
  \bibinfo {author} {\bibfnamefont {Zheng}\ \bibnamefont {Xu}}, \bibinfo
  {author} {\bibfnamefont {David}\ \bibnamefont {Jacobs}}, \ and\ \bibinfo
  {author} {\bibfnamefont {Tom}\ \bibnamefont {Goldstein}},\ }\bibfield
  {title} {\enquote {\bibinfo {title} {Stabilizing adversarial nets with
  prediction methods},}\ }\href@noop {} {\bibfield  {journal} {\bibinfo
  {journal} {arXiv preprint arXiv:1705.07364}\ } (\bibinfo {year}
  {2017})}\BibitemShut {NoStop}%
\bibitem [{\citenamefont {Foster}(2019)}]{foster2019generative}%
  \BibitemOpen
  \bibfield  {author} {\bibinfo {author} {\bibfnamefont {David}\ \bibnamefont
  {Foster}},\ }\href@noop {} {\emph {\bibinfo {title} {Generative deep
  learning: teaching machines to paint, write, compose, and play}}}\ (\bibinfo
  {publisher} {O'Reilly Media},\ \bibinfo {year} {2019})\BibitemShut {NoStop}%
\bibitem [{\citenamefont {Szegedy}\ \emph {et~al.}(2015)\citenamefont
  {Szegedy}, \citenamefont {Vanhoucke}, \citenamefont {Ioffe},\ and\
  \citenamefont {Shlens}}]{SzegedyRethinkingVision}%
  \BibitemOpen
  \bibfield  {author} {\bibinfo {author} {\bibfnamefont {Christian}\
  \bibnamefont {Szegedy}}, \bibinfo {author} {\bibfnamefont {Vincent}\
  \bibnamefont {Vanhoucke}}, \bibinfo {author} {\bibfnamefont {Sergey}\
  \bibnamefont {Ioffe}}, \ and\ \bibinfo {author} {\bibfnamefont {Jon}\
  \bibnamefont {Shlens}},\ }\href@noop {} {\emph {\bibinfo {title} {{Rethinking
  the Inception Architecture for Computer Vision}}}},\ \bibinfo {type} {Tech.
  Rep.}\ (\bibinfo {year} {2015})\BibitemShut {NoStop}%
\bibitem [{\citenamefont {Sadanandan}\ \emph {et~al.}(2017)\citenamefont
  {Sadanandan}, \citenamefont {Ranefall}, \citenamefont {Le~Guyader},\ and\
  \citenamefont {W{\"{a}}hlby}}]{Sadanandan2017AutomatedSegmentation}%
  \BibitemOpen
  \bibfield  {author} {\bibinfo {author} {\bibfnamefont {Sajith~Kecheril}\
  \bibnamefont {Sadanandan}}, \bibinfo {author} {\bibfnamefont {Petter}\
  \bibnamefont {Ranefall}}, \bibinfo {author} {\bibfnamefont {Sylvie}\
  \bibnamefont {Le~Guyader}}, \ and\ \bibinfo {author} {\bibfnamefont
  {Carolina}\ \bibnamefont {W{\"{a}}hlby}},\ }\bibfield  {title} {\enquote
  {\bibinfo {title} {{Automated Training of Deep Convolutional Neural Networks
  for Cell Segmentation}},}\ }\href {\doibase 10.1038/s41598-017-07599-6}
  {\bibfield  {journal} {\bibinfo  {journal} {Scientific Reports}\ }\textbf
  {\bibinfo {volume} {7}},\ \bibinfo {pages} {1--7} (\bibinfo {year}
  {2017})}\BibitemShut {NoStop}%
\bibitem [{\citenamefont {Al-Kofahi}\ \emph {et~al.}(2018)\citenamefont
  {Al-Kofahi}, \citenamefont {Zaltsman}, \citenamefont {Graves}, \citenamefont
  {Marshall},\ and\ \citenamefont {Rusu}}]{Al-Kofahi2018AImages}%
  \BibitemOpen
  \bibfield  {author} {\bibinfo {author} {\bibfnamefont {Yousef}\ \bibnamefont
  {Al-Kofahi}}, \bibinfo {author} {\bibfnamefont {Alla}\ \bibnamefont
  {Zaltsman}}, \bibinfo {author} {\bibfnamefont {Robert}\ \bibnamefont
  {Graves}}, \bibinfo {author} {\bibfnamefont {Will}\ \bibnamefont {Marshall}},
  \ and\ \bibinfo {author} {\bibfnamefont {Mirabela}\ \bibnamefont {Rusu}},\
  }\bibfield  {title} {\enquote {\bibinfo {title} {{A deep learning-based
  algorithm for 2-D cell segmentation in microscopy images}},}\ }\href
  {\doibase 10.1186/s12859-018-2375-z} {\bibfield  {journal} {\bibinfo
  {journal} {BMC Bioinformatics}\ }\textbf {\bibinfo {volume} {19}},\ \bibinfo
  {pages} {365} (\bibinfo {year} {2018})}\BibitemShut {NoStop}%
\bibitem [{\citenamefont {Song}\ \emph {et~al.}(2017)\citenamefont {Song},
  \citenamefont {Tan}, \citenamefont {Jiang}, \citenamefont {Cheng},
  \citenamefont {Ni}, \citenamefont {Chen}, \citenamefont {Lei},\ and\
  \citenamefont {Wang}}]{Song2017AccurateImages}%
  \BibitemOpen
  \bibfield  {author} {\bibinfo {author} {\bibfnamefont {Youyi}\ \bibnamefont
  {Song}}, \bibinfo {author} {\bibfnamefont {Ee~Leng}\ \bibnamefont {Tan}},
  \bibinfo {author} {\bibfnamefont {Xudong}\ \bibnamefont {Jiang}}, \bibinfo
  {author} {\bibfnamefont {Jie~Zhi}\ \bibnamefont {Cheng}}, \bibinfo {author}
  {\bibfnamefont {Dong}\ \bibnamefont {Ni}}, \bibinfo {author} {\bibfnamefont
  {Siping}\ \bibnamefont {Chen}}, \bibinfo {author} {\bibfnamefont {Baiying}\
  \bibnamefont {Lei}}, \ and\ \bibinfo {author} {\bibfnamefont {Tianfu}\
  \bibnamefont {Wang}},\ }\bibfield  {title} {\enquote {\bibinfo {title}
  {{Accurate cervical cell segmentation from overlapping clumps in pap smear
  images}},}\ }\href {\doibase 10.1109/TMI.2016.2606380} {\bibfield  {journal}
  {\bibinfo  {journal} {IEEE Transactions on Medical Imaging}\ }\textbf
  {\bibinfo {volume} {36}},\ \bibinfo {pages} {288--300} (\bibinfo {year}
  {2017})}\BibitemShut {NoStop}%
\bibitem [{\citenamefont {Akram}\ \emph {et~al.}(2016)\citenamefont {Akram},
  \citenamefont {Kannala}, \citenamefont {Eklund},\ and\ \citenamefont
  {Heikkil{\"{a}}}}]{Akram2016CellAnalysis}%
  \BibitemOpen
  \bibfield  {author} {\bibinfo {author} {\bibfnamefont {Saad~Ullah}\
  \bibnamefont {Akram}}, \bibinfo {author} {\bibfnamefont {Juho}\ \bibnamefont
  {Kannala}}, \bibinfo {author} {\bibfnamefont {Lauri}\ \bibnamefont {Eklund}},
  \ and\ \bibinfo {author} {\bibfnamefont {Janne}\ \bibnamefont
  {Heikkil{\"{a}}}},\ }\bibfield  {title} {\enquote {\bibinfo {title} {{Cell
  segmentation proposal network for microscopy image analysis}},}\ }in\ \href
  {\doibase 10.1007/978-3-319-46976-8{\_}3} {\emph {\bibinfo {booktitle}
  {Lecture Notes in Computer Science (including subseries Lecture Notes in
  Artificial Intelligence and Lecture Notes in Bioinformatics)}}},\ Vol.\
  \bibinfo {volume} {10008 LNCS}\ (\bibinfo  {publisher} {Springer Verlag},\
  \bibinfo {year} {2016})\ pp.\ \bibinfo {pages} {21--29}\BibitemShut {NoStop}%
\bibitem [{\citenamefont {Arbelle}\ and\ \citenamefont
  {Raviv}(2017)}]{Arbelle2017MicroscopyNetworks}%
  \BibitemOpen
  \bibfield  {author} {\bibinfo {author} {\bibfnamefont {Assaf}\ \bibnamefont
  {Arbelle}}\ and\ \bibinfo {author} {\bibfnamefont {Tammy~Riklin}\
  \bibnamefont {Raviv}},\ }\bibfield  {title} {\enquote {\bibinfo {title}
  {{Microscopy Cell Segmentation via Adversarial Neural Networks}},}\ }\href
  {http://arxiv.org/abs/1709.05860} {\bibfield  {journal} {\bibinfo  {journal}
  {Proceedings - International Symposium on Biomedical Imaging}\ }\textbf
  {\bibinfo {volume} {2018-April}},\ \bibinfo {pages} {645--648} (\bibinfo
  {year} {2017})}\BibitemShut {NoStop}%
\bibitem [{\citenamefont {Hatipoglu}\ and\ \citenamefont
  {Bilgin}(2017)}]{Hatipoglu2017CellRelationships}%
  \BibitemOpen
  \bibfield  {author} {\bibinfo {author} {\bibfnamefont {Nuh}\ \bibnamefont
  {Hatipoglu}}\ and\ \bibinfo {author} {\bibfnamefont {Gokhan}\ \bibnamefont
  {Bilgin}},\ }\bibfield  {title} {\enquote {\bibinfo {title} {{Cell
  segmentation in histopathological images with deep learning algorithms by
  utilizing spatial relationships}},}\ }\href {\doibase
  10.1007/s11517-017-1630-1} {\bibfield  {journal} {\bibinfo  {journal}
  {Medical and Biological Engineering and Computing}\ }\textbf {\bibinfo
  {volume} {55}},\ \bibinfo {pages} {1829--1848} (\bibinfo {year}
  {2017})}\BibitemShut {NoStop}%
\bibitem [{\citenamefont {Arbelle}\ and\ \citenamefont
  {Raviv}(2018)}]{Arbelle2018MicroscopyNetworks}%
  \BibitemOpen
  \bibfield  {author} {\bibinfo {author} {\bibfnamefont {Assaf}\ \bibnamefont
  {Arbelle}}\ and\ \bibinfo {author} {\bibfnamefont {Tammy~Riklin}\
  \bibnamefont {Raviv}},\ }\bibfield  {title} {\enquote {\bibinfo {title}
  {{Microscopy Cell Segmentation via Convolutional LSTM Networks}},}\ }\href
  {http://arxiv.org/abs/1805.11247} {\bibfield  {journal} {\bibinfo  {journal}
  {Proceedings - International Symposium on Biomedical Imaging}\ }\textbf
  {\bibinfo {volume} {2019-April}},\ \bibinfo {pages} {1008--1012} (\bibinfo
  {year} {2018})}\BibitemShut {NoStop}%
\bibitem [{\citenamefont {Lugagne}\ \emph {et~al.}(2020)\citenamefont
  {Lugagne}, \citenamefont {Lin},\ and\ \citenamefont
  {Dunlop}}]{Lugagne2020DeLTA:Learning}%
  \BibitemOpen
  \bibfield  {author} {\bibinfo {author} {\bibfnamefont {Jean-Baptiste}\
  \bibnamefont {Lugagne}}, \bibinfo {author} {\bibfnamefont {Haonan}\
  \bibnamefont {Lin}}, \ and\ \bibinfo {author} {\bibfnamefont {Mary~J.}\
  \bibnamefont {Dunlop}},\ }\bibfield  {title} {\enquote {\bibinfo {title}
  {{DeLTA: Automated cell segmentation, tracking, and lineage reconstruction
  using deep learning}},}\ }\href {\doibase 10.1371/journal.pcbi.1007673}
  {\bibfield  {journal} {\bibinfo  {journal} {PLOS Computational Biology}\
  }\textbf {\bibinfo {volume} {16}},\ \bibinfo {pages} {e1007673} (\bibinfo
  {year} {2020})}\BibitemShut {NoStop}%
\bibitem [{\citenamefont {Raza}\ \emph {et~al.}(2017)\citenamefont {Raza},
  \citenamefont {Cheung}, \citenamefont {Epstein}, \citenamefont {Pelengaris},
  \citenamefont {Khan},\ and\ \citenamefont
  {Rajpoot}}]{Raza2017MIMO-Net:Images}%
  \BibitemOpen
  \bibfield  {author} {\bibinfo {author} {\bibfnamefont {Shan~E.Ahmed}\
  \bibnamefont {Raza}}, \bibinfo {author} {\bibfnamefont {Linda}\ \bibnamefont
  {Cheung}}, \bibinfo {author} {\bibfnamefont {David}\ \bibnamefont {Epstein}},
  \bibinfo {author} {\bibfnamefont {Stella}\ \bibnamefont {Pelengaris}},
  \bibinfo {author} {\bibfnamefont {Michael}\ \bibnamefont {Khan}}, \ and\
  \bibinfo {author} {\bibfnamefont {Nasir~M.}\ \bibnamefont {Rajpoot}},\
  }\bibfield  {title} {\enquote {\bibinfo {title} {{MIMO-Net: A multi-input
  multi-output convolutional neural network for cell segmentation in
  fluorescence microscopy images}},}\ }in\ \href {\doibase
  10.1109/ISBI.2017.7950532} {\emph {\bibinfo {booktitle} {Proceedings -
  International Symposium on Biomedical Imaging}}}\ (\bibinfo  {publisher}
  {IEEE Computer Society},\ \bibinfo {year} {2017})\ pp.\ \bibinfo {pages}
  {337--340}\BibitemShut {NoStop}%
\bibitem [{\citenamefont {Ma}\ \emph {et~al.}(2018)\citenamefont {Ma},
  \citenamefont {Ban}, \citenamefont {Huang}, \citenamefont {Chen},
  \citenamefont {Liu},\ and\ \citenamefont {Zhi}}]{Ma2018DeepImages}%
  \BibitemOpen
  \bibfield  {author} {\bibinfo {author} {\bibfnamefont {Boyuan}\ \bibnamefont
  {Ma}}, \bibinfo {author} {\bibfnamefont {Xiaojuan}\ \bibnamefont {Ban}},
  \bibinfo {author} {\bibfnamefont {Haiyou}\ \bibnamefont {Huang}}, \bibinfo
  {author} {\bibfnamefont {Yulian}\ \bibnamefont {Chen}}, \bibinfo {author}
  {\bibfnamefont {Wanbo}\ \bibnamefont {Liu}}, \ and\ \bibinfo {author}
  {\bibfnamefont {Yonghong}\ \bibnamefont {Zhi}},\ }\bibfield  {title}
  {\enquote {\bibinfo {title} {{Deep Learning-Based Image Segmentation for
  Al-La Alloy Microscopic Images}},}\ }\href {\doibase 10.3390/sym10040107}
  {\bibfield  {journal} {\bibinfo  {journal} {Symmetry}\ }\textbf {\bibinfo
  {volume} {10}},\ \bibinfo {pages} {107} (\bibinfo {year} {2018})}\BibitemShut
  {NoStop}%
\bibitem [{\citenamefont {Azimi}\ \emph {et~al.}(2018)\citenamefont {Azimi},
  \citenamefont {Britz}, \citenamefont {Engstler}, \citenamefont {Fritz},\ and\
  \citenamefont {M{\"{u}}cklich}}]{Azimi2018AdvancedMethods}%
  \BibitemOpen
  \bibfield  {author} {\bibinfo {author} {\bibfnamefont {Seyed~Majid}\
  \bibnamefont {Azimi}}, \bibinfo {author} {\bibfnamefont {Dominik}\
  \bibnamefont {Britz}}, \bibinfo {author} {\bibfnamefont {Michael}\
  \bibnamefont {Engstler}}, \bibinfo {author} {\bibfnamefont {Mario}\
  \bibnamefont {Fritz}}, \ and\ \bibinfo {author} {\bibfnamefont {Frank}\
  \bibnamefont {M{\"{u}}cklich}},\ }\bibfield  {title} {\enquote {\bibinfo
  {title} {{Advanced steel microstructural classification by deep learning
  methods}},}\ }\href {\doibase 10.1038/s41598-018-20037-5} {\bibfield
  {journal} {\bibinfo  {journal} {Scientific Reports}\ }\textbf {\bibinfo
  {volume} {8}},\ \bibinfo {pages} {2128} (\bibinfo {year} {2018})}\BibitemShut
  {NoStop}%
\bibitem [{\citenamefont {Lateef}\ and\ \citenamefont
  {Ruichek}(2019)}]{lateef2019survey}%
  \BibitemOpen
  \bibfield  {author} {\bibinfo {author} {\bibfnamefont {Fahad}\ \bibnamefont
  {Lateef}}\ and\ \bibinfo {author} {\bibfnamefont {Yassine}\ \bibnamefont
  {Ruichek}},\ }\bibfield  {title} {\enquote {\bibinfo {title} {Survey on
  semantic segmentation using deep learning techniques},}\ }\href@noop {}
  {\bibfield  {journal} {\bibinfo  {journal} {Neurocomputing}\ }\textbf
  {\bibinfo {volume} {338}},\ \bibinfo {pages} {321--348} (\bibinfo {year}
  {2019})}\BibitemShut {NoStop}%
\bibitem [{\citenamefont {Li}\ \emph {et~al.}(2017)\citenamefont {Li},
  \citenamefont {Zeng}, \citenamefont {Peng},\ and\ \citenamefont
  {Ji}}]{Li2017DeepReconstruction}%
  \BibitemOpen
  \bibfield  {author} {\bibinfo {author} {\bibfnamefont {Rongjian}\
  \bibnamefont {Li}}, \bibinfo {author} {\bibfnamefont {Tao}\ \bibnamefont
  {Zeng}}, \bibinfo {author} {\bibfnamefont {Hanchuan}\ \bibnamefont {Peng}}, \
  and\ \bibinfo {author} {\bibfnamefont {Shuiwang}\ \bibnamefont {Ji}},\
  }\bibfield  {title} {\enquote {\bibinfo {title} {{Deep Learning Segmentation
  of Optical Microscopy Images Improves 3-D Neuron Reconstruction}},}\ }\href
  {\doibase 10.1109/TMI.2017.2679713} {\bibfield  {journal} {\bibinfo
  {journal} {IEEE Transactions on Medical Imaging}\ }\textbf {\bibinfo {volume}
  {36}},\ \bibinfo {pages} {1533--1541} (\bibinfo {year} {2017})}\BibitemShut
  {NoStop}%
\bibitem [{\citenamefont {{\c{C}}i{\c{c}}ek}\ \emph {et~al.}(2016)\citenamefont
  {{\c{C}}i{\c{c}}ek}, \citenamefont {Abdulkadir}, \citenamefont {Lienkamp},
  \citenamefont {Brox},\ and\ \citenamefont
  {Ronneberger}}]{Cicek20163DAnnotation}%
  \BibitemOpen
  \bibfield  {author} {\bibinfo {author} {\bibfnamefont {Özgün}\ \bibnamefont
  {{\c{C}}i{\c{c}}ek}}, \bibinfo {author} {\bibfnamefont {Ahmed}\ \bibnamefont
  {Abdulkadir}}, \bibinfo {author} {\bibfnamefont {Soeren~S.}\ \bibnamefont
  {Lienkamp}}, \bibinfo {author} {\bibfnamefont {Thomas}\ \bibnamefont {Brox}},
  \ and\ \bibinfo {author} {\bibfnamefont {Olaf}\ \bibnamefont {Ronneberger}},\
  }\bibfield  {title} {\enquote {\bibinfo {title} {{3D U-net: Learning dense
  volumetric segmentation from sparse annotation}},}\ }in\ \href {\doibase
  10.1007/978-3-319-46723-8{\_}49} {\emph {\bibinfo {booktitle} {Lecture Notes
  in Computer Science (including subseries Lecture Notes in Artificial
  Intelligence and Lecture Notes in Bioinformatics)}}},\ Vol.\ \bibinfo
  {volume} {9901 LNCS}\ (\bibinfo  {publisher} {Springer Verlag},\ \bibinfo
  {year} {2016})\ pp.\ \bibinfo {pages} {424--432}\BibitemShut {NoStop}%
\bibitem [{\citenamefont {Kleesiek}\ \emph {et~al.}(2016)\citenamefont
  {Kleesiek}, \citenamefont {Urban}, \citenamefont {Hubert}, \citenamefont
  {Schwarz}, \citenamefont {Maier-Hein}, \citenamefont {Bendszus},\ and\
  \citenamefont {Biller}}]{Kleesiek2016DeepStripping}%
  \BibitemOpen
  \bibfield  {author} {\bibinfo {author} {\bibfnamefont {Jens}\ \bibnamefont
  {Kleesiek}}, \bibinfo {author} {\bibfnamefont {Gregor}\ \bibnamefont
  {Urban}}, \bibinfo {author} {\bibfnamefont {Alexander}\ \bibnamefont
  {Hubert}}, \bibinfo {author} {\bibfnamefont {Daniel}\ \bibnamefont
  {Schwarz}}, \bibinfo {author} {\bibfnamefont {Klaus}\ \bibnamefont
  {Maier-Hein}}, \bibinfo {author} {\bibfnamefont {Martin}\ \bibnamefont
  {Bendszus}}, \ and\ \bibinfo {author} {\bibfnamefont {Armin}\ \bibnamefont
  {Biller}},\ }\bibfield  {title} {\enquote {\bibinfo {title} {{Deep MRI brain
  extraction: A 3D convolutional neural network for skull stripping}},}\ }\href
  {\doibase 10.1016/j.neuroimage.2016.01.024} {\bibfield  {journal} {\bibinfo
  {journal} {NeuroImage}\ }\textbf {\bibinfo {volume} {129}},\ \bibinfo {pages}
  {460--469} (\bibinfo {year} {2016})}\BibitemShut {NoStop}%
\bibitem [{\citenamefont {Betzig}\ \emph {et~al.}(2006)\citenamefont {Betzig},
  \citenamefont {Patterson}, \citenamefont {Sougrat}, \citenamefont
  {Lindwasser}, \citenamefont {Olenych}, \citenamefont {Bonifacino},
  \citenamefont {Davidson}, \citenamefont {Lippincott-Schwartz},\ and\
  \citenamefont {Hess}}]{Betzig2006ImagingResolution}%
  \BibitemOpen
  \bibfield  {author} {\bibinfo {author} {\bibfnamefont {Eric}\ \bibnamefont
  {Betzig}}, \bibinfo {author} {\bibfnamefont {George~H.}\ \bibnamefont
  {Patterson}}, \bibinfo {author} {\bibfnamefont {Rachid}\ \bibnamefont
  {Sougrat}}, \bibinfo {author} {\bibfnamefont {O.~Wolf}\ \bibnamefont
  {Lindwasser}}, \bibinfo {author} {\bibfnamefont {Scott}\ \bibnamefont
  {Olenych}}, \bibinfo {author} {\bibfnamefont {Juan~S.}\ \bibnamefont
  {Bonifacino}}, \bibinfo {author} {\bibfnamefont {Michael~W.}\ \bibnamefont
  {Davidson}}, \bibinfo {author} {\bibfnamefont {Jennifer}\ \bibnamefont
  {Lippincott-Schwartz}}, \ and\ \bibinfo {author} {\bibfnamefont {Harald~F.}\
  \bibnamefont {Hess}},\ }\bibfield  {title} {\enquote {\bibinfo {title}
  {{Imaging intracellular fluorescent proteins at nanometer resolution}},}\
  }\href {\doibase 10.1126/science.1127344} {\bibfield  {journal} {\bibinfo
  {journal} {Science}\ }\textbf {\bibinfo {volume} {313}},\ \bibinfo {pages}
  {1642--1645} (\bibinfo {year} {2006})}\BibitemShut {NoStop}%
\bibitem [{\citenamefont {Wu}\ \emph {et~al.}(2019)\citenamefont {Wu},
  \citenamefont {Luo}, \citenamefont {Chaudhari}, \citenamefont {Rivenson},
  \citenamefont {Calis}, \citenamefont {de~Haan},\ and\ \citenamefont
  {Ozcan}}]{Wu2019}%
  \BibitemOpen
  \bibfield  {author} {\bibinfo {author} {\bibfnamefont {Yichen}\ \bibnamefont
  {Wu}}, \bibinfo {author} {\bibfnamefont {Yilin}\ \bibnamefont {Luo}},
  \bibinfo {author} {\bibfnamefont {Gunvant}\ \bibnamefont {Chaudhari}},
  \bibinfo {author} {\bibfnamefont {Yair}\ \bibnamefont {Rivenson}}, \bibinfo
  {author} {\bibfnamefont {Ayfer}\ \bibnamefont {Calis}}, \bibinfo {author}
  {\bibfnamefont {Kevin}\ \bibnamefont {de~Haan}}, \ and\ \bibinfo {author}
  {\bibfnamefont {Aydogan}\ \bibnamefont {Ozcan}},\ }\bibfield  {title}
  {\enquote {\bibinfo {title} {Bright-field holography: cross-modality deep
  learning enables snapshot 3d imaging with bright-field contrast using a
  single hologram},}\ }\href {\doibase 10.1038/s41377-019-0139-9} {\bibfield
  {journal} {\bibinfo  {journal} {Light: Science {\&} Applications}\ }\textbf
  {\bibinfo {volume} {8}},\ \bibinfo {pages} {25} (\bibinfo {year}
  {2019})}\BibitemShut {NoStop}%
\bibitem [{\citenamefont {Rivenson}\ \emph {et~al.}(2019)\citenamefont
  {Rivenson}, \citenamefont {Liu}, \citenamefont {Wei}, \citenamefont {Zhang},
  \citenamefont {de~Haan},\ and\ \citenamefont
  {Ozcan}}]{Rivenson2019PhaseStain:Learning}%
  \BibitemOpen
  \bibfield  {author} {\bibinfo {author} {\bibfnamefont {Yair}\ \bibnamefont
  {Rivenson}}, \bibinfo {author} {\bibfnamefont {Tairan}\ \bibnamefont {Liu}},
  \bibinfo {author} {\bibfnamefont {Zhensong}\ \bibnamefont {Wei}}, \bibinfo
  {author} {\bibfnamefont {Yibo}\ \bibnamefont {Zhang}}, \bibinfo {author}
  {\bibfnamefont {Kevin}\ \bibnamefont {de~Haan}}, \ and\ \bibinfo {author}
  {\bibfnamefont {Aydogan}\ \bibnamefont {Ozcan}},\ }\bibfield  {title}
  {\enquote {\bibinfo {title} {{PhaseStain: the digital staining of label-free
  quantitative phase microscopy images using deep learning}},}\ }\href
  {\doibase 10.1038/s41377-019-0129-y} {\bibfield  {journal} {\bibinfo
  {journal} {Light: Science and Applications}\ }\textbf {\bibinfo {volume}
  {8}},\ \bibinfo {pages} {2047--7538} (\bibinfo {year} {2019})}\BibitemShut
  {NoStop}%
\bibitem [{\citenamefont {Masoudi}\ \emph {et~al.}(2019)\citenamefont
  {Masoudi}, \citenamefont {Razi}, \citenamefont {Wright}, \citenamefont
  {Gatlin},\ and\ \citenamefont {Bagci}}]{MasoudiInstance-LevelTracking}%
  \BibitemOpen
  \bibfield  {author} {\bibinfo {author} {\bibfnamefont {Samira}\ \bibnamefont
  {Masoudi}}, \bibinfo {author} {\bibfnamefont {Afsaneh}\ \bibnamefont {Razi}},
  \bibinfo {author} {\bibfnamefont {Cameron H~G}\ \bibnamefont {Wright}},
  \bibinfo {author} {\bibfnamefont {Jesse~C}\ \bibnamefont {Gatlin}}, \ and\
  \bibinfo {author} {\bibfnamefont {Ulas}\ \bibnamefont {Bagci}},\ }\href@noop
  {} {\emph {\bibinfo {title} {IEEE transactions on medical imaging}}},\
  \bibinfo {type} {Tech. Rep.}\ (\bibinfo {year} {2019})\BibitemShut {NoStop}%
\bibitem [{\citenamefont {Zelger}\ \emph {et~al.}(2018)\citenamefont {Zelger},
  \citenamefont {Kaser}, \citenamefont {Rossboth}, \citenamefont {Velas},
  \citenamefont {Sch{\"{u}}tz},\ and\ \citenamefont
  {Jesacher}}]{Zelger2018Three-dimensionalLearning}%
  \BibitemOpen
  \bibfield  {author} {\bibinfo {author} {\bibfnamefont {P}~\bibnamefont
  {Zelger}}, \bibinfo {author} {\bibfnamefont {K}~\bibnamefont {Kaser}},
  \bibinfo {author} {\bibfnamefont {B}~\bibnamefont {Rossboth}}, \bibinfo
  {author} {\bibfnamefont {L}~\bibnamefont {Velas}}, \bibinfo {author}
  {\bibfnamefont {G~J}\ \bibnamefont {Sch{\"{u}}tz}}, \ and\ \bibinfo {author}
  {\bibfnamefont {A}~\bibnamefont {Jesacher}},\ }\bibfield  {title} {\enquote
  {\bibinfo {title} {{Three-dimensional localization microscopy using deep
  learning}},}\ }\href {\doibase 10.1364/OE.26.033166} {\bibfield  {journal}
  {\bibinfo  {journal} {Optical Express}\ } (\bibinfo {year} {2018}),\
  10.1364/OE.26.033166}\BibitemShut {NoStop}%
\bibitem [{\citenamefont {Franchini}\ and\ \citenamefont
  {Krevor}(2020)}]{Franchini2020CutSetups}%
  \BibitemOpen
  \bibfield  {author} {\bibinfo {author} {\bibfnamefont {Simon}\ \bibnamefont
  {Franchini}}\ and\ \bibinfo {author} {\bibfnamefont {Samuel}\ \bibnamefont
  {Krevor}},\ }\bibfield  {title} {\enquote {\bibinfo {title} {{Cut, overlap
  and locate: a deep learning approach for the 3D localization of particles in
  astigmatic optical setups}},}\ }\href {\doibase 10.1007/s00348-020-02968-w}
  {\bibfield  {journal} {\bibinfo  {journal} {Experiments in Fluids}\ }\textbf
  {\bibinfo {volume} {61}},\ \bibinfo {pages} {140} (\bibinfo {year}
  {2020})}\BibitemShut {NoStop}%
\bibitem [{\citenamefont {Wollmann}\ \emph {et~al.}(2019)\citenamefont
  {Wollmann}, \citenamefont {Ritter}, \citenamefont {Dohrke}, \citenamefont
  {Lee}, \citenamefont {Bartenschlager},\ and\ \citenamefont
  {Rohr}}]{Wollmann2019Detnet:Images}%
  \BibitemOpen
  \bibfield  {author} {\bibinfo {author} {\bibfnamefont {T.}~\bibnamefont
  {Wollmann}}, \bibinfo {author} {\bibfnamefont {C.}~\bibnamefont {Ritter}},
  \bibinfo {author} {\bibfnamefont {J.~N.}\ \bibnamefont {Dohrke}}, \bibinfo
  {author} {\bibfnamefont {J.~Y.}\ \bibnamefont {Lee}}, \bibinfo {author}
  {\bibfnamefont {R.}~\bibnamefont {Bartenschlager}}, \ and\ \bibinfo {author}
  {\bibfnamefont {K.}~\bibnamefont {Rohr}},\ }\bibfield  {title} {\enquote
  {\bibinfo {title} {{Detnet: Deep neural network for particle detection in
  fluorescence microscopy images}},}\ }in\ \href {\doibase
  10.1109/ISBI.2019.8759234} {\emph {\bibinfo {booktitle} {Proceedings -
  International Symposium on Biomedical Imaging}}},\ Vol.\ \bibinfo {volume}
  {2019-April}\ (\bibinfo  {publisher} {IEEE Computer Society},\ \bibinfo
  {year} {2019})\ pp.\ \bibinfo {pages} {517--520}\BibitemShut {NoStop}%
\bibitem [{\citenamefont {Ritter}\ \emph {et~al.}(2020)\citenamefont {Ritter},
  \citenamefont {Wollmann}, \citenamefont {Lee}, \citenamefont
  {Bartenschlager},\ and\ \citenamefont {Rohr}}]{Ritter2020DeepImages}%
  \BibitemOpen
  \bibfield  {author} {\bibinfo {author} {\bibfnamefont {C.}~\bibnamefont
  {Ritter}}, \bibinfo {author} {\bibfnamefont {T.}~\bibnamefont {Wollmann}},
  \bibinfo {author} {\bibfnamefont {J.~Y.}\ \bibnamefont {Lee}}, \bibinfo
  {author} {\bibfnamefont {R.}~\bibnamefont {Bartenschlager}}, \ and\ \bibinfo
  {author} {\bibfnamefont {K.}~\bibnamefont {Rohr}},\ }\bibfield  {title}
  {\enquote {\bibinfo {title} {{Deep Learning Particle Detection for
  Probabilistic Tracking in Fluorescence Microscopy Images}},}\ }in\ \href
  {\doibase 10.1109/ISBI45749.2020.9098598} {\emph {\bibinfo {booktitle}
  {Proceedings - International Symposium on Biomedical Imaging}}},\ Vol.\
  \bibinfo {volume} {2020-April}\ (\bibinfo  {publisher} {IEEE Computer
  Society},\ \bibinfo {year} {2020})\ pp.\ \bibinfo {pages}
  {977--980}\BibitemShut {NoStop}%
\bibitem [{\citenamefont {Oktay}\ and\ \citenamefont
  {Gurses}(2019)}]{Oktay2019AutomaticImages}%
  \BibitemOpen
  \bibfield  {author} {\bibinfo {author} {\bibfnamefont {Ayse~Betul}\
  \bibnamefont {Oktay}}\ and\ \bibinfo {author} {\bibfnamefont {Anıl}\
  \bibnamefont {Gurses}},\ }\bibfield  {title} {\enquote {\bibinfo {title}
  {{Automatic detection, localization and segmentation of nano-particles with
  deep learning in microscopy images}},}\ }\href {\doibase
  10.1016/j.micron.2019.02.009} {\bibfield  {journal} {\bibinfo  {journal}
  {Micron}\ }\textbf {\bibinfo {volume} {120}},\ \bibinfo {pages} {113--119}
  (\bibinfo {year} {2019})}\BibitemShut {NoStop}%
\bibitem [{\citenamefont {Spilger}\ \emph {et~al.}(2020)\citenamefont
  {Spilger}, \citenamefont {Imle}, \citenamefont {Lee}, \citenamefont {Muller},
  \citenamefont {Fackler}, \citenamefont {Bartenschlager},\ and\ \citenamefont
  {Rohr}}]{Spilger2020ADetections}%
  \BibitemOpen
  \bibfield  {author} {\bibinfo {author} {\bibfnamefont {Roman}\ \bibnamefont
  {Spilger}}, \bibinfo {author} {\bibfnamefont {Andrea}\ \bibnamefont {Imle}},
  \bibinfo {author} {\bibfnamefont {Ji~Young}\ \bibnamefont {Lee}}, \bibinfo
  {author} {\bibfnamefont {Barbara}\ \bibnamefont {Muller}}, \bibinfo {author}
  {\bibfnamefont {Oliver~T.}\ \bibnamefont {Fackler}}, \bibinfo {author}
  {\bibfnamefont {Ralf}\ \bibnamefont {Bartenschlager}}, \ and\ \bibinfo
  {author} {\bibfnamefont {Karl}\ \bibnamefont {Rohr}},\ }\bibfield  {title}
  {\enquote {\bibinfo {title} {{A Recurrent Neural Network for Particle
  Tracking in Microscopy Images Using Future Information, Track Hypotheses, and
  Multiple Detections}},}\ }\href {\doibase 10.1109/TIP.2020.2964515}
  {\bibfield  {journal} {\bibinfo  {journal} {IEEE Transactions on Image
  Processing}\ }\textbf {\bibinfo {volume} {29}},\ \bibinfo {pages}
  {3681--3694} (\bibinfo {year} {2020})}\BibitemShut {NoStop}%
\bibitem [{\citenamefont {Granik}\ \emph {et~al.}(2019)\citenamefont {Granik},
  \citenamefont {Weiss}, \citenamefont {Nehme}, \citenamefont {Levin},
  \citenamefont {Chein}, \citenamefont {Perlson}, \citenamefont {Roichman},\
  and\ \citenamefont {Shechtman}}]{Granik2019Single-ParticleLearning}%
  \BibitemOpen
  \bibfield  {author} {\bibinfo {author} {\bibfnamefont {Naor}\ \bibnamefont
  {Granik}}, \bibinfo {author} {\bibfnamefont {Lucien~E.}\ \bibnamefont
  {Weiss}}, \bibinfo {author} {\bibfnamefont {E.}~\bibnamefont {Nehme}},
  \bibinfo {author} {\bibfnamefont {Maayan}\ \bibnamefont {Levin}}, \bibinfo
  {author} {\bibfnamefont {Michael}\ \bibnamefont {Chein}}, \bibinfo {author}
  {\bibfnamefont {Eran}\ \bibnamefont {Perlson}}, \bibinfo {author}
  {\bibfnamefont {Yael}\ \bibnamefont {Roichman}}, \ and\ \bibinfo {author}
  {\bibfnamefont {Yoav}\ \bibnamefont {Shechtman}},\ }\bibfield  {title}
  {\enquote {\bibinfo {title} {{Single-Particle Diffusion Characterization by
  Deep Learning}},}\ }\href {\doibase 10.1016/j.bpj.2019.06.015} {\bibfield
  {journal} {\bibinfo  {journal} {Biophysical Journal}\ }\textbf {\bibinfo
  {volume} {117}},\ \bibinfo {pages} {185--192} (\bibinfo {year}
  {2019})}\BibitemShut {NoStop}%
\bibitem [{\citenamefont {Bo}\ \emph {et~al.}(2019)\citenamefont {Bo},
  \citenamefont {Schmidt}, \citenamefont {Eichhorn},\ and\ \citenamefont
  {Volpe}}]{Bo2019MeasurementNetworks}%
  \BibitemOpen
  \bibfield  {author} {\bibinfo {author} {\bibfnamefont {Stefano}\ \bibnamefont
  {Bo}}, \bibinfo {author} {\bibfnamefont {Falko}\ \bibnamefont {Schmidt}},
  \bibinfo {author} {\bibfnamefont {Ralf}\ \bibnamefont {Eichhorn}}, \ and\
  \bibinfo {author} {\bibfnamefont {Giovanni}\ \bibnamefont {Volpe}},\
  }\bibfield  {title} {\enquote {\bibinfo {title} {{Measurement of anomalous
  diffusion using recurrent neural networks}},}\ }\href {\doibase
  10.1103/PhysRevE.100.010102} {\bibfield  {journal} {\bibinfo  {journal}
  {Physical Review E}\ }\textbf {\bibinfo {volume} {100}} (\bibinfo {year}
  {2019}),\ 10.1103/PhysRevE.100.010102}\BibitemShut {NoStop}%
\bibitem [{\citenamefont {Kowalek}\ \emph {et~al.}(2019)\citenamefont
  {Kowalek}, \citenamefont {Loch-Olszewska},\ and\ \citenamefont
  {Szwabi{\'{n}}ski}}]{Kowalek2019ClassificationApproach}%
  \BibitemOpen
  \bibfield  {author} {\bibinfo {author} {\bibfnamefont {Patrycja}\
  \bibnamefont {Kowalek}}, \bibinfo {author} {\bibfnamefont {Hanna}\
  \bibnamefont {Loch-Olszewska}}, \ and\ \bibinfo {author} {\bibfnamefont
  {Janusz}\ \bibnamefont {Szwabi{\'{n}}ski}},\ }\bibfield  {title} {\enquote
  {\bibinfo {title} {{Classification of diffusion modes in single-particle
  tracking data: Feature-based versus deep-learning approach}},}\ }\href
  {\doibase 10.1103/PhysRevE.100.032410} {\bibfield  {journal} {\bibinfo
  {journal} {Physical Review E}\ }\textbf {\bibinfo {volume} {100}} (\bibinfo
  {year} {2019}),\ 10.1103/PhysRevE.100.032410}\BibitemShut {NoStop}%
\bibitem [{\citenamefont {Midtvedt}\ \emph
  {et~al.}(2020{\natexlab{b}})\citenamefont {Midtvedt}, \citenamefont
  {Helgadottir}, \citenamefont {Argun}, \citenamefont {Pineda}, \citenamefont
  {Midtvedt},\ and\ \citenamefont {Volpe}}]{deeptrackgithub}%
  \BibitemOpen
  \bibfield  {author} {\bibinfo {author} {\bibfnamefont {Benjamin}\
  \bibnamefont {Midtvedt}}, \bibinfo {author} {\bibfnamefont {Saga}\
  \bibnamefont {Helgadottir}}, \bibinfo {author} {\bibfnamefont {Aykut}\
  \bibnamefont {Argun}}, \bibinfo {author} {\bibfnamefont {Jes\'us}\
  \bibnamefont {Pineda}}, \bibinfo {author} {\bibfnamefont {Daniel}\
  \bibnamefont {Midtvedt}}, \ and\ \bibinfo {author} {\bibfnamefont {Giovanni}\
  \bibnamefont {Volpe}},\ }\href@noop {} {\enquote {\bibinfo {title}
  {Deeptrack-2.0},}\ }\bibinfo {howpublished}
  {\url{https://github.com/softmatterlab/DeepTrack-2.0}} (\bibinfo {year}
  {2020}{\natexlab{b}})\BibitemShut {NoStop}%
\bibitem [{\citenamefont {Midtvedt}\ \emph
  {et~al.}(2020{\natexlab{c}})\citenamefont {Midtvedt}, \citenamefont
  {Helgadottir}, \citenamefont {Argun}, \citenamefont {Pineda}, \citenamefont
  {Midtvedt},\ and\ \citenamefont {Volpe}}]{appgithub}%
  \BibitemOpen
  \bibfield  {author} {\bibinfo {author} {\bibfnamefont {Benjamin}\
  \bibnamefont {Midtvedt}}, \bibinfo {author} {\bibfnamefont {Saga}\
  \bibnamefont {Helgadottir}}, \bibinfo {author} {\bibfnamefont {Aykut}\
  \bibnamefont {Argun}}, \bibinfo {author} {\bibfnamefont {Jes\'us}\
  \bibnamefont {Pineda}}, \bibinfo {author} {\bibfnamefont {Daniel}\
  \bibnamefont {Midtvedt}}, \ and\ \bibinfo {author} {\bibfnamefont {Giovanni}\
  \bibnamefont {Volpe}},\ }\href@noop {} {\enquote {\bibinfo {title}
  {Deeptrack-2.0-app},}\ }\bibinfo {howpublished}
  {\url{https://github.com/softmatterlab/DeepTrack-2.0-app}} (\bibinfo {year}
  {2020}{\natexlab{c}})\BibitemShut {NoStop}%
\bibitem [{\citenamefont {Chollet}\ \emph {et~al.}(2015)\citenamefont {Chollet}
  \emph {et~al.}}]{chollet2015keras}%
  \BibitemOpen
  \bibfield  {author} {\bibinfo {author} {\bibfnamefont {Fran\c{c}ois}\
  \bibnamefont {Chollet}} \emph {et~al.},\ }\href@noop {} {\enquote {\bibinfo
  {title} {Keras},}\ }\bibinfo {howpublished} {\url{https://keras.io}}
  (\bibinfo {year} {2015})\BibitemShut {NoStop}%
\bibitem [{\citenamefont {LeCun}\ \emph {et~al.}(2010)\citenamefont {LeCun},
  \citenamefont {Cortes},\ and\ \citenamefont {Burges}}]{lecun2010mnist}%
  \BibitemOpen
  \bibfield  {author} {\bibinfo {author} {\bibfnamefont {Yann}\ \bibnamefont
  {LeCun}}, \bibinfo {author} {\bibfnamefont {Corinna}\ \bibnamefont {Cortes}},
  \ and\ \bibinfo {author} {\bibfnamefont {CJ}~\bibnamefont {Burges}},\
  }\bibfield  {title} {\enquote {\bibinfo {title} {Mnist handwritten digit
  database},}\ }\href@noop {} {\bibfield  {journal} {\bibinfo  {journal} {ATT
  Labs [Online]. Available: http://yann.lecun.com/exdb/mnist}\ }\textbf
  {\bibinfo {volume} {2}} (\bibinfo {year} {2010})}\BibitemShut {NoStop}%
\bibitem [{\citenamefont {Midtvedt}\ \emph
  {et~al.}(2020{\natexlab{d}})\citenamefont {Midtvedt}, \citenamefont {Eklund},
  \citenamefont {Ols{\'e}n}, \citenamefont {Midtvedt}, \citenamefont
  {Swenson},\ and\ \citenamefont {H{\"o}{\"o}k}}]{Midtvedt2020}%
  \BibitemOpen
  \bibfield  {author} {\bibinfo {author} {\bibfnamefont {Daniel}\ \bibnamefont
  {Midtvedt}}, \bibinfo {author} {\bibfnamefont {Fredrik}\ \bibnamefont
  {Eklund}}, \bibinfo {author} {\bibfnamefont {Erik}\ \bibnamefont
  {Ols{\'e}n}}, \bibinfo {author} {\bibfnamefont {Benjamin}\ \bibnamefont
  {Midtvedt}}, \bibinfo {author} {\bibfnamefont {Jan}\ \bibnamefont {Swenson}},
  \ and\ \bibinfo {author} {\bibfnamefont {Fredrik}\ \bibnamefont
  {H{\"o}{\"o}k}},\ }\bibfield  {title} {\enquote {\bibinfo {title} {Size and
  refractive index determination of subwavelength particles and air bubbles by
  holographic nanoparticle tracking analysis},}\ }\href {\doibase
  10.1021/acs.analchem.9b04101} {\bibfield  {journal} {\bibinfo  {journal}
  {Analytical Chemistry}\ }\textbf {\bibinfo {volume} {92}},\ \bibinfo {pages}
  {1908--1915} (\bibinfo {year} {2020}{\natexlab{d}})}\BibitemShut {NoStop}%
\bibitem [{\citenamefont {Pontén}\ and\ \citenamefont
  {Saksela}(1967)}]{PontenU2OS1967}%
  \BibitemOpen
  \bibfield  {author} {\bibinfo {author} {\bibfnamefont {J.}~\bibnamefont
  {Pontén}}\ and\ \bibinfo {author} {\bibfnamefont {E.}~\bibnamefont
  {Saksela}},\ }\bibfield  {title} {\enquote {\bibinfo {title} {{{T}wo
  established in vitro cell lines from human mesenchymal tumours}},}\
  }\href@noop {} {\bibfield  {journal} {\bibinfo  {journal} {Int. J. Cancer}\
  }\textbf {\bibinfo {volume} {2}},\ \bibinfo {pages} {434--447} (\bibinfo
  {year} {1967})}\BibitemShut {NoStop}%
\bibitem [{\citenamefont {Ljosa}\ \emph {et~al.}(2012)\citenamefont {Ljosa},
  \citenamefont {Sokolnicki},\ and\ \citenamefont {Carpenter}}]{Ljosa2012}%
  \BibitemOpen
  \bibfield  {author} {\bibinfo {author} {\bibfnamefont {Vebjorn}\ \bibnamefont
  {Ljosa}}, \bibinfo {author} {\bibfnamefont {Katherine~L.}\ \bibnamefont
  {Sokolnicki}}, \ and\ \bibinfo {author} {\bibfnamefont {Anne~E.}\
  \bibnamefont {Carpenter}},\ }\bibfield  {title} {\enquote {\bibinfo {title}
  {Annotated high-throughput microscopy image sets for validation},}\ }\href
  {\doibase 10.1038/nmeth.2083} {\bibfield  {journal} {\bibinfo  {journal}
  {Nature Methods}\ }\textbf {\bibinfo {volume} {9}},\ \bibinfo {pages}
  {637--637} (\bibinfo {year} {2012})}\BibitemShut {NoStop}%
\bibitem [{\citenamefont {Gerhard}\ \emph {et~al.}(2013)\citenamefont
  {Gerhard}, \citenamefont {Funke}, \citenamefont {Martel}, \citenamefont
  {Cardona},\ and\ \citenamefont {Fetter}}]{gerhard2013segmented}%
  \BibitemOpen
  \bibfield  {author} {\bibinfo {author} {\bibfnamefont {Stephan}\ \bibnamefont
  {Gerhard}}, \bibinfo {author} {\bibfnamefont {Jan}\ \bibnamefont {Funke}},
  \bibinfo {author} {\bibfnamefont {Julien}\ \bibnamefont {Martel}}, \bibinfo
  {author} {\bibfnamefont {Albert}\ \bibnamefont {Cardona}}, \ and\ \bibinfo
  {author} {\bibfnamefont {Richard}\ \bibnamefont {Fetter}},\ }\bibfield
  {title} {\enquote {\bibinfo {title} {Segmented anisotropic sstem dataset of
  neural tissue},}\ }\href {\doibase 10.6084/m9.figshare.856713.v1} {\bibfield
  {journal} {\bibinfo  {journal} {figshare}\ } (\bibinfo {year} {2013}),\
  10.6084/m9.figshare.856713.v1}\BibitemShut {NoStop}%
\bibitem [{\citenamefont {He}\ \emph {et~al.}(2016)\citenamefont {He},
  \citenamefont {Zhang}, \citenamefont {Ren},\ and\ \citenamefont
  {Sun}}]{He2016DeepRecognition}%
  \BibitemOpen
  \bibfield  {author} {\bibinfo {author} {\bibfnamefont {Kaiming}\ \bibnamefont
  {He}}, \bibinfo {author} {\bibfnamefont {Xiangyu}\ \bibnamefont {Zhang}},
  \bibinfo {author} {\bibfnamefont {Shaoqing}\ \bibnamefont {Ren}}, \ and\
  \bibinfo {author} {\bibfnamefont {Jian}\ \bibnamefont {Sun}},\ }\bibfield
  {title} {\enquote {\bibinfo {title} {{Deep residual learning for image
  recognition}},}\ }in\ \href {\doibase 10.1109/CVPR.2016.90} {\emph {\bibinfo
  {booktitle} {Proceedings of the IEEE Computer Society Conference on Computer
  Vision and Pattern Recognition}}},\ Vol.\ \bibinfo {volume} {2016-December}\
  (\bibinfo  {publisher} {IEEE Computer Society},\ \bibinfo {year} {2016})\
  pp.\ \bibinfo {pages} {770--778}\BibitemShut {NoStop}%
\bibitem [{\citenamefont {Isola}\ \emph {et~al.}(2017)\citenamefont {Isola},
  \citenamefont {Zhu}, \citenamefont {Zhou},\ and\ \citenamefont
  {Efros}}]{isola2017image}%
  \BibitemOpen
  \bibfield  {author} {\bibinfo {author} {\bibfnamefont {Phillip}\ \bibnamefont
  {Isola}}, \bibinfo {author} {\bibfnamefont {Jun-Yan}\ \bibnamefont {Zhu}},
  \bibinfo {author} {\bibfnamefont {Tinghui}\ \bibnamefont {Zhou}}, \ and\
  \bibinfo {author} {\bibfnamefont {Alexei~A}\ \bibnamefont {Efros}},\
  }\bibfield  {title} {\enquote {\bibinfo {title} {Image-to-image translation
  with conditional adversarial networks},}\ }in\ \href@noop {} {\emph {\bibinfo
  {booktitle} {Proceedings of the IEEE conference on computer vision and
  pattern recognition}}}\ (\bibinfo {year} {2017})\ pp.\ \bibinfo {pages}
  {1125--1134}\BibitemShut {NoStop}%
\end{thebibliography}

%

\end{document}